\documentclass[notitlepage,nofootinbib,12pt]{revtex4-2}
\usepackage{amsfonts}
\usepackage{amsmath}
\usepackage{amssymb}
\usepackage{amsthm}
\usepackage{bm}
\usepackage{braket}
\usepackage{calc}
\usepackage{cancel}
\usepackage{color}
\usepackage{dsfont}
\usepackage[inline, shortlabels]{enumitem}
\usepackage{fancybox}
\usepackage{fancyhdr}
\usepackage{fixmath}
\usepackage{float}
\usepackage{graphicx}
\usepackage{hyperref}
\usepackage{mathrsfs}
\usepackage{mathtools}
\usepackage{slashed}
\usepackage{tcolorbox}										
\usepackage{tensor}
\usepackage{textcomp}
\usepackage{thmtools}
\usepackage{transparent}
\usepackage{verbatim}
\usepackage{xcolor}
\usepackage{xparse}
\renewcommand\thesection{\arabic{section}}
\renewcommand\thesubsection{\thesection.\arabic{subsection}}

\numberwithin{figure}{section}
\numberwithin{equation}{section}

\newcommand{\mbd}[1]{\mathbold{#1}}

\newcommand{\de}{\delta}
\newcommand{\De}{\Delta}

\newcommand{\f}{\frac}

\newcommand{\tr}{{\rm tr}}

\newcommand{\PD}{\partial}

\def\beq{\begin{equation}}
\def\eeq{\end{equation}}
\def\bea{\begin{eqnarray}}
\def\eea{\end{eqnarray}}
\def\eq#1{{Eq.~(\ref{#1})}}

\def\beq{\begin{equation}}
\def\eeq{\end{equation}}
\def\bea{\begin{eqnarray}}
\def\eea{\end{eqnarray}}

\def\beq{\begin{equation}}
\def\eeq{\end{equation}}
\def\bea{\begin{eqnarray}}
\def\eea{\end{eqnarray}}
\def\eq#1{{Eq.~(\ref{#1})}}

\newcommand{\Tr}{{\rm Tr}}

\makeatletter
\newsavebox{\@brx}
\newcommand{\llangle}[1][]{\savebox{\@brx}{\(\m@th{#1\langle}\)}%
  \mathopen{\copy\@brx\kern-0.5\wd\@brx\usebox{\@brx}}}
\newcommand{\rrangle}[1][]{\savebox{\@brx}{\(\m@th{#1\rangle}\)}%
  \mathclose{\copy\@brx\kern-0.5\wd\@brx\usebox{\@brx}}}
\makeatother

\theoremstyle{definition}

\theoremstyle{remark}

\declaretheoremstyle[
spaceabove=6pt, spacebelow=6pt,
headfont=\normalfont\bfseries,
notefont=\mdseries, notebraces={(}{)},
bodyfont=\normalfont,
postheadspace=1em,
headpunct={},
qed=$\Box$,
numbered=no
]{solstyle}

\newcommand{\prf}[1]{\begin{description}[leftmargin=0cm]\item[Proof]{ #1 $\blacksquare$}\end{description}}

\newcommand{\thry}[2]{\begin{description}[leftmargin=0cm]\item[Theorem.]{\textbf{\lowercase {(#1)}}}{ #2 }\end{description}}

\newlist{mylist}{enumerate}{1}
\setlist[mylist,1]{label=\textbf{\thesection.\arabic*},ref={\thesection.\arabic*}}
\makeatletter
\def\p@subsection{}
\makeatother
\begin{document}
\title{Topological quantization of Fractional Quantum Hall conductivity}
\author{J. Miller}
\affiliation{Ariel University, Ariel, 40700, Israel}

\author{ M.A.Zubkov }
\affiliation{Ariel University, Ariel, 40700, Israel}

\date{\today}
\begin{abstract}\noindent 
	We consider the quantum Hall effect (QHE) in a  system of interacting electrons.  
	Our formalism is valid for  systems in the presence of an external magnetic field,
	as well as for  systems with a nontrivial band topology.
	That is, the expressions for the conductivity derived are valid for both the ordinary QHE and for the intrinsic anomalous QHE. 
	The expression for the conductivity applies to external fields that may vary in an arbitrary way, 
	and takes into account disorder. 
	It is assumed that the ground state of the system is degenerate. We represent the QHE conductivity as $\frac{e^2}{h} \times \frac{\cal N}{K}$, where $K$ is the degeneracy of the ground state,
	while $\cal N$ is the topological invariant composed of the Wigner - transformed multi - leg Green functions. 
	$\cal N$ takes discrete values, which gives rise to quantization  of the fractional QHE conductivity. 
\end{abstract}
\maketitle
\tableofcontents
\section{Introduction}\noindent
The quantum Hall effect (QHE) is a phenomenon observed in electrons confined to a plane in a magnetic field. 
It
is perhaps one of the most tangible observations of quantum theory in experiment.
Originally the Hall conductivity 
was found experimentally to take integer values of the
inverse of the \emph{quantum of resistivity} or the \emph{Klitzing constant},
equal to $2\pi\hbar/e^2$
\cite{vonKlitzing:1980pdk}.
Granted the quantization of physical quantities on the atomic scale should not be surprising,
but the Hall conductivity is a macroscopic quantity in a system involving many particles.
This observation of the Hall conductivity being quantized 
can be explained theoretically by the role of topology in quantum many-body systems.
\par 
Later it was discovered that the QHE has two starkly different types:
the first is the \emph{integer quantum Hall effect} (IQHE) discussed above. The second is the \emph{fractional quantum Hall effect} (FQHE),
a phenomenon
where  the Hall conductivity can take very specific fractional values 
of the conductivity quantum. 
The most prominent fractions found experimentally
are $1/3$ and $2/5$,  but  many dozens of different fractions  have
been observed. 
Such fractional quantization of the conductivity can be accounted for by interactions between
electrons.
\par 
To explain the QHE in theoretical terms Thouless, Kohomoto, Nightingale and den Nijs (TKNN) 
derived a formula called the \emph{TKNN formula}
for the quantized Hall conductivity in their seminal paper \cite{Thouless:1982zz}.
The TKNN formula contains
an integer factor in front of the conductivity quantum,  
given by a sum of
\emph{Chern numbers} commonly referred to as the \emph{TKNN invariant}.
A pedagogical overview of the theory
can be found in references \cite{Tong:QHE,Girvin:99,goerbig2009quantum,Witten:2015aoa,Nayak:04}.
\par 
The TKNN formula
is the statement that the Hall conductivity is a topological invariant
of the system \cite{Thouless:1982zz}, 
proposed for systems subject to a
constant external magnetic field.
In this case the invariant is the TKNN invariant, related to the Hall conductivity by the TKNN formula.
In 
\cite{Thouless:1982zz} the Hall conductivity for lattice models has been expressed as an integral of the Berry curvature 
over the magnetic Brillouin zone.  
The nontrivial topology makes only integer
multiples of the Hall conductivity possible.
\par 
The TKNN invariant has two major drawbacks:
(i)
it is not defined 
for systems where interactions occur, and 
(ii) it can only be applied to systems subject to a 
constant magnetic field, or homogeneous Chern insulators.
The first is overcome through an alternative form of the TKNN invariant applicable to Chern insulators,  expressed in terms of the  two point Green
function.
In this approach the topological invariant for  systems with
interactions is obtained.
The simplest such topological invariant composed of the two point Green function is responsible for the stability of the Fermi surface in   $3 + 1 D$
systems, and has been shown to be admissible for interacting systems. Nonetheless it is still not valid for non-homogeneous systems.\par 
Progress has been made towards this goal.
It has been shown  in references \cite{Volovik:2003fe,Ishikawa:1986wx,volovik:90} 
that in the absence of electron interactions the TKNN invariant for the intrinsic anomalous QHE (AQHE) is expressible in terms of the
momentum space Green  function,
and importantly, 
this expression is unchanged when the given system is modified smoothly. 
While this representation was derived originally only for non-interacting systems, 
it has since been suggested \cite{Ishikawa:1986wx, volovik:90} that it can be generalized to describe interactions
simply by replacing 
the non-interacting two point Green function 
with the  full two point Green function that includes  corrections due to interactions.\par 
This has now been proven in the framework of $2+1D$ QED \cite{Coleman:1985zi,Lee:1985pg}. The corresponding property
is now referred to as non-renormalization of the parity anomaly in $2 + 1D$ QED by higher order terms in perturbation theory.
Recently \cite{Zhang:2019mpf} the influence of interactions on the AQHE  conductivity in  tight-binding models of
the $2 + 1D$ topological insulator and $3 
+ 1D$ Weyl semi-metals has been investigated. 
Several types of interactions were considered including contact four-fermion interactions, Yukawa and Coulomb
interactions.
It was shown that the Hall conductivity for the insulator is the topological invariant, given by a formula \cite{Ishikawa:1986wx,volovik:90} composed of the
full two-point Green's function of the interaction model.\par 
A number of new results were obtained  for the Hall conductivity in non-homogeneous systems,
in particular for systems subject to
a varying magnetic field. 
A new formula has been suggested \cite{Zubkov:2019amq}
for the Hall conductivity, constituting a topological
invariant containing the Wigner transformed two point Green  functions. 
This idea has since been generalized \cite{Fialkovsky:2019nso} to condensed-matter systems
with $Z_2$ invariance (Graphene in particular)
in the presence of elastic deformations. 
Even more, in \cite{Zhang:2019zqa} it was proved that in the presence of
interactions, the Hall conductivity is still given by the expression proposed in \cite{Zubkov:2019amq}
but with the two point Green  function replaced with that which includes interactions.
Similar methods can be used to describe the QHE in $3+1\ D$ systems.
\par 
Similar methods can be used to describe the QHE in $3+1\ D$ systems,
which opens the door to a number of research goals addressed in this article.
The first is to apply these methods to the QHE in Weyl semi-metals. 
The machinery developed for the representation of the QHE current in terms of the topological invariant composed of the Wigner transformed Green functions,
has also been extended to the chiral separation effect (CSE) \cite{Suleymanov:2020wcb}. 
However, the question about the role of interactions in the CSE still remains open. The family of non - dissipative transport effects contains more members, such as the chiral torsional effect, chiral magnetic effect, chiral vortical effect, Hall viscosity, and more. An additional  research goal is to construct the topological representation for the conductivities of these effects in terms of the Wigner - transformed Green functions. A similar representation for the fractional Hall effect also awaits investigation. In the latter case it might be necessary to build  more involved topological invariants,  composed of multi - leg Green functions. Such complicated topological invariants may also be relevant for 
considering various other topological phenomena in QCD.  \par  
First, to summarize some background theory. 
In the presence of a magnetic field the Hall conductivity is given by 
\cite{Tong:QHE}
\begin{equation}
\sigma_H = \frac{\cal N}{2\pi},
\end{equation}
where {${\cal N}$ is related to the number of filled Landau states}. (Here the conductivity is expressed in units of $e^2/\hbar$.) A similar expression for the intrinsic QHE conductivity in topological insulators 
is derived in 
\cite{Volovik:88,Ishikawa:1986wx,Volovik1} in terms of the { two-point Green function $G(p)$ (in the absence of interactions)}:
\begin{eqnarray}
{\cal N}
=  {-} \frac{ \epsilon_{ijk}}{  \,3!\,4\pi^2}\, \int d^3p\, {\rm Tr}\,
{G}(p ) \frac{\partial {G}^{-1}(p )}{\partial p_i}  \frac{\partial  {G}(p )}{\partial p_j}  \frac{\partial  {G}^{-1}(p )}{\partial p_k}
.
\label{N-0}
\end{eqnarray}
In \cite{Fialkovsky:2019dmc} the expression in 
(\ref{N-0}) was generalized to {include interactions in the case of a varying magnetic field}.
In that expression the non-homogeneous nature of the system is characterized by the full two-point Green function expressed in terms of the Wigner symbol
${G}_{W}(x,p )$.
Its explicit form
is
\begin{eqnarray}
{\cal N}
=  {-} \frac{T \epsilon_{ijk}}{ A \,3!\,4\pi^2}\, \int {{d}^3x} \int  {{d}^3p}
\, {\rm tr}\, {G}_{W}(x,p )\star 
\frac{\partial {Q}_{W}(x,p )}{\partial p^i} \star \frac{\partial  {G}_{W}(x,p )}{\partial p^j} \star \frac{\partial  {Q}_{W}(x,p )}{\partial p^k}\ ,
\label{calM2d230I}
\end{eqnarray}
where $T \to 0$ is temperature, $A$ is the area of the system, ${G}_{W}(x,p )$ is the Wigner transformation of the two-point Green's function ${\hat G} = \hat{Q}^{-1}$, while $Q_W$ is the Wigner transformation of $\hat{Q}$. The star product $\star$ entering the above expression is the Moyal product of the conventional Wigner - Weyl calculus.\par 
In \cite{Zhang:2019zqa} it is proved that in the presence of interactions the IQHE  conductivity is given by the expression of 
\cite{Zubkov:2019amq}, where the complete interacting two-point Green   function is substituted. It makes heavy use of the version of the Wigner-Weyl calculus used in these notes, which is described
fully in 
\cite{Suleymanov:2018hkm}. However, this treatment is not valid for the  FQHE . 
\par 
The absence of  correction terms to the IQHE due to Coulomb interactions and impurities (in the presence of a constant magnetic field) has been widely discussed some time ago in refs. \cite{2,3,4,5} (see also \cite{6,7,8,9,10}). In particular, in \cite{11} the systems with both   inter - electron interactions  and   disorder were considered, and the corresponding topological expression for the Hall conductivity was derived. It may be applied both to the IQHE and to the  FQHE. 
Although the expression given in \cite{11} was not applied for a practical calculation of the Hall conductivity, its topological nature itself
is proof that the FQHE in the presence of a constant magnetic field is robust with respect to  smooth modifications of the system.
This proof is important for a more practical consideration of materials with the FQHE.
Still, a substantial gap remains   between the relevant theoretical models and  real experiments in which  magnetic fields are never precisely homogeneous. 
Rather  variations of the magnetic field are always present. For the latter case, a theoretical proof that the FQHE conductivity is robust with respect to smooth modifications of the system, has still not  been given.
In this article we fill this gap  and present this very proof.\par 
In our approach we use a specific version of the Wigner - Weyl (WW) calculus developed earlier for  field theoretical models of solid state physics. 
Originally the  WW  formalism was formulated 
by Groenewold \cite{Groenewold:1946kp} and  Moyal \cite{Moyal:1949sk}
as a way
of expressing 
results of quantum mechanics in terms of classical functions in phase space instead of operators.
A  transformation from a given operator to a classical function 
exists in general called the Weyl transformation.
Later the WW formalism  was applied to quantum field theory (QFT) and condensed matter physics.
This WW calculus allows us to express the FQHE conductivity through a certain topological invariant composed of  multi - particle Green functions. 
A number of results from the WW formalism are assumed. 
For a full discussion and derivation of these results the reader is recommended to consult \cite{Suleymanov:2018hkm}. 
A summary of the background and essential results are given in \S\ref{sec_ids_weyl_symbols}.
\par 
\section{Statement of the main result}\noindent
We consider a system that has a varying number of particles but fixed chemical potential. 
A number of identities that involve creation and annihilation operators are used in this section.
Their derivations can be found in 
 Appendix \ref{sec_multi_particle_states}.  \par 
The Hamiltonian operator for the whole interacting system is
\begin{equation}
\hat H = \int d^2x\,a^{\dagger}(x) {\mathscr H}_0\,a(x)
+
\int d^2x\,d^2y\,a^{\dagger}
(x)  a(x){\mathscr V}(x-y)a^\dagger (y) a(y) + \Delta\ .\label{top_inv_var_partcl_numb_3_}
\end{equation}
Here ${\mathscr H}_0$ is the one - particle Hamiltonian  
defined with respect to the Fermi level, i.e. it is equal to the true one particle Hamiltonian minus a chemical potential, $\mu$.
The term ${\mathscr V}(x-y)$ is a potential term representing an inter - particle interaction.
If $\Delta$ is a constant, its presence in $\hat H$ does not affect observable quantities.
With this freedom, $\Delta$ is  chosen in a way that the ground state of the system has negative energy while all excited states carry positive energy values.
It is easily verified that 
\begin{align}
\hat H a^\dagger(x_1)\ldots  a^\dagger(x_N) | \emptyset \rangle
=&
\left(\sum^N_{a=1} {\mathscr H}_0(x_a)+\sum^N_{a,b=1} {\mathscr V}(x_a-x_b) +\Delta \right)a^\dagger(x_1)\ldots  a^\dagger(x_N) | \emptyset \rangle\ .
\label{top_inv_var_partcl_numb_5C_B}
\end{align}
Note that the particle-number operator, $\hat{N}$ commutes with the Hamiltonian. Therefore, 
$\hat H$ and $\hat N$ share  common  eigenstates. As a result, the ground state 
in particular  
corresponds to a definite value for the number of particles in the state. 
The ground state may be degenerate. However, at least in  non - marginal cases, a degenerate ground state  does not correspond to different eigenvalues for $\hat N$. 
\par 
The statement immediately below is the main result of this paper:
For a system with a Hamiltonian of the form of Eq. (\ref{top_inv_var_partcl_numb_3_}), the Hall conductivity in the units of $e^2/\hbar$ averaged over the system area $A$ is  
\begin{equation}\sigma_{xy} = \frac{\cal N}{2 \pi K}\ ,\label{rez1}
\end{equation}
where $K$ is the degeneracy of the ground state while ${\cal N}$ is a topologically invariant quantity given by
\begin{align}
&\!\!\!\!{\cal N} =  {-\frac{1}{2 A }} \sum_{N=0,\ldots} \frac{1}{(2\pi)^{2N} }\, \sum^N_{b,c=1 } \int  d\omega \left(\prod^N_{a=1} d^2p_{ a} \, d^2x_{ a} \right)   \epsilon^{jk} 
\nonumber\\
&\!\!\!\!
{\rm tr}\left[  {G}^{(N)}_W(\omega,\{p_a\},\{x_a\})\star \frac{\partial {Q}^{(N)}_W(\omega,\{p_a\},\{x_a\} )}{\partial \omega} 
\star \frac{\partial  {G}^{(N)}_W( \omega,\{p_a\},\{x_a\} )}{\partial p^j_{ b}} \star \frac{\partial  {Q}^{(N)}_W(\omega,\{p_a\},\{x_a\} )}{\partial p_{ c }^k} \right]
\nonumber\\ &\!\!\!\!=
{-\frac{1}{2 A}} \frac{1}{(2\pi)^{2N_0} }\, \sum^{N_0}_{b,c=1 } \int  d\omega \left(\prod^{N_0}_{a=1} d^2p_{ a} \, d^2x_{ a} \right)   \epsilon^{jk} 
\nonumber\\
&\!\!\!\!
{\rm tr}\left[  {G}^{(N_0)}_W(\omega,\{p_a\},\{x_a\})\star \frac{\partial {Q}^{(N_0)}_W(\omega,\{p_a\},\{x_a\} )}{\partial \omega} 
\star \frac{\partial  {G}^{(N_0)}_W( \omega,\{p_a\},\{x_a\} )}{\partial p^j_{ b}} \star \frac{\partial  {Q}^{(N_0)}_W(\omega,\{p_a\},\{x_a\} )}{\partial p_{ c }^k} \right].
\label{calM2d238_0}
\end{align}
Here, $N_0$ is the number of particles in the ground state of the system.
The $\star$ operator is defined as
\begin{align}
&A_W(\{x_a\},\{p_a\}) \star B_W(\{x_a\},\{p_a\}) \nonumber\\=&
A_W(\{x_a\},\{p_a\})\exp\left[\frac{i}{2} \sum^N_{a=1}\sum^2_{i=1}\left( 
\overleftarrow{\f\PD{\PD x_a^i}}\ \overrightarrow{\f\PD{\PD p^i_{a}}}
-
\overleftarrow{\f\PD{\PD p_a^i}}\ \overrightarrow{\f\PD{\PD x^i_{a}}}
\right )\right] B_W(\{x_a\},\{p_a\})\ .
\label{fxd_numb_partcls_8_}
\end{align} 
The Weyl symbols ${Q}^{(N)}_W(\omega,\{p_a\},\{x_a\})$ and ${G}^{(N)}_W(\omega,\{p_a\},\{x_a\})$ that appear in (\ref{calM2d238_0})
are functions of $2N+1$ variables
$\omega,p_1,x_1,\ldots,p_N,x_N$. 
Specifically,
${Q}^{(N)}_W(\omega,\{p_a\},\{x_a\})$  is the Weyl symbol of the operator $\hat Q^{(N)}$ defined by 
\begin{equation}
Q^{(N)}_W(\omega,\{p_a\},\{x_a\}) =\frac1{N!}
\int\left( \prod^N_{a=1} dq_ a e^{i q_a x_a}\right)
\langle  \left\{ p_a+q_a/2\right\}  |\hat{Q}^{(N)} | \left\{p_a - q_a/2\right\} \rangle
\end{equation}
where 
$\ket {\left\{p_a\right\}}$ denotes  the multi-particle state   defined by
\begin{equation}
\ket{\left\{p_a\right\}}\equiv 
a_1^\dagger (p_1)\dots a_N^\dagger (p_N)\ket{\emptyset }\ ,
\label{fxd_numb_partcls_5_}
\end{equation} 
and the operator $\hat Q^{(N)}$ is defined by
\begin{equation} 
\hat Q^{(N)} = (i \omega - \hat H)\hat \Pi_N\ ,
\label{proof_var_num_0}
\end{equation}
with $\hat H$ given explicitly in (\ref{top_inv_var_partcl_numb_3_})
being the field - theoretical Hamiltonian. 
Its matrix elements $ \langle  \{p_a\}  | \hat H |  \{q_a\}\rangle$ are between  states with $N$ particles having momenta that belong to the sets 
$ \{p_a\}$ and $ \{q_a\}$. 
Here
$\hat \Pi_N$ is the projection operator onto $N$ particle states defined by 
\begin{equation}\hat \Pi_N=\frac1{N!}\int dp_1 ... dp_N \ket{\{p_a\}}\bra{\{p_a\}}\ ,\label{projectio_operator}\end{equation}
or equivalently
\begin{equation}
\hat \Pi_N=\frac1{N!}\int dx_1\dots dx_N \, a^\dagger(x_1)\dots a^\dagger(x_N)\ket{0}\bra{0}a (x_N)\dots a (x_1)\ .
\label{the_form_of_PiN_}
\end{equation}
${G}^{(N)}_W(\omega,\{p_a\},\{x_a\})$  is the Weyl symbol of the operator $\hat G^{(N)}$ defined by 
\begin{equation}
G^{(N)}_W(\omega,\{p_a\},\{x_a\}) =\frac1{N!} \int \left(\prod^N_{a=1} dq_ a e^{iq_a x_a}\right)
\langle  \left\{ p_a+q_a/2\right\}  |\hat{G}^{(N)} | \left\{p_a - q_a/2\right\} \rangle
\end{equation}
where 
\begin{equation} 
\hat G^{(N)} = \frac1{i \omega - \hat H}\hat \Pi_N\ .
\label{proof_var_num_0_again}
\end{equation}
\par 
\section{Fixed number of different particles}
\label{fixed}\noindent
\subsection{Derivation of the expression for Hall conductance}\label{sec_der_expr_HC}\noindent 
In this section  the Hall conductivity of
a system in the presence of a varying magnetic field is discussed. 
We seek an expression for the Hall conductivity  for a system of $N$ \emph{different}  particles. 
By different it is meant that the particles themselves are different, and to that extent 
neither symmeterization or anti-symmeterization is applied to the state.
To that degree the results obtained in this section are intermediate,
however the techniques developed  
are crucial for obtaining  the main result in the next section, where a system of identical fermions is considered. 
In all expressions from now on,  $\hbar=c=1$ is assumed unless stated   explicitly otherwise.
\par 
Let  the operator  $\hat Q$ be defined as
\begin{equation}
\hat{Q} = i \omega - \hat H,\label{fxd_numb_partcls_1}
\end{equation}
where $\hat H$ is the multi - particle Hamiltonian inclusive of interaction terms:
\begin{equation}
\hat H= \sum^N_{a=1}\hat H_0(x_a,-i\partial_{x_a})
+\frac{1}{2}\sum^N_{\substack{a,b=1\\a \ne b}} V(x_a-x_b)\,,
\label{fxd_numb_partcls_2}
\end{equation}
where $\hat H_0$ is the free-particle Hamiltonian and indices  $a,b=1,\dots, N$ label the particles themselves.
We assume that the ground state (either degenerate or unique) 
corresponds to a negative value of energy,
while all excited states have positive values of energy.
This may always be achieved simply by adding a constant to the single particle Hamiltonian $\hat{H}_0$ that appears in Eq. (\ref{fxd_numb_partcls_2}).  
The inverse operator of $\hat Q$ is
\begin{equation}
\hat{G} = \frac{1}{i \omega - \hat H}
\label{fxd_numb_partcls_3}
\end{equation}
where the notation on the right of Eq.~(\ref{fxd_numb_partcls_3}) is intended to denote the inverse of the operator $i\omega-\hat H$.
\par 
The Wigner transformation of the operator $\hat{Q}$ is defined as a function of the $2N+1$ variables 
$\omega, \{p_a\}, \{x_a\}$
($a=1,...,N$)
in terms of its matrix elements in momentum space as
\begin{equation}
Q_W(\omega,\{p_a\},\{x_a\}) = \int \left(\prod\limits_{a=1}^N dq_a \ e^{iq_a x_a}\right) 
\langle   \{p_a +\tfrac{q_a}{2}\}|\hat{Q} | \{p_a -\tfrac{q_a }{2}\} \rangle\ .
\label{fxd_numb_partcls_4}
\end{equation}
Here, 
$\ket{\{p_a-\f{q_a}{2}\}}$ refers to a state comprised of $N$ different fermions, defined by
\begin{equation}
\ket{\{p_a\}}\equiv 
a_1^\dagger (p_1)\dots a_N^\dagger (p_N)\ket{0}\ ,
\label{fxd_numb_partcls_5}
\end{equation}
where the suffix ${}_1,{}_2,\dots,{}_N$ labels the particle,
following the convention in \cite{Dirac_permutation_notation}.
The operators themselves are creation and annihilation  operators of a single fermion that  satisfy the familiar anticommutation relations
\begin{equation}
\{a_r(p),a_s^\dagger(p')\}=\de(p-p')\delta_{rs}\ ,\quad  \{a_r(p),a_s(p')\}=\{a_r^\dagger(p),a_s^\dagger(p')\}=0\ . \label{fxd_numb_partcls_5A}\end{equation}
In a precisely analogous way, the Wigner symbol of $\hat G$ is
\begin{equation}
G_W(\omega,\{p_a\},\{x_a\}) = \int\left( \prod\limits_{a=1}^N dq_a\  e^{ i  q_a x_a }\right) 
\langle   \{p_a+\tfrac{q_a}{2}\}|\hat{G} |\{ p_a-\tfrac{q_a}{2}\} \rangle\ .
\label{fxd_numb_partcls_6}
\end{equation}\par 
The goal of this paper  is two fold.
Firstly,
to  show that the Hall conductivity averaged over the system area $A$  is given by 
\begin{equation}\sigma_{xy} = \frac{\cal N}{2 \pi K},\label{fxd_numb_partcls_6A}\end{equation}
where $K$ is the degeneracy of the ground state, ${\cal N}$ is given by
\begin{align}
{\cal N} =&   {-\frac{1}{2 A}} \frac{1}{\left(2\pi\right)^{2N} }\,\sum^N_{b,c=1}  \int d\omega\left(\prod^N_{a=1} d^2p_a \, d^2x_a\right)
\nonumber\\&
\epsilon^{jk}\, {\rm tr}\bigg[  {G}_W(\omega,\{p_a\},\{x_a\})\star \frac{\partial {Q}_W(\omega,\{p_a\},\{x_a\})}{\partial \omega} \star \frac{\partial  {G}_W( \omega,\{p_a\},\{x_a\})}{\partial p_{ b}^j} \nonumber\\
&
\star \frac{\partial  {Q}_W(\omega,\{p_a\},\{x_a\})}{\partial p_c^k} \bigg]
\label{fxd_numb_partcls_7}
\end{align}
and 
\begin{equation}
\star
=
\exp\left(\f i{2}\sum^N_{a=1}\overleftrightarrow{\De}_a\right),\qquad 
\overleftrightarrow{\De}_a=
\overleftarrow{\f\PD{\PD x_a^i}}\ \overrightarrow{\f\PD{\PD p_{a,i}}}
-
\overleftarrow{\f\PD{\PD p_a^i}}\ \overrightarrow{\f\PD{\PD x_{a,i}}}\ ,\qquad (i=1,2)\ .
\label{fxd_numb_partcls_8}\end{equation}
The second goal is to show  that ${\cal N}$   is topologically invariant.\par 
The identity $G_W\star Q_W=1$ 
that shall be proven below,  together with the product rule of differentiation and the 
commutative property of derivatives, allows ordinary derivatives and the $\star$ operator to be interchanged.
In particular,
\begin{equation}
\frac\PD{\PD p_b} G_W\star Q_W+ G_W\star\frac{\PD Q_W}{\PD p_b} =0\qquad (b=1,\dots,N)\ ,
\end{equation} 
such  that the following relation holds:
\begin{equation}\frac{\PD G_W}{\PD p_b}=-G_W\star\frac{\PD Q_W}{\PD p_b}\star G_W
 ,\qquad (b=1,\dots,N)
 \ .\label{proof1_1} \end{equation}
By substituting (\ref{proof1_1}) in (\ref{fxd_numb_partcls_7}) it is obtained that
\begin{align}
{\cal N} =&    {\frac{1}{2A}}\frac{1}{(2\pi)^{2N} }\,\sum^N_{b,c=1}    \int  d\omega\left(\prod^N_{a=1} d^2p_a \, d^2x_a\right)    
\nonumber\\&\epsilon^{jk}\, {\rm tr}
\bigg(   {G}_W(\omega,\{p_a\},\{x_a\})\star \frac{\partial {Q}_W(\omega,\{p_a\},\{x_a\})}{\partial \omega} \star 
G_W(\omega,\{p_a\},\{x_a\})
\nonumber\\&
\star 
\frac{\partial  {Q}_W( \omega,\{p_a\},\{x_a\})}{\partial p_{ b}^j} 
\star G_W(\omega,\{p_a\},\{x_a\})
\star \frac{\partial  {Q}_W(\omega,\{p_a\},\{x_a\})}{\partial p_c^k}  \bigg).
\label{proof1_2}
\end{align}\par 
By (\ref{fxd_numb_partcls_1}) and (\ref{fxd_numb_partcls_4}),
\begin{equation}
\frac{\partial Q_W(\omega,\{p_a\},\{x_a\})}{\partial p_b^j} = \int \left(\prod\limits_{a=1}^N dq_a \ e^{iq_a x_a}\right) 
\frac{\partial}{\partial p_b^j}\langle   \{p_a +\tfrac{q_a}{2}\}|i\omega-\hat H | \{p_a -\tfrac{q_a }{2}\} \rangle\ .
\label{dQwdp}
\end{equation}
The following identities from the standard bra-ket formalism 
may be invoked:
\begin{equation}
-i\frac\PD{\PD p_j^b}{\ket{p}}{}=
\hat x^j_b{\ket{p}}{},\qquad 
-i\frac\PD{\PD p_j^b}
{\bra{p}}{}
=
-{\bra{p}}{}\hat x^j_b\ .
\label{proof1_7}
\end{equation}
For an explanation of the origins the identities quoted in Eqs.~(\ref{proof1_7}), the reader may consult 
ref. \cite{Dirac:67}  for example). 
Accordingly (\ref{dQwdp}) becomes
\begin{align}
\frac{\partial Q_W(\omega,\{p_a\},\{x_a\})}{\partial p_b^j} = &\int \left(\prod\limits_{a=1}^N dq_a \ e^{iq_a x_a}\right) 
i\langle   \{p_a +\tfrac{q_a}{2}\}|[\hat x^j_b,\hat H ]| \{p_a -\tfrac{q_a }{2}\} \rangle
\label{dQwdp_0}\\
=& -\int \left(\prod\limits_{a=1}^N dq_a \ e^{iq_a x_a}\right) 
\langle   \{p_a +\tfrac{q_a}{2}\}|\hat J^j_b| \{p_a -\tfrac{q_a }{2}\} \rangle\  .
\label{dQwdp_1}
\end{align} 
where in the last step the relation
$ [\hat x_b^j,\hat H]=i\hat J^j_b$ was substituted, where
$\hat J^j_b$ is the operator associated with the $j$ component of the electric current. 
\emph{Note that in our calculations we define electric current in  units of electric charge $e$, and we use natural units where $c=\hbar = 1$.}
\par 
Let (\ref{dQwdp_1}) be the definition of the Weyl symbol $J^j_{b\, W}$, such that
(\ref{proof1_2}) can be cast in the form
\begin{align}
{\cal N} =&    {\frac{1}{2 A}}\frac{1}{(2\pi)^{2N} }\,\sum^N_{b,c=1}    \int  d\omega\left(\prod^N_{a=1} d^2p_a \, d^2x_a\right)    
\nonumber\\&\epsilon^{jk}\, {\rm tr}
\bigg(   {G}_W(\omega,\{p_a\},\{x_a\})\star \frac{\partial {Q}_W(\omega,\{p_a\},\{x_a\})}{\partial \omega} \star 
G_W(\omega,\{p_a\},\{x_a\})
\nonumber\\&
\star 
J^j_{b\, W} 
\star G_W(\omega,\{p_a\},\{x_a\})
\star J^k_{c\, W}  \bigg).
\label{proof1_3}
\end{align} 
After invoking the relation 
\begin{equation}
A_W(x,p) \star B_W(x,p) :=(AB)_W(x,p)=A_W(x,p) 
\exp\left(\frac{i}{2} \left( \overleftarrow{\partial_x}\overrightarrow{\partial_p}
-\overleftarrow{\partial_p}\overrightarrow{\partial_x}\right )\right) B_W(x,p)\ 
\label{proof1_2_A}
\end{equation} 
we obtain
\begin{equation}
{\cal N} =  {\frac{1}{ {2A}(2  \pi)^{2N}} }\sum^N_{b,c=1} \int   d\omega \left(\prod^N_{a=1} d^2p_a \, d^2x_a\right)
\epsilon^{jk} {\rm tr}\!
\left( \hat G \ \frac{\partial \hat Q}{\partial \omega}\  \hat G
\ \hat J_b^{j}
\ \hat G
\ \hat J_c^{k}  \right)_W.
\label{proof1_3A}
\end{equation}
Next, by  substituting the formal definition of a Weyl symbol
we find that
\begin{align}
{\cal N} = & {\frac{1}{ {2A}(2 \pi)^{2N} }}\,\sum^N_{b,c=1} \int   d\omega   \left(\prod^N_{a=1} d^2p_a\, d^2q_a\, d^2x_a\, e^{iq_a x_a}\right) 
\nonumber\\
&\epsilon^{jk} {\rm tr}
{\bra{\{p_a +\frac{q_a }{2}\}} 
\hat G \, \frac{\partial \hat Q}{\partial \omega}\,  \hat G\, \hat J_b^j \, \hat G\, \hat J_c^k
\ket{\{p_a -\frac{q_a }{2}}\}} \ ,
\label{proof1_4}
\end{align}
where $\ket{\{p_a+\frac {q_a}{2}\}}$ is the $N$ fermion state defined in 
(\ref{fxd_numb_partcls_5}) 
but with $\{p_a\}$ replaced with $\{p_a+q_a/2\}$.
Eq.~(\ref{proof1_4}) can be expressed using a complete set of antisymmetric $N$ fermion states, using the result proved in (\ref{n_ferm_state_17}), as
\begin{align}
&{\cal N} =  \frac{1}{ {2A}(2 \pi)^{2N} }\sum^N_{b,c=1} \int   d\omega   \left(\prod^N_{a=1} d^2p_a\, d^2q_a\, d^2x_a\, 
d^2p_{a,1}\,d^2p_{a,2}\,d^2p_{a,3}\,e^{iq_a x_a}\right)
\nonumber\\&
\epsilon^{jk} \,{\rm tr}
{\bra{\{p_a+\tfrac{q_a}{2}\}}\hat G \frac{\partial \hat Q}{\partial \omega} \hat G \ket{\{p_{a,1}\}}} 
{\bra{\{p_{a, 1}\}}\hat  J_b^j\ket{\{p_{a,2}\}}} 
{\bra{\{p_{a,2}\}}\hat G\ket{\{p_{a, 3}\}}} 
{\bra{\{p_{a,3}\}}\hat  J_c^k  \ket{\{p_a -\tfrac{q_a }{2}\}}} \ .
\label{proof1_5}
\end{align}
The outcome from  evaluating the $x$ integrals is a product of $\delta$ functions,
namely one factor of $(2\pi)\delta(q_a)$ corresponding to each integrand labeled by  $a$.
These $\delta$ functions make each $q_a$ integral trivial. In all, after evaluating the $x_a$ and subsequently 
the $q_a$ integrals, 
(\ref{proof1_5}) reduces to
\begin{align}
{\cal N} 
=&    \frac{1}{2A}   \sum^N_{b,c=1} \int   d\omega   \left(\prod^N_{a=1} d^2p_a\, 
d^2p_{a,1}\,d^2p_{a,2}\,d^2p_{a,3}\right)
\nonumber\\&
\epsilon^{jk}{\rm tr}
{\bra{\{p_a \}}\hat G \frac{\partial \hat Q}{\partial \omega} \hat G \ket{\{p_{a,1}\}}} 
{\bra{\{p_{a, 1}\}}\hat  J_b^j\ket{\{p_{a,2}\}}} 
{\bra{\{p_{a,2}\}}\hat G\ket{\{p_{a, 3}\}}} 
{\bra{\{p_{a,3}\}}\hat  J_c^k  \ket{\{p_a  \}}} \ .
\label{proof1_9}
\end{align}\par 
The next steps are first to plug in the explicit forms  in (\ref{fxd_numb_partcls_1}) and (\ref{fxd_numb_partcls_3}),
from which $\PD\hat Q/\PD\omega=i$.
Subsequently
complete sets of eigenstates of $\hat H$ are inserted into the expression,
assuming that each set is discrete and belongs to discrete 
eigenvalues.
This yields
\begin{align}
{\cal N} 
=&  \frac{i}{2A} \sum^N_{b,c=1} \int   d\omega   \left(\prod^N_{a=1} d^2p_a\, 
d^2p_{a,1}\,d^2p_{a,2}\,d^2p_{a,3}\,\right)
\nonumber\\&
\epsilon^{jk}\, {\rm tr}   
\sum_{E ,E^{\prime}  ,E^{\prime\prime}  }
\bra{\{p_a\} }
\frac1{i\omega-\hat H}
\ket{E }
\bra{E }
\frac1{i\omega-\hat H}
\ket{E^{\prime\prime}  }
\braket{E^{\prime\prime}  |p_{1}}
\bra{\{p_{a, 1}\}} \hat J^j_b\ket{\{p_{ a,2}\}}
\nonumber\\&
\bra{\{p_{a,2}\}}
\frac1{i\omega-\hat H}
\ket{E^{\prime}  }
\braket{E^{\prime}  |\{p_{a,3}\}}
\bra{p_{a,3}} \hat J^k_c\ket{\{p_a\}}\ .
\label{proof1_10}
\end{align}
By their very definition of being eigenstates of $\hat H$ it stands to reason that 
the inverse operator $(i\omega-\hat H)^{-1}$, denoted as $1/(i\omega-\hat H)$ in the expression above,
has the eigenvalue equation
$(i\omega-\hat H)^{-1}\ket{E}=\dfrac1{i\omega-E}\ket{E}$.
To that extent (\ref{proof1_10}) becomes 
\begin{align}
{\cal N} 
=&
\frac{i}{2 A} \sum^N_{b,c=1} \int   d\omega   \left(\prod^N_{a=1} d^2p_a\, 
d^2p_{a,1}\,d^2p_{a,2}\,d^2p_{a,3}\,\right)
\sum_{E ,E^{\prime}  ,E^{\prime\prime}  }
\frac1{i\omega-E }\frac1{i\omega-E^{\prime }  }\frac1{i\omega-E^{\prime\prime}  }
\nonumber\\&
\epsilon^{jk}\,{\rm tr}
\braket{\{p_a\} |E }
\braket{E |E^{\prime\prime}  }
\braket{E^{\prime\prime}  |\{p_{a,1}\}}
\bra{\{p_{a,1}\}} \hat J^j_b \ket{\{p_{a,2}\}}
\nonumber\\
&\braket{\{p_{a,2}\}|E^{\prime}  }
\braket{E^{\prime}  |\{p_{a,3}\}}
 \bra{\{p_{a,3}\}} \hat J^k_c \ket{\{p_a\}} \ ,
\label{proof1_11}
\end{align}
and since the trace operator  allows the freedom to change the order of inner products cyclically,  this can be written equally  as
\begin{align}
{\cal N} 
=&
\frac{i}{2 A} \sum^N_{b,c=1} \int   d\omega   \left(\prod^N_{a=1} d^2p_a\, 
d^2p_{a,1}\,d^2p_{a,2}\,d^2p_{a,3}\,\right)
\sum_{E ,E^{\prime}  ,E^{\prime\prime}  }
\frac1{i\omega-E }\frac1{i\omega-E^{\prime }  }\frac1{i\omega-E^{\prime\prime}  }
\nonumber\\&
\epsilon^{jk}\,{\rm tr}
\braket{E |E^{\prime\prime}  }
\braket{E^{\prime\prime}  |\{p_{a,1}\}}
\bra{\{p_{a,1}\}} \hat J^j_b \ket{\{p_{a,2}\}}
\nonumber\\
&\braket{\{p_{a,2}\}|E^{\prime}  }
\braket{E^{\prime}  |\{p_{a,3}\}}
 \bra{\{p_{a,3}\}} \hat J^k_c \ket{\{p_a\}} 
 \braket{\{p_a\} |E }\ .
\label{proof1_12}
\end{align}
The integrals are simplified using the identity derived in
(\ref{n_ferm_state_17}),
which fixes each of the intermediate
outer products to be the identity operator, namely
\begin{equation}\begin{aligned}
\int\left( \prod^N_{a=1}dp_a\right){\ket{\{p_a\} }\bra{\{p_a\}}}=&1 \ ,\\
\int \left(\prod^N_{a=1}dp_{a,1}\right)\ket{\{p_{a,1}\} }\bra{\{p_{a,1}\}}=&1 \ ,\\ 
\int \left(\prod^N_{a=1}dp_{a,2}\right)\ket{\{p_{a,2}\} }\bra{\{p_{a,2}\}} =&1 \ ,\\
\int \left(\prod^N_{a=1}dp_{a,3}\right)\ket{\{p_{a,3}\} }\bra{\{p_{a,3}\}} =&1 \ .
\end{aligned}\end{equation}
As well
the relation
$\braket{E |E^{\prime\prime}  }=\delta_{E ,E^{\prime\prime}  }$ eliminates the sum over $E^{\prime\prime}  $.
Putting everything together,
\begin{equation}
{\cal N} 
=  \frac{i}{2 A} \sum^N_{b,c=1}  
\sum_{E ,E^{\prime}  }
\int   d\omega
\frac1{\left(i\omega-E \right)^2\left(i\omega-E^{\prime}\right)  }
{\rm tr} \,  \epsilon^{jk} 
\bra{E } \hat J^j_b \ket{E^{\prime}  }\bra{E^{\prime}  } \hat J^k_c \ket{E }\ .
\label{proof1_13}
\end{equation}\par 
The $\omega$ integral, with  the line of the integration range $ [-\infty,\infty]$,
is evaluated by extending $l$ to be a closed semi-circle, $C$ in the upper complex plane.
Two possibilities arise:
(i)
$C$
encloses both of the points $iE$ and $iE'$ on the imaginary axis
if $E>0$ and $E'>0$, or (ii) $C$ encloses  only one of them, if say $E>0$ $E'<0$, or the converse.
If $E<0$ and $E'<0$, 
integrating over the variable  $-\omega$ (namely minus $\omega$) instead,
produces the same integral described in (i) with the same contour $C$.
In both cases the integral can be evaluated using Cauchy's integral formula.
For case (i) the integral vanishes, but for case (ii) there is a non-zero contribution.
The result is
\begin{equation}
{\cal N} 
=  -\frac{2\pi   i}{A}   \sum^N_{b,c=1} 
\sum_{E ,E^{\prime}  }
\frac{\theta( {-}E)\theta(E')}{\left(E -E^{\prime}  \right)^2}
\epsilon^{jk}  \,
{\rm tr}
\bra{E } \hat J^j_b \ket{E^{\prime }  }\bra{E^{\prime }  }\hat J^k_c \ket{E } \ .
\label{proof1_14}
\end{equation} 
Just as the presence  of the term $\theta(E)\theta(-E')$ indicates,
the integral was solved assuming that $E>0$, $E'<0$.
Had the converse been assumed, 
precisely the same formula in   (\ref{proof1_14})
would hold, seeing as interchanging $E$ and $E'$, given the trace operator in front, 
leaves the expression unaltered.
\par 
Appropriately the sum over the labels $b$ and $c$  gets absorbed by replacing
$\sum^N_{b=1}\hat J^j_b=\hat J^j$
and
$\sum^N_{c=1}\hat J^k_c=\hat J^k$, resulting in 
\begin{align}
{\cal N} 
=& - \frac{2\pi   i}{A}     
\sum_{E ,E^{\prime}  }
\frac{\theta( -E)\theta(E')}{\left(E -E^{\prime}  \right)^2}
\epsilon^{jk}  \,
{\rm tr}
\braket{E|\hat J^j|E^{\prime }  }\braket{E^{\prime } |\hat J^k|E } 
\label{proof1_16}\\[1em]
=&  -\frac{ 2\pi  i   }{K A}
\sum_{E ,E^{\prime}  }
\frac{\theta( -E)\theta(E')}{\left(E -E^{\prime}  \right)^2}
\,
{\rm tr}
\left(\braket{E|\hat J^x|E^{\prime }  }\braket{E^{\prime } |\hat J^y|E } -\braket{E|\hat J^y|E^{\prime }  }\braket{E^{\prime } |\hat J^x|E }\right)\ 
\label{proof1_17}
\end{align} 
Here, the ket $\ket{E}$ with $E<0$ is an $N$ fermion eigenstate of $\hat H$.\par 
There are two separate cases to consider. 
The first applied when there is  only one such state: the ground state of the system. In this case $K=1$.
Let this state be denoted by $\ket{0}$ instead, and let the eigenvalue of $\hat H$
that it belongs to be denoted by $E_0$, such that $\hat H\ket{0}=E_0\ket{0}$.
Correspondingly (\ref{proof1_17}) reads 
\begin{equation}{\cal N} 
=  {-} \frac{2\pi   i}{A}   
\sum_{E }
\frac1{\left(E -E_0  \right)^2}
\,
{\rm tr}
\left(\braket{0|\hat J^x|E   }\braket{E  |\hat J^y|0} -\braket{0|\hat J^y|E   }\braket{E  |\hat J^x|0}\right)\ .
\label{proof1_18}
\end{equation} 
But, as stated in (\ref{fxd_numb_partcls_6A}), $\sigma_{xy}= {\cal N}/2\pi$
(in natural units), thus
\begin{equation}
\sigma_{xy} 
=    \frac{i}{A}  
\sum_{E }
\frac1{\left(E -E_0  \right)^2}
\,
{\rm tr}
\left(\braket{0|\hat J^y|E   }\braket{E  |\hat J^x|0} -\braket{0|\hat J^x|E   }\braket{E  |\hat J^y|0}\right)\ ,
\label{proof1_19}
\end{equation} 
which is the conventional expression for Hall conductivity.\par 
Now  assume that there are $K$ degenerate ground states. In this case the system at zero temperature does not remain in a pure quantum state. 
Instead, the true state is described by a density matrix. 
In such a state, with a diagonal density matrix 
(corresponding to the probabilities of all ground states being equal) 
the conventional expression for the Hall conductivity of Eq.~(\ref{proof1_19}) has the modified form
\begin{equation}
\sigma_{xy} 
=    \frac{i}{K A}  
\sum_n\sum_{E}
\frac1{\left(E -E_0  \right)^2}
\,
{\rm tr}
\left(\braket{n|\hat J^y|E   }\braket{E  |\hat J^x|n} -\braket{n|\hat J^x|E   }\braket{E  |\hat J^y|n}\right)\ .
\label{proof1_19}
\end{equation} 
Here the sum is over the degenerate ground states $|n\rangle$.
This expression can be rewritten as $\sigma_{xy} = \dfrac{\cal N}{2\pi K} $ with $\cal N$ given by Eq. (\ref{proof1_17}).
\subsection{Proof of topological invariance}\label{sec_prf_top_inv}\noindent 
In this section the fact that Eq.(\ref{fxd_numb_partcls_7}) is a topological invariant is proved.
Let it be written  in  compact form as
\begin{align}
 {\cal N} =&  {-} \frac{1}{2 A \left(2\pi\right)^{2N} }
\int  d\omega\int \left(\prod_{a=1}^N d^2p_{a}  d^2x_{a}\right){\rm tr}
\epsilon^{3jk}  
\nonumber\\&
 \bigg[ {G}_W(\omega,\{p_a\},\{x_a\})\star \frac{\partial {Q}_W(\omega,\{p_a\},\{x_a\})}{\partial \omega} 
\star \sum_a\frac{\partial  {G}_W( \omega,\{p_a\},\{x_a\})}{\partial p^j_{ a}} \nonumber\\&\star \sum_b\frac{\partial  {Q}_W(\omega,\{p_a\},\{x_a\})}{\partial p^k_{b}} \bigg]\ .
\label{calM2d23'}
\end{align}
We introduce a convenient  notation
\begin{equation}
D_3=\frac{\partial}{\partial \omega}\ ,\qquad  D_i=\sum_a\frac{\partial}{\partial p^i_{ a }},\quad(i=1,2),
\end{equation}
and write Eq.(\ref{calM2d23'}) as
\begin{align}
{\cal N} = & {-} \frac{1}{6 A \left(2\pi\right)^{2N} }
\int  d\omega\int \left(\prod_{a=1}^N d^2p_{a}  d^2x_{a}\right)
\nonumber\\&
{\rm tr}\, 
\epsilon^{ijk}   \bigg[ {G}_W(\omega,\{p_a\},\{x_a\})\star D_i {Q}_W(\omega,\{p_a\},\{x_a\})
\star D_j {G}_W( \omega,\{p_a\},\{x_a\}) 
\nonumber\\&
\star D_k  {Q}_W(\omega,\{p_a\},\{x_a\}) \bigg]\ .
\label{calM2d23''}
\end{align}
It is instructive to write (\ref{calM2d23''})  using the identity
$D_j {G}_W( \omega,\{p_a\},\{x_a\}) = -G_W\star D_j Q_W\star G_W$,
to obtain a more 
symmetric form for ${\cal N}$ 
as
\begin{align}
{\cal N} = &  
\frac{1}{6 A \left(2\pi\right)^{2N} }
\int  d\omega\int \left(\prod_{a=1}^N d^2p_{a}  d^2x_{a}\right)
\nonumber\\&
{\rm tr}\,
\epsilon^{ijk}   
{G}_W( \omega,\{p_a\},\{x_a\})\star D_i {Q}_W( \omega,\{p_a\},\{x_a\})
\star   G_W( \omega,\{p_a\},\{x_a\})\star D_j Q_W( \omega,\{p_a\},\{x_a\})
\nonumber\\&
\star G_W( \omega,\{p_a\},\{x_a\})     \star D_k {Q}_W( \omega,\{p_a\},\{x_a\}) 
\nonumber\\[1em]
= &   
\frac{1}{6 A \left(2\pi\right)^{2N}}
\int  d\omega\int \left(\prod_{a=1}^N d^2p_{a}  d^2x_{a}\right)
{\rm tr}\,
\epsilon^{ijk}  \bigg[  K_{i,W}( \omega,\{p_a\},\{x_a\})\star K_{j,W}( \omega,\{p_a\},\{x_a\})
\nonumber\\
&
\star   K_{k,W}( \omega,\{p_a\},\{x_a\}) \bigg]
\ ,
\label{prf_top_inv_1}
\end{align}
where $K_{i,W}\equiv {G}_W( \omega,\{p_a\},\{x_a\})\star D_i {Q}_W( \omega,\{p_a\},\{x_a\})$.
Now it is straightforward to apply an  arbitrary variation of the Green function
$G \rightarrow G + \delta G$.
The resulting variation of ${\cal N}$ is
\begin{align}
\delta {\cal N} 
=&  
\frac{3}{6 A \left(2\pi\right)^{2N}}
\int  d\omega\int \left(\prod_{a=1}^N d^2p_{a}  d^2x_{a}\right)
{\rm tr}\,
\epsilon^{ijk}  \bigg[  K_{i,W}( \omega,\{p_a\},\{x_a\})\star K_{j,W}( \omega,\{p_a\},\{x_a\})
\nonumber\\&
\star   K_{k,W}( \omega,\{p_a\},\{x_a\}) \bigg]
\nonumber\\[1em]
=&   
\frac{3}{6 A \left(2\pi\right)^{2N}}
\int  d\omega\int \left(\prod_{a=1}^N d^2p_{a}  d^2x_{a}\right)
{\rm tr}\,
\epsilon^{ijk}  \bigg[\delta  K_{i,W}( \omega,\{p_a\},\{x_a\})\star K_{j,W}( \omega,\{p_a\},\{x_a\})
\nonumber\\&
\star   K_{k,W}( \omega,\{p_a\},\{x_a\}) \bigg]
\end{align}
where $\delta  K_{i,W}( \omega,\{p_a\},\{x_a\})=\de G_W\star D_iQ_W+G_W\star D_i\de Q_W
=- G_W\star \de Q_W\star G_W\star D_iQ_W+G_W\star D_i\de Q_W
$.
Putting everything together we find that 
\begin{align}
\delta {\cal N} 
=&   - \frac{3}{6 A \left(2\pi\right)^{2N}}
\int  d\omega\int \left(\prod_{a=1}^N d^2p_{a}  d^2x_{a}\right)
{\rm tr}\,
\epsilon^{ijk} 
\nonumber\\& 
\bigg[
\left(-G_W\star \de Q_W\star G_W\star D_iQ_W+G_W\star D_i\de Q_W\right)\star G_W\star D_jQ_W\star G_W\star D_kQ_W\ .
\bigg]\label{prf_top_inv_2}
\end{align}
The trace can be re-ordered as
{\small
\[\begin{aligned}
&{\rm tr}\,
\epsilon^{ijk} 
\left[
\left(-G_W\star \de Q_W\star G_W\star D_iQ_W+G_W\star D_i\de Q_W\right)\star G_W\star D_jQ_W\star G_W\star D_kQ_W
\right]\\
=&
{\rm tr}\,
\epsilon^{ijk} 
\left[
\left(-\de Q_W\star G_W\star D_iQ_W+ D_i\de Q_W\right)\star G_W\star D_jQ_W\star G_W\star D_kQ_W\star G_W
\right]
\\
=&
{\rm tr}\,
\epsilon^{ijk} 
\left[
\left(-\de Q_W\star G_W\star D_iQ_W\star G_W+ D_i\de Q_W\star G_W\right)\star D_jQ_W\star G_W\star D_kQ_W\star G_W
\right]
\\
=&
-{\rm tr}\,
\epsilon^{ijk} 
\left[
\left(\de Q_W \star D_iG_W+ D_i\de Q_W\star G_W\right)\star D_jQ_W\star D_kG_W
\right]
\\
=&
-{\rm tr}\,
\epsilon^{ijk} 
\left[
D_i\left(\de Q_W \star  G_W\right)\star D_jQ_W\star D_kG_W
\right]
\\
=&
-{\rm tr}\,
\epsilon^{ijk} 
D_i\left[
\left(\de Q_W \star  G_W\right)\star D_jQ_W\star D_kG_W
\right]\ .
\end{aligned}\]
}
Finally, by  substituting this back into (\ref{prf_top_inv_2}) we end up with
\begin{align}
\delta {\cal N} 
=&  +
\frac{3}{6 A \left(2\pi\right)^{2N} }
\int  d\omega\int \left(\prod_{a=1}^N d^2p_{a}  d^2x_{a}\right)
{\rm tr}\,
\epsilon^{ijk} 
D_i\bigg[
\left(\de Q_W \star  G_W\right)\star D_jQ_W\star D_kG_W
\bigg].
\label{prf_top_inv_3}
\end{align}
This is of course zero, since  the integrand comprises a total derivative.
In conclusion,
\begin{equation}
\de{\cal N}=0\ .\label{prf_top_inv_4}
\end{equation}
The implication is that ${\cal N}$ is topologically invariant.
\section{System with varying number of identical particles and fixed chemical potential}
\label{var}\noindent
\subsection{Derivation of the topological expression}\label{sec_derivatn_of_toplgcl_exprssn}\noindent 
Suppose that the system has a varying number of particles but a fixed chemical potential. 
If the ground state of the system is non-degenerate, the Hall conductivity is still given by the familiar Kubo expression
\begin{equation}
\sigma_{12}=\frac{i}{A}\sum_{n\ne 0}\frac{\langle 0 | \hat  J_2 |n\rangle \langle n|\hat J_1| 0 \rangle
-\langle 0 | \hat J_1 |n\rangle \langle n| \hat J_2 | 0 \rangle}{(E_n-E_{0})^2}\,.\label{kubo}
\end{equation}
If the ground state is degenerate, then the linear response of the system 
(remaining in thermal equilibrium at zero temperature) to an external electric field gives rise to the following expression for the Hall conductivity: 
\begin{equation}
\sigma_{12}=\frac{i}{KA}\sum_{k=1}^K\sum_{n}\frac{\langle 0_k | \hat  J_2 |n\rangle \langle n|\hat J_1| 0_k \rangle
-\langle 0_k | \hat J_1 |n\rangle \langle n| \hat J_2 | 0_k \rangle}{(E_n-E_{0})^2}\,.\label{kubo_d}
\end{equation}
Here the sum $\sum\limits_{k=1}^K$ is over the $K$ degenerate ground states $| 0_k \rangle$,
while the sum $\sum\limits_{n}$ is over excited states of the system.
\par 
For a sufficiently weak magnetic field,
\begin{equation}
\hat J_i=\frac{1}{i}[ \hat x _i,\hat H] \ .\label{top_inv_var_partcl_numb_1}
\end{equation}
Here the Hamiltonian operator for the case of varying particle number is
\begin{equation}
\hat H = \int d^2x\,a^{\dagger}(x)( { H}_0-\mu)\,a(x)
+
\int d^2x\,d^2y\,a^{\dagger}(x)  a(x){\mathscr V}(x-y)a^\dagger (y) a(y)
+\Delta \ \label{top_inv_var_partcl_numb_3}
\end{equation}
with  $\Delta$  a constant term  chosen in such a way that the ground states (i.e. the states with minimal values of the total energy) have negative energy,
while all excited states belong to energy eigenvalues  that are positive.
The position operator is
\begin{equation}
\hat{x}^i = \int d^2x \,a^{\dagger}(x ) x^i a(x ) \ .\label{top_inv_var_partcl_numb_2}
\end{equation}
The physical meaning of this operator is that it is a measure of spatial inhomogeneity. 
Namely, for a system  in which particles are distributed homogeneously in space, its value is equal to zero. 
At the same time, the value of this operator is nonzero if the particles are distributed in a non - uniform way. 
In the marginal case, when $N$ particles are placed around the coordinates of vector $X^i$, the corresponding eigenvalue of $\hat{x}^i$ is close to $N X^i$.  
\par 
In the following we denote 
$$
{\mathscr H}_0 = {H}_0-\mu
$$
where $\mu$ is the chemical potential.
These forms for $\hat H$ and $\hat x$ in
(\ref{top_inv_var_partcl_numb_3}) and (\ref{top_inv_var_partcl_numb_2}) 
are justified as follows.
From the  expression  in (\ref{top_inv_var_partcl_numb_3}),
the operator $\hat H$
acts on states comprised of $N$ quanta of energy
as described by (\ref{top_inv_var_partcl_numb_5C_A}) below:
\begin{align}
\left(\hat H -\Delta\right)a^\dagger(x_1)\ldots  a^\dagger(x_N) | \emptyset \rangle 
=& 
\int d^2x\,a^{\dagger}(x) {\mathscr H}_0(x)\,a(x)a^\dagger(x_1)\ldots  a^\dagger(x_N) | \emptyset \rangle
\nonumber\\
&
+
\int d^2x\,d^2y\,a^{\dagger}(x)  a(x){\mathscr V}(x-y)a^\dagger (y) a(y)a^\dagger(x_1)\ldots  a^\dagger(x_N) | \emptyset \rangle\ .
\label{top_inv_var_partcl_numb_5C_A}
\end{align}
The right-hand side can be recast by re-arranging the order of creation and annihilation operators.
Repeated use of the anticommutation relation in (\ref{fxd_numb_partcls_5A}), namely
$\{a(x),a^\dagger(x_1)\}=\de(x-x_1)$,
leads to
\begin{align}
&a^{\dagger}(x) \,a(x)a^\dagger(x_1)\ldots  a^\dagger(x_N)\ket{\emptyset}
\nonumber\\[1em]
=&\de(x-x_1)
a^{\dagger}(x) a^\dagger(x_2)\ldots  a^\dagger(x_N)\ket{\emptyset}
-\de(x-x_2)
a^{\dagger}(x) a^\dagger(x_1)a^\dagger(x_3)\ldots  a^\dagger(x_N)\ket{\emptyset}
+\cdots
\nonumber\\[1em]
&\cdots +(-1)^{N-1}\de(x-x_N)
a^{\dagger}(x) a^\dagger(x_1)a^\dagger(x_2)\ldots  a^\dagger(x_{N-1})\ket{\emptyset}
\nonumber\\[2em]
=&\left(\sum^N_{a=1}\de(x-x_a)(-1)^{a-1}\right)
a^{\dagger}(x_1) a^\dagger(x_2)\ldots  a^\dagger(x_N)
\ket{\emptyset}\ .
\label{3.6}
\end{align}
This in turn implies that
\begin{align}
&a^{\dagger}(x) \,a(x)a^{\dagger}(y) \,a(y)a^\dagger(x_1)\ldots  a^\dagger(x_N)\ket{\emptyset}
\nonumber\\
=&\left(\sum^N_{a,b=1}\de(x-x_a)\de(y-x_b)(-1)^{a+b}\right)
a^{\dagger}(x_1) a^\dagger(x_2)\ldots  a^\dagger(x_N)
\ket{\emptyset}\ .
\label{3.6_again}
\end{align}
Thus, by substituting (\ref{3.6}) and (\ref{3.6_again}) in 
(\ref{top_inv_var_partcl_numb_5C_A}) the outcome is the eigenvalue equation
\begin{align}
\hat H a^\dagger(x_1)\ldots  a^\dagger(x_N) | \emptyset \rangle
=&
\left(\sum^N_{a=1} {\mathscr H}_0(x_a)+\sum^N_{a,b=1} {\mathscr V}(x_a-x_b)+\Delta \right)a^\dagger(x_1)\ldots  a^\dagger(x_N) | \emptyset \rangle\ .
\label{top_inv_var_partcl_numb_5C_B}
\end{align}\par 
Based on the expression in (\ref{top_inv_var_partcl_numb_2}) the operator $\hat x$ acts on $N$ particle  states  as
\begin{align}
\hat x^ia^\dagger(x_1) \dots a^\dagger(x_N) | \emptyset \rangle=&\int dx\, x^i a^\dagger (x) a(x)a^\dagger(x_1)\dots  a^\dagger(x_N)
| \emptyset \rangle
\nonumber\\
=&  
\int dx\, x^i\sum^N_{a=1}(-1)^a\delta(x-x_a) a^\dagger(x_1)\dots  a^\dagger(x_N)
| \emptyset \rangle
\nonumber\\
=&
\sum^N_{a=1}x^i_a a^\dagger(x_1)\dots  a^\dagger(x_N)
| \emptyset \rangle\ ,\label{3.9}
\end{align}
where in the second step the identity in (\ref{3.6}) was used.
Eq.~(\ref{3.9}) can be re-cast as
\begin{equation}
\hat x^ia^\dagger(x_1) \dots a^\dagger(x_N) | \emptyset \rangle
=
x^i a^\dagger(x_1)\dots  a^\dagger(x_N)
| \emptyset \rangle\ ,\label{3.9_again}
\end{equation}
where $x^i\equiv\sum^N_{a=1}x^i_a$ is the $i$ component of the vector sum of the position vectors of all of the $N$ particles.
\par 
By substituting (\ref{top_inv_var_partcl_numb_1}) in (\ref{kubo}) we arrive at
\begin{equation}
\sigma_{12}=\frac{1}{iKA}\sum^K_{k=1}\sum_{n\ne 0_k}\frac{\langle 0_k | [ \hat{x}_2,{\hat {H}}] |n\rangle\langle n|  [ \hat{x}_1,{\hat {H}}] | 0_k\rangle
-\langle 0_k |  [  \hat{x}_1,{\hat {H}}] |n\rangle\langle n| [ \hat{x}_2,{\hat {H}}] | 0_k\rangle}{(E_n-E_0)^2}\,.\label{sig1}
\end{equation}
Say that the ground states of the system, $|0_k\rangle$ is a sum over states containing  $N = 0, 1, 2, \ldots $ particles,
with the form
\begin{equation}
|0_k \rangle  \equiv \sum_N \frac1{\sqrt{N!}}\int d^2x_1\ldots d^2 x_N\psi_N^{(0_k)}(x_1,\ldots,x_N) a^\dagger(x_1)\ldots  a^\dagger(x_N) | \emptyset \rangle
\label{top_inv_var_partcl_numb_4}
\end{equation}
where  $| \emptyset \rangle $ denotes the vacuum state, in which there are no particles at all.
Since the number operator commutes with the Hamiltonian, it is reasonable to suppose 
that $\psi_N^{(0_k)}(x_1,\ldots,x_N)$ is non-zero only for the value of $N = N_{0_k}$. 
Moreover, $N_{0_k}$ does not depend on $k$ except when the marginal case is encountered, 
when a particle can be added to the system without changing the energy of the system.
Excited states are decomposed in a similar fashion:
\begin{equation}
|n \rangle  \equiv \sum_N\frac1{\sqrt{N!}} \int d^2x_1\ldots d^2 x_N\psi_N^{(n)}(x_1,\ldots,x_N) a^\dagger(x_1)\ldots  a^\dagger(x_N) | \emptyset \rangle.
\label{top_inv_var_partcl_numb_5}
\end{equation}
In the same way, only one value of $N$  contributes to this sum. 
Nevertheless in the discussion to follow, sums over $N$ are retained in the expressions
in order to have expressions in a forms that are easily generalized to the cases  where the Hamiltonian  does not conserve particle number.
\par
The normalization condition  $\braket{n|n}=1$ is assumed, implying that
\begin{align}
&\sum_{N,N'}\frac1{\sqrt{N!}\sqrt{N'!}} 
\int d^2x_1\ldots d^2 x_N\ 
d^2x'_1\ldots d^2 x'_{N'}
\psi_{N'}^{(n)\dagger}(x_1',\ldots,x_{N'}')\psi_N^{(n)}(x_1,\ldots,x_N) 
\nonumber\\&
\langle\emptyset |a(x'_{N'})\ldots  a(x'_{1}) a^\dagger(x_1)\ldots  a^\dagger(x_N) | \emptyset \rangle\nonumber\\=&1\ .
\label{top_inv_var_partcl_numb_5A}
\end{align}
We invoke (\ref{n_ferm_state_6}) in order to write this as
\begin{align*}
&\sum_{N}\frac1{N!}
\int d^2x_1\ldots d^2 x_N\ 
d^2x'_1\ldots d^2 x'_{N}
\psi_{N }^{(n)\dagger}(x_1',\ldots,x_N')\psi_N^{(n)}(x_1,\ldots,x_N) 
\nonumber\\& \sum_{{i_1\dots i_N}}
\epsilon^{i_1\dots i_N}\delta(x_1-x'_{i_1})\dots \delta(x_N-x'_{i_N})\nonumber\\=&1
\label{top_inv_var_partcl_numb_5B}
\end{align*}
or equally
\begin{equation}
\sum_{N}\frac1{N!}\sum_{{i_1\dots i_N}}
\int d^2x_1\ldots d^2 x_N\ 
\psi_{N}^{(n)\dagger}(x_{i_1},\ldots,x_{i_N})\psi_N^{(n)}(x_1,\ldots,x_N) 
\epsilon^{i_1\dots i_N} 
=1
\ .
\label{top_inv_var_partcl_numb_5C}
\end{equation}
The factor $1/N !$ is cancelled by the antisymmetric sum over $N !$ identical terms, through the contraction
with the Levi-Civita symbol, to finally yield
\begin{equation}
\sum_{N} 
\int d^2x_1\ldots d^2 x_N\ 
\psi_{N}^{(n)\dagger}(x_1,\ldots,x_N)\psi_N^{(n)}(x_1,\ldots,x_N) 
=1
\ .
\end{equation}
\par 
The total Fock space,  $ \mbd H$ of the system may be decomposed into a direct sum of sub-spaces $\mbd {  H}^{(N)}$,
each containing a fixed number of particles as
\begin{equation}
{\mbd   H} = {  \mbd H}^{(0)} \cup \ldots{ \mbd  H}^{(N)} \cup \ldots 
\end{equation}
The functions $\psi ^{(n)}_N$ are defined on ${\mbd  H}^{(N)}$.  
In a case where only one value of $N$ contributes to $|n\rangle$, the latter may be denoted as  $\ket{\psi^{(n)}_N}$. In line with
the convention of notation in standard quantum mechanics, the coordinate representation of 
the functions $\psi^{(n)}_N(x_1,\ldots,x_N)$ may be expressed as the inner product of  $\ket{\psi^{(n)}_N}$
with a basis of coordinate eigenstates as
\begin{equation}
\psi _N^{(n)}(x_{ 1},\ldots,x_{N}) =\frac1{\sqrt{N!}} \langle  x_{ 1},\ldots,x_{N} |\psi ^{(n)}_N\rangle\ .
\label{3.16}
\end{equation}
\par 
In the framework of this structure of the Fock space, the goal 
is to derive
a new expression for the Hall conductivity, starting from (\ref{top_inv_var_partcl_numb_4}), as a sum over terms where each term
is the contribution coming from a state with $N$ particles. For this purpose let the 
$N$ -particle Hamiltonian be defined as
\begin{equation}\label{HCP}
\hat{\mathscr {H}}_N = \sum_a ({ H}_0(x_{ a },-i\partial_{x_{ a}})-\mu)
+\frac{1}{2}\sum_{a \ne b} {\mathscr V}(x_a-x_b)+\Delta \, ,
\end{equation}
such that (\ref{top_inv_var_partcl_numb_5C_B}) reads
\begin{equation}\label{3.18}
\hat Ha^\dagger(x_1)\dots a^\dagger(x_N)
| \emptyset \rangle
=
\hat{\mathscr {H}}_N
a^\dagger(x_1)\dots a^\dagger(x_N)
| \emptyset \rangle\ .
\end{equation}\par 
By a similar set of steps as \S\ref{sec_der_expr_HC},
it can be shown that 
$\sigma_{12} = \dfrac{{\cal N}}{2 \pi K}$
where ${\cal N}$ is given by
\begin{align}
{\cal N} = & {-\frac{1}{2A}}\sum_{N=0,\ldots} \frac{1}{(2\pi)^{2N} }\, \sum^N_{b,c=1 } \int  d\omega \left(\prod^N_{a=1} d^2p_{ a} \, d^2x_{ a} \right)   \epsilon^{jk} 
\nonumber\\
&
{\rm tr}\Big[  {G}^{(N)}_W(\omega,\{p_{ a }\},\{x_{ a }\})\star \frac{\partial {Q}^{(N)}_W(\omega,\{p_a\},\{x_a\} )}{\partial \omega} 
\star \frac{\partial  {G}^{(N)}_W( \omega,\{p_a\},\{x_a\} )}{\partial p^j_{ b}} \star \frac{\partial  {Q}^{(N)}_W(\omega,\{p_a\},\{x_a\} )}{\partial p_{ c }^k} \Big]
\nonumber\\
= & {-\frac{1}{2A}} \frac{1}{(2\pi)^{2N_0} }\, \sum^{N_0}_{b,c=1 } \int  d\omega \left(\prod^{N_0}_{a=1} d^2p_{ a} \, d^2x_{ a} \right)   \epsilon^{jk} 
\nonumber\\
&
{\rm tr}\Big[  {G}^{(N_0)}_W(\omega,\{p_{ a }\},\{x_{ a }\})\star \frac{\partial {Q}^{(N_0)}_W(\omega,\{p_a\},\{x_a\} )}{\partial \omega} 
\star \frac{\partial  {G}^{(N_0)}_W( \omega,\{p_a\},\{x_a\} )}{\partial p^j_{ b}}
\nonumber\\
&\star \frac{\partial  {Q}^{(N_0)}_W(\omega,\{p_a\},\{x_a\} )}{\partial p_{ c }^k} \Big]
\label{calM2d238}
\end{align}
where  $\star$ is given by (\ref{fxd_numb_partcls_8}),
while ${Q}^{(N)}_W(\omega, \{p_a\},\{x_a\})$ and ${G}^{(N)}_W(\omega,\{p_a\},\{x_a\})$ are functions of $2N+1$ variables
$\omega,p_1,x_1,\ldots,p_N,x_N$. These functions are the Weyl symbols of the corresponding operators:
\begin{equation}
Q^{(N)}_W(\omega,\{p_{a}\},\{x_{ a}\}) =\frac1{N!}
\int\left( \prod^N_{a=1} dq_ a e^{i q_a x_a}\right)
\langle   \{p_a+q_a/2\}  |\hat{Q}^{(N)} | \{p_a - q_a/2 \}\rangle\label{def_QN_W}
\end{equation}
where  the multi-particle state $\ket{ \{p_a\}}$  is defined above in 
(\ref{fxd_numb_partcls_5}), and similarly
\begin{equation}
G^{(N)}_W(\omega,\{p_{a}\},\{x_{ a}\}) =\frac1{N!} \int \left(\prod^N_{a=1} dq_ a e^{iq_a x_a}\right)
\langle   \{p_a+q_a/2\}  |\hat{G}^{(N)} | \{p_a - q_a/2 \}\rangle\label{def_GN_W}
\end{equation}
where 
\begin{equation} 
\hat Q^{(N)} = (i \omega - \hat H)\hat \Pi_N,\qquad \hat G^{(N)} = \frac1{i \omega - \hat H}\hat \Pi_N,
\label{proof_var_num_0}
\end{equation}
and where 
\begin{equation}\hat \Pi_N=\frac1{N!}\int\left( \prod^N_{a=1} dp_ a \right)\ket{\{p_a\}} \bra{\{p_a\}}\  \label{projectio_operator}\end{equation}
is the projector onto $N$-particle states,
with $\hat H$ given explicitly in (\ref{top_inv_var_partcl_numb_3}),
being the field - theoretical Hamiltonian. 
Its matrix elements $ \langle  \{p_a\}  | \hat H |  \{q_a\}\rangle$ are between  states with $N$ particles  having momenta that belong to the sets 
$ \{p_a\}$ and $ \{q_a\}$.
The proof that (\ref{sig1}) is equivalent to (\ref{calM2d238}) is the topic of \S\ref{sec_proof_3.28_is_equivalent_to_3.29}. The proof that the given expression for $\cal N$ is
a
topological invariant closely follows the proof given in Sect. \ref{sec_prf_top_inv} for the case of different particles. 
The presence of identical particles results in extra factors $1/N!$ and an  antisymmetric basis of states, 
but this does not affect the logic behind the derivation.
\par 
\subsection{The proof of the statement that (\ref{sig1}) is equivalent to  (\ref{calM2d238})}
\label{sec_proof_3.28_is_equivalent_to_3.29}\noindent 
The proof of this statement that 
Eq.~(\ref{sig1}) is equivalent to  Eq.~(\ref{calM2d238})
proceeds along analogous lines to the argument in 
\S\ref{sec_der_expr_HC} as mentioned above.
\par 
From the definition of $G_W$ given in (\ref{fxd_numb_partcls_3}) with (\ref{fxd_numb_partcls_6}), its derivative with respect to $p_b^j$ is
\begin{equation}
\frac{\partial G_W(\omega,\{p_a\},\{x_a\})}{\partial p_b^j} = \frac1{N!}\int \left(\prod\limits_{a=1}^N dq_a \ e^{iq_a x_a}\right) 
\frac{\partial}{\partial p_b^j}\langle   \{p_a +\tfrac{q_a}{2}\}| (i\omega-\hat H )^{-1} | \{p_a -\tfrac{q_a }{2}\} \rangle\ .
\label{dGwdp}
\end{equation}
For a one - particle state we have 
\cite{Dirac:67} 
\begin{equation}
-i\frac\PD{\PD p_j}{\ket{p}}{}=
\hat x^j{\ket{p}}{},\qquad 
-i\frac\PD{\PD p_j}
{\bra{p}}{}
=
-{\bra{p}}{}\hat x^j\ ,
\label{proof1_7_yet_again}
\end{equation}
At the same time a multi - particle state has the form
\begin{equation}
|\{p\}\rangle = \frac{1}{\sqrt{N!}} \sum_{i_1 \ldots  i_N} \epsilon^{i_1\ldots i_N} |p_{i_1}\rangle \otimes \ldots  \otimes |p_{i_N}\rangle = a^\dagger_{1}  \ldots  a^\dagger_{N} | \emptyset \rangle\ .\label{eq_a_multi-particle_state}
\end{equation}
The action of an annihilation operator on a multi-particle state of the form (\ref{eq_a_multi-particle_state}) is
\begin{equation}
a_l|\{p\}\rangle = \frac{1}{\sqrt{N!}}\sum_{k=1\ldots N} (-1)^{k+1} \sum_{i_1 \ldots  i_N} \epsilon^{i_1\ldots i_N} |p_{i_1}\rangle \otimes \ldots \otimes \langle p_l|p_{i_k}\rangle \otimes \ldots  \otimes |p_{i_N}\rangle 
\end{equation}
while a creation operator acts on (\ref{eq_a_multi-particle_state})  as  
\begin{equation}
a^\dagger_l|\{p\}\rangle = \frac{1}{\sqrt{(N+1)!}}\sum_{k=1\ldots N} (-1)^{k+1} \sum_{i_1 \ldots  i_N} \epsilon^{i_1\ldots i_N} |p_{i_1}\rangle \otimes \ldots \otimes | p_l \rangle \otimes |p_{i_k}\rangle \otimes \ldots  \otimes |p_{i_N}\rangle =a^\dagger_l a^\dagger_{1}  \ldots  a^\dagger_{N} | \emptyset \rangle
\ ,
\end{equation}
(here $|p\rangle \otimes |\emptyset\rangle \equiv |p\rangle$).
It follows that a derivative acts on (\ref{eq_a_multi-particle_state}) as
\begin{equation}
\sum_{b=1\ldots N} \frac{\partial}{\partial p_b^j}|\{p\}\rangle = \frac{1}{\sqrt{N!}}\sum_{k=1\ldots N} \sum_{i_1 \ldots  i_N} \epsilon^{i_1\ldots i_N} \ldots  \otimes \frac{\partial}{\partial p_{i_k}^j}|p_{i_k}\rangle \otimes \ldots  \otimes |p_{i_N}\rangle = i \hat{x}^j |\{p\}\rangle
\end{equation}
The last  equality is established as follows:
\begin{eqnarray}
i \hat{x}^j |\{p\}\rangle & = & 
\frac{1}{\sqrt{N!}}\int dp a^\dagger(p) \left(-i\frac{\partial}{\partial p^j}\right)  a(p) \sum_{i_1 \ldots  i_N} 
\epsilon^{i_1\ldots i_N} \ldots  \otimes |p_{i_k}\rangle \otimes \ldots  \otimes |p_{i_N}\rangle \nonumber\\ 
&=& \frac{1}{\sqrt{N!}}\int dp a^\dagger(p) \left(-i\frac{\partial}{\partial p^j}\right) \sum_{k=1\ldots N} (-1)^{k+1} \sum_{i_1 \ldots  i_N}
\epsilon^{i_1\ldots i_N} \ldots  \otimes \langle p |p_{i_k}\rangle \otimes \ldots  \otimes |p_{i_N}\rangle \nonumber\\ 
&=& \frac{1}{\sqrt{N!}}\int dp a^\dagger(p) \sum_{k=1\ldots N} (-1)^{k+1} \sum_{i_1 \ldots  i_N} 
\epsilon^{i_1\ldots i_N} \ldots  \otimes \left( i\frac{\partial}{\partial p^j_{i_k}}\right)\langle p |p_{i_k}\rangle \otimes \ldots  \otimes |p_{i_N}\rangle \nonumber\\
&=& \frac{1}{\sqrt{N!}}  \sum_{k=1\ldots N} (-1)^{k+1} \left( i\frac{\partial}{\partial p^j_{i_k}}\right)a^\dagger(p_{i_k}) \sum_{i_1 \ldots  i_N} 
\epsilon^{i_1\ldots i_N} \ldots  \otimes  |p_{i_{k-1}}\rangle\otimes |p_{i_{k+1}}\rangle \otimes \ldots  \otimes |p_{i_N}\rangle \nonumber\\ 
& = & \frac{1}{\sqrt{N!}}\sum_{k=1\ldots N} \sum_{i_1 \ldots  i_N} 
\epsilon^{i_1\ldots i_N} \ldots  \otimes \frac{\partial}{\partial p_{i_k}^j}|p_{i_k}\rangle \otimes \ldots  \otimes |p_{i_N}\rangle\ .
\end{eqnarray}
After summation over $b$ Eq. (\ref{dGwdp}) becomes
\begin{align}
\sum_{b=1\ldots N}\frac{\partial G_W(\omega,\{p_a\},\{x_a\})}{\partial p_b^j} = &\frac1{N!}\int \left(\prod\limits_{a=1}^N dq_a \ e^{iq_a x_a}\right) 
i\langle   \{p_a +\tfrac{q_a}{2}\}|\, [\, \hat x^j\, , \,(i\omega-\hat H )^{-1} \,]\, | \{p_a -\tfrac{q_a }{2}\} \rangle
\label{dGwdp_0}
\end{align} 
Based on the identity 
\begin{equation}[\hat B,\hat A^{-1}]\hat A=-\hat A^{-1}[\hat B, \hat A] \qquad\Leftrightarrow\qquad 
 [\hat B,\hat A^{-1}]=-\hat A^{-1}[\hat B, \hat A] \hat A^{-1}\ ,
 \label{proof1_20}\end{equation}
then
\begin{equation}[\, \hat x^j\, , \,(i\omega-\hat H )^{-1} \,]
=-(i\omega-\hat H )^{-1}
[\, \hat x^j\, , \,(i\omega-\hat H )  \,]\, (i\omega-\hat H )^{-1}\label{proof1_21}
,\end{equation}
or equivalently
\begin{equation}
 [\, \hat x^j\, , \,(i\omega-\hat H )^{-1} \,]
=-\hat G\, 
[\, \hat x^j\, , \,\hat Q  \,]\, \hat G
=\hat G\, 
[\, \hat x^j\, , \,\hat H  \,]\, \hat G\ .\label{proof1_22}
\end{equation}
By substituting (\ref{proof1_22})
in (\ref{dGwdp_0})
we obtain
\begin{align}
\sum_{b=1...N}\frac{\partial G_W(\omega,\{p_a\},\{x_a\})}{\partial p_b^j} = &\frac1{N!}\int \left(\prod\limits_{a=1}^N dq_a \ e^{iq_a x_a}\right) 
i\langle   \{p_a +\tfrac{q_a}{2}\}|\, \hat G\, 
[\, \hat x^j\, , \,\hat H  \,]\, \hat G\, | \{p_a -\tfrac{q_a }{2}\} \rangle
\label{dGwdp_1}
\\
=&i\left(\hat G\, 
[\, \hat x^j\, , \,\hat H  \,]\, \hat G\right)_W
\label{dGwdp_2}
\end{align} 
where the term $\left(\hat G\, 
[\, \hat x^j\, , \,\hat H  \,]\, \hat G\right)_W$ 
in Eq.~(\ref{dGwdp_2}) is defined to be the Weyl symbol of the operator
$\hat G\, 
[\, \hat x^j\, , \,\hat H  \,]\, \hat G $.
Using a similar argument  it may be  derived 
that
\begin{align}
\sum_{b=1...N}\frac{\partial Q_W(\omega,\{p_a\},\{x_a\})}{\partial p_b^j} = &\frac1{N!}\int \left(\prod\limits_{a=1}^N dq_a \ e^{iq_a x_a}\right) 
i\langle   \{p_a +\tfrac{q_a}{2}\}|\,   
[\, \hat x^j\, , \,\hat H  \,] \, | \{p_a -\tfrac{q_a }{2}\} \rangle
\label{dQwdp_2}
\\
=&i \left( 
[\, \hat x^j\, , \,\hat H  \,]\,  \right)_W
\label{dQwdp_3}
\end{align} 
where the term $\left( 
[\, \hat x^j\, , \,\hat H  \,]\,  \right)_W$ 
in Eq.~(\ref{dQwdp_3}) is defined to be the Weyl symbol of the operator
$ 
[\, \hat x^j\, , \,\hat H  \,]\,  $.\par 
We conclude that
\begin{equation}
\sum_{b=1...N}\frac{\partial G_W(\omega,\{p_a\},\{x_a\})}{\partial p_b^j}  = - G_W(\omega,\{p_a\},\{x_a\}) \star \sum_{b=1...N}\frac{\partial Q_W(\omega,\{p_a\},\{x_a\})}{\partial p_b^j} \star G_W(\omega,\{p_a\},\{x_a\})\ .
\end{equation}
This means, in particular, that
\begin{equation}
\sum_{b=1...N}\frac{\partial 1_W(\omega,\{p_a\},\{x_a\})}{\partial p_b^j}  = 0\ .
\end{equation}
The last identity is verified by  calculating the Weyl symbol of unity operator:
\begin{align}
1_W(\omega,\{p_a\},\{x_a\}) = &\frac1{N!}\int \left(\prod\limits_{a=1}^N dq_a \ e^{iq_a x_a}\right) 
\langle   \{p_a +\tfrac{q_a}{2}\}|\,   
\{p_a -\tfrac{q_a }{2}\} \rangle\ .
\label{1wdp_2}
\end{align} 
For its derivative we obtain
\begin{align}
\sum_{b=1...N}\frac{\partial 1_W(\omega,\{p_a\},\{x_a\})}{\partial p_b^j} = &\frac1{N!}\int \left(\prod\limits_{a=1}^N dq_a \ e^{iq_a x_a}\right) 
i\langle   \{p_a +\tfrac{q_a}{2}\}|\,   
[\, \hat x^j\, , \,\hat 1  \,] \, | \{p_a -\tfrac{q_a }{2}\} \rangle
\label{d1wdp_2}
\\
=&i \left( 
[\, \hat x^j\, , \,\hat 1  \,]\,  \right)_W = 0\ .
\label{d1wdp_3}
\end{align} \par 
Now we substitute Eqs.~(\ref{dGwdp_2}) and ~(\ref{dQwdp_3})
in  (\ref{calM2d238}) to obtain
\begin{align}
{\cal N} 
= &  -\frac{1}{2 A} \sum_{N=0,\ldots} \frac{1}{(2\pi)^{2N} }\,  \int  d\omega \left(\prod^{N}_{a=1} d^2p_{ a} \, d^2x_{ a} \right)   \epsilon^{jk} 
\nonumber\\
&
{\rm tr}\Big[  {G}^{(N)}_W(\omega,p_{ a },x_{ a })\star \frac{\partial {Q}^{(N)}_W(\omega,\{p_a\},\{x_a\} )}{\partial \omega} 
\star \left(\hat G\, 
[\, \hat x^j\, , \,\hat H  \,]\, \hat G\right)_W
\star \left( 
[\, \hat x^k\, , \,\hat H  \,]\,  \right)_W \Big]
\label{proof1_2_A}
\end{align}
Next (\ref{moyal_prod_6}) can be invoked to replace the star product of Weyl symbols with a single Weyl symbol 
corresponding to a product of operators as
\begin{align}
{\cal N} =&    {\frac{1}{2 A}}\sum_{N=0,\dots}\, \frac{1}{(2  \pi)^{2N} }  \int   d\omega \left(\prod^N_{a=1} d^2p_a \, d^2x_a\right)
\epsilon^{jk} 
\nonumber\\
&
\left( \hat G^{(N)} \ \frac{\partial \hat Q^{(N)}}{\partial \omega}\  \hat G^{(N)}
\ i[\hat x^j,\hat H]  
\ \hat G^{(N)}
\ i[\hat x^k,\hat H]  \right)_W.
\label{proof1_4_A}
\end{align}
Next we substitute the formal definition of a Weyl symbol given above in (\ref{fxd_numb_partcls_4}),
(where in the case of varying particle number $\ket{\{p_a\pm\tfrac {q_a}{2}\}}
=
a_{1\pm}^\dagger\dots a_{N\pm}^\dagger\ket{0}$)
to find
\begin{align}
{\cal N} =&   \frac{1}{2 A} \sum_{N=0,\dots}\, \frac{1}{(2 \pi)^{2N}N! }\,\int   d\omega   \left(\prod^N_{a=1} d^2p_a\, d^2q_a\, d^2x_a\, e^{iq_a x_a}\right) 
\nonumber\\&
\epsilon^{jk}  
\bra{\{p_a +\tfrac{q_a }{2}\}} 
 \hat G^{(N)} \ \frac{\partial \hat Q^{(N)}}{\partial \omega}\  \hat G^{(N)}
\ i[\hat x^j,\hat H]  
\ \hat G^{(N)}
\ i[\hat x^k,\hat H] 
\ket{\{p_a +\tfrac{q_a }{2}\}} \ .
\label{proof1_5_A}
\end{align}
This can be expressed using a complete set of antisymmetric $N$ fermion states, using the result proved in (\ref{n_ferm_state_17}).
Even more, this particular result is valid for varying particle numbers, 
due to the fact that
all states are deliberately expressed in terms of creation and annihilation operators.
The implication, after invoking  (\ref{n_ferm_state_17}),
is that 
\begin{align}
{\cal N} =&   \frac{1}{2 A}  \sum_{N=0,\dots}\,\frac{1}{(2 \pi)^{2N}}\frac1{N!^4}  \int   d\omega   \left(\prod^N_{a=1} d^2p_a\, d^2q_a\, d^2x_a\, 
d^2p_{a,1}\,d^2p_{a,2}\,d^2p_{a,3}\,e^{iq_a x_a}\right)
\nonumber\\&
\epsilon^{jk} \,
\bra{\{p_a+\tfrac{q_a}{2}\}}\hat G^{(N)} \frac{\partial \hat Q^{(N)}}{\partial \omega} \hat G^{(N)} \ket{\{p_{a,1}\}}
\bra{\{p_{a, 1}\}}i[\hat x^j,\hat H] \ket{\{p_{a,2}\}}
\nonumber\\
&\bra{\{p_{a,2}\}}\hat G^{(N)}\ket{\{p_{a,3}\}}
\bra{\{p_{a,3}\}}\ i[\hat x^k,\hat H]   \ket{\{p_a -\tfrac{q_a }{2}\}}\ .
\label{proof1_5A_A}
\end{align}
The $x$ integrals yield a product of $\delta$ functions, namely one factor of $(2\pi)\delta(q_a )$ for each $a$, which render
the $q_a$ integrals trivial. To that extent (\ref{proof1_5A_A}) reduces to
\begin{align}
&{\cal N} =    {\frac{1}{2 A}}\sum_{N=0,1,\dots} \frac1{N!^4}  \int   d\omega   \left(\prod^N_{a=1} d^2p_a\, 
d^2p_{a,1}\,d^2p_{a,2}\,d^2p_{a,3}\,\right)\epsilon^{jk}\, {\rm tr}  \nonumber\\&
\bra{\{p_a \}}\hat G^{(N)} \frac{\partial \hat Q^{(N)}}{\partial \omega} \hat G^{(N)} \ket{\{p_{a,1}\}}
\bra{\{p_{a, 1}\}}i[\hat x^j,\hat H] \ket{\{p_{a,2}\}}
\nonumber\\
&\bra{\{p_{a,2}\}}\hat G^{(N)}\ket{\{p_{a,3}\}}
\bra{\{p_{a,3}\}}\ i[\hat x^k,\hat H]   \ket{\{p_a  \}}\ .
\label{proof1_6_A}
\end{align}\par
By Substituting the explicit forms in (\ref{proof_var_num_0}), 
from which $\PD\hat Q^{(N)}/\PD\omega=i$,
and inserting complete sets of eigenstates of $\hat H$ assuming that each set is discrete belonging to discrete 
eigenvalues, we obtain
\begin{align}
{\cal N} 
=&  {-\frac{i}{2 A}} \sum_{N=0,1,\dots}\frac1{N!^4}  \int   d\omega   \left(\prod^N_{a=1} d^2p_a\, 
d^2p_{a,1}\,d^2p_{a,2}\,d^2p_{a,3}\,\right)
\nonumber\\&
\epsilon_{jk}\, {\rm tr}   
\sum_{E ,E^{\prime}  ,E^{\prime\prime}  }
\bra{\{p_a\} }
\frac1{i\omega-\hat H}\hat\Pi_N
 \ket{E }
\bra{E } 
\frac1{i\omega-\hat H}\hat\Pi_N
\ket{E^{\prime\prime}  }
\braket{E^{\prime\prime}  |\{p_{a,1\}}}
\bra{\{p_{a, 1}\}}[\hat x^j\ ,\ \hat H]\hat\Pi_N\ket{\{p_{a, 2}\}}
\nonumber\\&
\bra{\{p_{a,2}\}} 
\frac1{i\omega-\hat H}\hat\Pi_N
 \ket{E^{\prime}  }
\braket{E^{\prime}  |\{p_{a,3}\}} 
\bra{\{p_{a,3}\}}[\hat x^k\ , \ \hat H]\hat\Pi_N\ket{\{p_a\}}\ .
\label{proof1_10_A}
\end{align}
Here by $|E\rangle$ we denote the eigenstates of the Hamiltonian corresponding to the eigenvalue $E$. We assume here for simplicity that all eigenvalues are not degenerate. However, the extension to the case of degenerate eigenvalues is straightforward. 
For the next step of the argument it is necessary to show that the Hamiltonian $\hat H$ commutes with the projection operator onto $N$ particle states,
$\hat \Pi_N$.
The 
form of the projection operator   assumed is 
\begin{equation}
\hat \Pi_N=\frac1{N!}\int dx_1\dots dx_N \, a^\dagger(x_1)\dots a^\dagger(x_N)\ket{0}\bra{0}a (x_N)\dots a (x_1)\ .
\label{the_form_of_PiN}
\end{equation}
This is consistent with the requirement that, given a state $\ket{\psi}=\ket{x_1\dots x_{N'}}=a^\dagger(x_1)\dots
a^\dagger(x_N')\ket{0}$,
then $\hat \Pi_N\ket{\psi}=\delta_{NN'}\ket{\psi}$,
which is easily shown to be true by invoking theorem \ref{thm_inner_prod_multi_ferm_states_1}.
Based on (\ref{the_form_of_PiN}) and the form of the Hamiltonian in (\ref{top_inv_var_partcl_numb_3}),
then it can be shown that the two commute:
\begin{equation}
[\hat H,\hat \Pi_N]=0.\label{H_commutes_with_PiN}
\end{equation}
To show that they commute substitute their explicit forms:
\begin{align}
[\hat H,\hat \Pi_N]=&
\int dX\, {\mathscr H}_0(X) \frac1{N!} 
\int dx_1\dots dx_N \, 
[a^\dagger (X) a(X),a^\dagger(x_1)\dots a^\dagger(x_N)\ket{0}\bra{0}a (x_N)\dots a (x_1)]
\nonumber\\
&+
\int dX\,dY\, {\mathscr V}(X-Y) \frac1{N!} 
\int dx_1\dots dx_N \,\nonumber\\&
[a^\dagger (X) a(X)a^\dagger (Y) a(Y),a^\dagger(x_1)\dots a^\dagger(x_N)\ket{0}\bra{0}a (x_N)\dots a (x_1)]
\nonumber\\[2em]
=&
\int dX\, {\mathscr H}_0(X) \frac1{N!} 
\int dx_1\dots dx_N \, 
\bigg\{
a^\dagger (X) a(X)a^\dagger(x_1)\dots a^\dagger(x_N)\ket{0}\bra{0}a (x_N)\dots a (x_1)
\nonumber\\
&-
a^\dagger(x_1)\dots a^\dagger(x_N)\ket{0}\bra{0}a (x_N)\dots a (x_1)
a^\dagger (X) a(X)
\bigg\}
\nonumber\\[1em]
&+
\int dX\,dY\, {\mathscr V}(X-Y) \frac1{N!} 
\int dx_1\dots dx_N \,\nonumber\\&
\bigg\{
a^\dagger (X) a(X)
a^\dagger (Y) a(Y)
a^\dagger(x_1)\dots a^\dagger(x_N)\ket{0}\bra{0}a (x_N)\dots a (x_1)
\nonumber\\
&-
a^\dagger(x_1)\dots a^\dagger(x_N)\ket{0}\bra{0}a (x_N)\dots a (x_1)
a^\dagger (X) a(X)
a^\dagger (Y) a(Y)
\bigg\}
\label{Pi_commutes_with_H_1}
\end{align}
By substituting (\ref{Pi_commutes_with_H_2_appendix})--(\ref{Pi_commutes_with_H_5_appendix})
into (\ref{Pi_commutes_with_H_1}) and integrating over $X$ and $Y$ the result is
\begin{align}
[\hat H,\hat \Pi_N]
=&
\sum^N_{i=1}{\mathscr H}_0(x_i) \frac1{N!} 
\int dx_1\dots dx_N \,\nonumber\\&
\bigg\{
a^\dagger(x_1)\dots a^\dagger(x_N)\ket{0}\bra{0}a (x_N)\dots a (x_1)
-
a^\dagger(x_1)\dots a^\dagger(x_N)\ket{0}\bra{0}a (x_N)\dots a (x_1)
\bigg\}
\nonumber\\
&+
\sum^N_{\substack{i,j=1\\i\neq j}} {\mathscr V}(x_i-x_j) \frac1{N!} 
\int dx_1\dots dx_N \,\nonumber\\&
\bigg\{
a^\dagger(x_1)\dots a^\dagger(x_N)\ket{0}\bra{0}a (x_N)\dots a (x_1)
-
a^\dagger(x_1)\dots a^\dagger(x_N)\ket{0}\bra{0}a (x_N)\dots a (x_1)
\bigg\}
\nonumber\\[0.5em]
=0\ .
\label{Pi_commutes_with_H_6}
\end{align}
Even more, by their very definition of being eigenstates of $\hat H$, 
the inverse operator $(i\omega-\hat H)^{-1}$ [denoted in (\ref{proof1_10_A}) as $1/(i\omega-\hat H)$]
has the eigenvalue equation
$(i\omega-\hat H)^{-1}\ket{E}=\dfrac1{i\omega-E}\ket{E}$.
Importantly, the energy eigenstates correspond to  definite values of the particle number $N$. We denote this number by $N(E)$.
Hence, this fact and the fact that $\hat\Pi_N$ and $\hat H$ commute, mean that (\ref{proof1_10_A}) becomes 
\begin{align}
{\cal N} 
=& {-\frac{i}{2 A}} \sum_{N=0,1,\dots}\frac1{N!^4} \int   d\omega   \left(\prod^N_{a=1} d^2p_a\, 
d^2p_{a,1}\,d^2p_{a,2}\,d^2p_{a,3}\,\right)
\sum_{E ,E^{\prime}  ,E^{\prime\prime}  }
\frac1{i\omega-E }\frac1{i\omega-E^{\prime }  }\frac1{i\omega-E^{\prime\prime}  }
\nonumber\\&
\epsilon_{jk}\,{\rm tr}
 \braket{\{p_a\} |E }
\braket{E |E^{\prime\prime}  }
\braket{E^{\prime\prime}  |\{p_{a,1}\}}
\bra{\{p_{a,1\}}}[\hat x^j \ ,\ \hat H]\ket{\{p_{a,2}\}}
\nonumber\\&
\braket{\{p_{a,2}\}|E^{\prime}  }
\braket{E^{\prime}  |\{p_{a,3}\}}
\bra{\{p_{a,3}\}}[\hat x^k \ , \ \hat H]\ket{\{p_a\}}\ ,
\label{proof1_11_A}
\end{align}
and since the trace is unaffected by a change in order of inner products, this can equally be written as
\begin{align}
{\cal N} 
=&  {-\frac{i}{2 A}}  \sum_{N=0,1,\dots}\frac1{N!^4}  \int   d\omega   \left(\prod^N_{a=1} d^2p_a\, 
d^2p_{a,1}\,d^2p_{a,2}\,d^2p_{a,3}\,\right)
\sum_{E ,E^{\prime}  ,E^{\prime\prime}  }
\frac1{i\omega-E }\frac1{i\omega-E^{ \prime}  }\frac1{i\omega-E^{\prime\prime}  }
\nonumber\\&
{\rm tr}\,
\epsilon_{jk}
\braket{E |E^{\prime\prime}  }
\braket{E^{\prime\prime}  |
\{p_{a,1\}}}
\bra{\{p_{a,1}\}}[\hat x^j \ ,\ \hat H]\ket{\{p_{a,2}\}}
\nonumber\\&
\braket{\{p_{a,2}\}|E^{\prime}  }
\braket{E^{\prime}  |
\{p_{a,3}\}}
\bra{\{p_{a,3}\}}[\hat x^k \ , \ \hat H]\ket{\{p_a\}}
\braket{p |E } \ .
\label{proof1_12_A}
\end{align}
The integrals simplify through the identities in 
(\ref{n_ferm_state_17}),
and 
the relation
$\braket{E |E^{\prime\prime}  }=\delta_{E ,E^{\prime\prime}  }$ eliminates the  sum over $E^{\prime\prime}  $.
Ergo
\begin{equation}
{\cal N} 
=  {-\frac{i}{2 A}}\sum_{N=0,1,\dots}    
\sum_{E ,E^{\prime}  }
\int   d\omega
\frac1{\left(i\omega-E \right)^2\left(i\omega-E^{\prime}\right)  }
{\rm tr} \,  \epsilon_{jk} 
\bra{E }[\hat x^j \ ,\ \hat H]\ket{E^{\prime}  }\bra{E^{\prime}  }[\hat x^k \ , \ \hat H]\ket{E }\ .
\label{proof1_13_A}
\end{equation}\par 
Here, both states with energies $E$ and $E'$ correspond to the same value of particle number $N=N(E)$. 
The $\omega$ integral  over the integration range $ l= [-\infty,\infty]$,
is evaluated by deforming $l$ to the closed
contour being a semi circle, $C$ in the upper complex plane.
Two possibilities arise:
(i)
$C$
encloses both of the points $iE$ and $iE'$ on the imaginary axis
if $E>0$ and $E'>0$, or (ii) $C$ encloses only one of them, if say $E>0$ $E'<0$, or the converse.
If $E<0$ and $E'<0$, 
integrating over the variable  $-\omega$ (minus  $\omega$) instead,
produces the same integral described in (i) with the same contour $C$.
In both cases the integral can be evaluated using Cauchy's integral formula.
For case (i) the integral vanishes, but for case (ii) the contribution is non-zero.
The result is
\begin{equation}
{\cal N} 
= \frac{-i}{A}\sum_{N=0,1,\dots}2\pi  
\sum_{E ,E^{\prime}  }
\frac{\theta(E)\theta(-E')}{\left(E -E^{\prime}  \right)^2}
\epsilon^{jk}  \,
{\rm tr}
\bra{E }[\hat x^j \ ,\ \hat H]\ket{E^{\prime }  }\bra{E^{\prime }  }\hat x^k \ , \ \hat H]\ket{E } \ .
\label{proof1_14_A}
\end{equation} 
As the term $\theta(E)\theta(-E')$ itself indicates,
the integral was solved assuming that $E>0$, $E'<0$.
Had the converse been assumed, 
precisely the same formula in the form (\ref{proof1_14_A})
would hold, since interchanging $E$ and $E'$ (noting the trace operator in front) 
leaves the expression unaltered.\par
According to our choice of value for the constant term $\Delta$ entering the field Hamiltonian of Eq. (\ref{top_inv_var_partcl_numb_3}),
only the ground states have negative energy $E'<0$.
Let the ground states be denoted by $\ket{0_k}$, $k = 1,...,K$ instead, and let the eigenvalue of $\hat H$
that they belong to be denoted by $E_0$, such that $\hat H\ket{0_k}=E_0\ket{0_k}$.
Accordingly (\ref{proof1_14_A}) reads 
\begin{align}
{\cal N} 
=&  \frac{-i}{A} \sum^K_{k=1}  2\pi  
\sum_{E\ne E_0   }
\frac1{\left(E -E_0  \right)^2}
\epsilon_{ij}  \,
{\rm tr}
\bra{0_k }[\hat x^j \ ,\ \hat H]\ket{E}\bra{E   }\hat x^i \ , \ \hat H]\ket{0_k }
\ ,
\label{proof1_14_A_A}
\end{align} 
where in the last step the two inner products inside the trace were swapped, since this is the order that
the Hall conductivity is conventionally written.
The eigenstates $\ket{E}$ and can be identified with $\ket{n}$ 
in the coordinate representation defined above in (\ref{top_inv_var_partcl_numb_4}) and (\ref{top_inv_var_partcl_numb_5}):
\begin{equation}
{\cal N} 
=   -\frac{ 2\pi i}{A}\sum^K_{k=1}  
\sum_{n   }
\frac1{\left(E -E_0  \right)^2}
\epsilon_{ij}  \,
{\rm tr}
\bra{0_k }[\hat x^j\ ,\ \hat H]\ket{n  }\bra{n  }\hat x^i\ , \ \hat H]\ket{0_k } \ .
\label{proof1_14_A_B}
\end{equation}
 Eq.~(\ref{proof1_14_A_B})
is precisely analogous to the result in (\ref{sig1}).
\par 
This completes the proof. It is clear in the sum over $N$ in Eq. (\ref{calM2d238})  that only the term with $N=N_0$ remains. 
This is a direct consequence of our choice for the value of $\Delta$,
according to which only the ground state has negative energy.  
\subsection{The case of a non-interacting system}\noindent 
The aim of this subsection is to show that the expression derived in \eq{fxd_numb_partcls_7}
is equivalent to an analogous formula but with two-point Green functions instead of $N$-point Green functions.
\par 
It was shown in \S\ref{sec_proof_3.28_is_equivalent_to_3.29} that the expression for the Hall conductivity (\ref{proof1_2_A}), viz
 \begin{align}
{\cal N} =&   {\frac{1}{2A}} \sum_{N=0,\dots}\,\frac{1}{(2\pi)^{2N} }\, \sum^N_{b,c=1}    \int  d\omega\left(\prod^N_{a=1} d^2p_a \, d^2x_a\right)    
\nonumber\\&\epsilon^{jk}\, {\rm tr}
\bigg[   {G}^{(N)}_W(\omega,\{p_a\},\{x_a\})\star\! \frac{\partial {Q}^{(N)}_W(\omega,\{p_a\},\{x_a\})}{\partial \omega} \star 
G^{(N)}_W(\omega,\{p_a\},\{x_a\})
\star 
\frac{\partial  {Q}^{(N)}_W( \omega,\{p_a\},\{x_a\})}{\partial p_{ b}^j} 
\nonumber\\&
\star G^{(N)}_W(\omega,\{p_a\},\{x_a\})
\star \frac{\partial  {Q}^{(N)}_W(\omega,\{p_a\},\{x_a\})}{\partial p_c^k}  \bigg]
\label{proof1_2_A_yav_0}
\end{align}
with   $G^{(N)}$ and $Q^{(n)}_W$ given by 
(\ref{def_QN_W}--\ref{projectio_operator})  
and $\hat H$ given by (\ref{top_inv_var_partcl_numb_3}) 
without the interaction term, 
is equivalent to (\ref{proof1_14_A}).
To emphasize that the Hamiltonian in (\ref{proof1_14_A}) is the field theoretical Hamiltonian, let it be written in the notation
\begin{equation}
\mathbb H
=\sum_qa^\dagger_q a_q{\cal E}_q\label{field_theoretical_H}
\end{equation}
and let the field theoretical position operator be denoted by
\begin{equation}
\mathbb X_j
=\sum_{k,n}a_kX_{kn}{a^\dagger}_n\ .
\label{field_theoretical_X}
\end{equation}
The reader may consult \cite{Dirac_5} for an explanation of the origins of the right-hand side of Eq.~(\ref{field_theoretical_X}).
\par 
Here by $a_q$ we denote the annihilation operator corresponding to the one particle state with energy ${\cal E}_q$. For convenience in this section we do not define the field theoretical Hamiltonian 
with a chemical potential subtracted from ${\cal E}_q$. 
This  redefinition does not change the expressions given below. At any rate  in the absence of interactions,
we set $\Delta = 0$. The conductivity in (\ref{proof1_2_A_yav_0}) was shown in \S\ref{sec_proof_3.28_is_equivalent_to_3.29} 
to be equivalent to (\ref{proof1_14_A}), which in the notation of (\ref{field_theoretical_H}) and (\ref{field_theoretical_X}) is given by
\begin{align}
{\cal N} 
=&   - \frac{2\pi i}{A} 
\sum_{n\neq 0   }
\frac1{\left(E -E_0  \right)^2}
\epsilon_{ij}  \,
{\rm tr}
\llangle 0|[\mathbb X^j\ ,\ \mathbb  H]|n  \rrangle\llangle n |\mathbb X^i \ , \ \mathbb H]|0 \rrangle \ .
\label{non_interacting_20_yav}\\
=&   - \frac{2\pi i}{A} \sum_{n\ne 0}
\frac{\llangle 0 | [ \mathbb{X}_2,{\mathbb {H}}] |n\rrangle\llangle n|  [ \mathbb{X}_1,{\mathbb  {H}}] | 0\rrangle
-\llangle 0 |  [  \mathbb{X}_1,{\mathbb {H}}] |n\rrangle\llangle n| [ \mathbb{X}_2,{\mathbb {H}}] | 0\rrangle}{(E_n-E_0)^2}\ ,
\label{sig1_yav}
\end{align} 
where here, to distinguish from single particle bra and ket vectors,
$| 0\rrangle$ denotes the non - degenerate multi-particle ground state
\begin{equation} | 0\rrangle 
=
a_1^\dagger\dots a_N^\dagger |{\emptyset}\rrangle
\end{equation}
where  $|{\emptyset}\rrangle$ denotes the true vacuum   and
\begin{equation}  
a_k^\dagger=\int dx\ \psi_k(x)a^\dagger (x)\ .
\end{equation}
Here $\psi_k(x)$ is the wave function of the $k$th one - particle state. We enumerate  one - particle states in such a way that ${\cal E}_1 \le {\cal E}_2 \le ... \le {\cal E}_N < \mu$, where $\mu$ is the chemical potential. 
By $| n\rrangle$ we denote the excited multi - particle states that have the same number of particles, but  which have total energy larger than that of the ground state.  
 Using the anticommutation relations in (\ref{fxd_numb_partcls_5A}),
\begin{align}
\left[\sum_{k,n}a^\dagger_k X_{kn} a_n,\sum_qa^\dagger_q {\cal E}_q a_q\right]
=&\sum_{k,n}\sum_q
a^\dagger_k X_{kn} a_n 
a^\dagger_q {\cal E}_q a_q
-
a^\dagger_q {\cal E}_q a_q
a^\dagger_k X_{kn} a_n 
\nonumber\\
=&\sum_{k,n}\sum_qX_{kn}{\cal E}_q\left(
a^\dagger_k  a_n 
a^\dagger_q  a_q
-
a^\dagger_q   a_q
a^\dagger_k   a_n \right)
\nonumber\\
=&\sum_{k,n}\sum_qX_{kn}{\cal E}_q\left(
a^\dagger_k  \delta_{nq}  a_q
-
a^\dagger_k a^\dagger_q  a_n 
 a_q
-
a^\dagger_q   a_q
a^\dagger_k   a_n \right)
\nonumber\\
=&\sum_{k,n}\sum_q X_{kn}{\cal E}_q\left(
a^\dagger_k  \delta_{nq}  a_q
+
a^\dagger_q
a^\dagger_k   a_n 
 a_q
-
a^\dagger_q   a_q
a^\dagger_k   a_n \right)
\nonumber\\
=&\sum_{k,n}\sum_q X_{kn}{\cal E}_q\left(
a^\dagger_k  \delta_{nq}  a_q
-
a^\dagger_q
a^\dagger_k   a_qa_n 
-
a^\dagger_q   a_q
a^\dagger_k   a_n \right)
\nonumber\\
=&\sum_{k,n}\sum_q X_{kn}{\cal E}_q\left(
a^\dagger_k  \delta_{nq}  a_q
-
a^\dagger_q 
\delta_{kq}   a_n 
+
a^\dagger_qa_q
a^\dagger_k   a_n 
-
a^\dagger_q   a_q
a^\dagger_k   a_n \right)
\nonumber\\
=&\sum_{k,n}\sum_q X_{kn}{\cal E}_q\left(
a^\dagger_k  \delta_{nq}  a_q
-
a^\dagger_q 
\delta_{kq}   a_n \right)
\nonumber\\
=&\sum_{k,n}  \left(  X_{kn}{\cal E}_n
a^\dagger_k    a_n
-
X_{kn}{\cal E}_k
a^\dagger_k 
    a_n \right)\ .
\end{align}
The right-hand side is precisely the matrix elements of the commutator $[\hat x,\hat H]$,
where $\hat x$ and $\hat H$ are the ordinary one-particle position and Hamiltonian operators.
Hence
\begin{equation}
\left[\sum_{k,n}a^\dagger_k X_{kn} a_n,\sum_qa^\dagger_q {\cal E}_q a_q\right]
=\sum_{k,n}  
a^\dagger_k    [\hat x,\hat H]_{kn}
a_n  \ .\label{X-H_commutator_1}
\end{equation}  
After combining Eqs.(\ref{field_theoretical_H}), (\ref{field_theoretical_X}) and (\ref{X-H_commutator_1})
it follows that
\begin{equation}
\left[\mathbb X_i,\mathbb H\right]
=\sum_{k,n} a^\dagger_k    [\hat x_i,\hat H]_{kn}
a_n  \ .\label{X-H_commutator_2}
\end{equation}\par 
With the result (\ref{X-H_commutator_2}) an expression may be derived for 
$\llangle 0_k|\left[\mathbb X_i,\mathbb H\right]| n\rrangle$.
The ground state $|0_k\rrangle$ consisting of $N$ particles has the form
\begin{equation}
|0\rrangle= a^\dagger_1\ldots a_N^\dagger\ket{\emptyset}\label{N_particle_ground_state}\ .
\end{equation}
The state $| n\rrangle$
is taken to be the state that differs from the ground state 
(\ref{N_particle_ground_state}) 
by one out of the $N$ particles, say particle $j$ ($j=1,\dots N$), 
which gets excited and jumps to a state of higher energy.
Correspondingly
the creation operator $a^\dagger_j$
is replaced with $a^\dagger_l$, $l=N+1,\dots,\infty$,
such that $| n\rrangle$ has the form
\begin{equation}
|n\rrangle= a^\dagger_1\ldots a^\dagger_{j-1}a^\dagger_{j+1}\dots a_N^\dagger a_l^\dagger\ket{\emptyset}=:|l,j\rrangle \ ,
\qquad ( j=1,\dots,N\ ,\quad l\geq N+1)\ .
\label{N_particle_excited_state}
\end{equation}
Then, by (\ref{X-H_commutator_2}), (\ref{N_particle_ground_state})
and (\ref{N_particle_excited_state}),
\begin{align}
\llangle 0|\left[\mathbb X_i,\mathbb H\right]| n\rrangle
=&
\bra{\emptyset}
a_N\ldots a_1
\sum_{k,n} a^\dagger_k    [\hat x_i,\hat H]_{kn} a_n
a^\dagger_1\ldots a^\dagger_{j-1}a^\dagger_{j+1}\dots a_N^\dagger a_l^\dagger\ket{\emptyset}
\nonumber\\
=&
\sum_{k,n}     [\hat x_i,\hat H]_{kn}
\bra{\emptyset}
a_N\ldots a_1
a^\dagger_k a_n
a^\dagger_1\ldots a^\dagger_{j-1}a^\dagger_{j+1}\dots a_N^\dagger a_l^\dagger\ket{\emptyset}
\nonumber\\
=&
\sum_{k,n}    \bra{k} [\hat x_i,\hat H]\ket{n}
\bra{\emptyset}
a_N\ldots a_1
a^\dagger_k a_n
a^\dagger_1\ldots a^\dagger_{j-1}a^\dagger_{j+1}\dots a_N^\dagger a_l^\dagger\ket{\emptyset}
\ .
\label{X-H_commutator_3}
\end{align}
The operator $a^\dagger_l$ may be anticommuted to the left past the $N-1$ operators standing in front of it to obtain
\begin{align}
\llangle 0|\left[\mathbb X_i,\mathbb H\right]| n\rrangle
=& (-1)^{N-1}
\sum_{k,n}    \bra{k} [\hat x_i,\hat H]\ket{n}
\bra{\emptyset}
a_N\ldots a_1
a^\dagger_k a_n
a_l^\dagger
a^\dagger_1\ldots a^\dagger_{j-1}a^\dagger_{j+1}\dots a_N^\dagger \ket{\emptyset}
\ .
\label{X-H_commutator_4}
\end{align}
Note two important observations of the inner product  in (\ref{X-H_commutator_4}).
First, there stands one annihilation operator $a_j$ to the right of the bra of the vacuum state $\bra{0_k}$, so 
for it not to vanish there must be present one creation operator $a_j^\dagger$.
Since $j=1,\ldots,N$ and $l\geq N+1$, then $a^\dagger_l$ is never equal to $a^\dagger_j$
but 
$a^\dagger_k$
is equal to $a^\dagger_j$ 
corresponding to the $k=j$ term in the sum.
Secondly, 
the creation operator $a^\dagger_l$ will act on the vacuum to create one particle in the state $l$.
Consequently the inner product will vanish without the presence of one annihilation operator $a_l$,
which forces $a_n=a_l\delta_{nl}$.
Putting all this together, the non-vanishing contribution to 
(\ref{X-H_commutator_4}) is found to be
\begin{align}
\llangle 0|\left[\mathbb X_i,\mathbb H\right]| j,l\rrangle
=& (-1)^{N-1}
\sum_{k,n} \delta_{k,j}\delta_{n,l}   \bra{k} [\hat x_i,\hat H]\ket{n}
\bra{\emptyset}
a_N\ldots a_1
a^\dagger_k a_n
a_l^\dagger
a^\dagger_1\ldots a^\dagger_{j-1}a^\dagger_{j+1}\dots a_N^\dagger \ket{\emptyset}
\nonumber\\[1em]
=& (-1)^{N-1}
   \bra{j} [\hat x_i,\hat H]\ket{l}
\bra{\emptyset}
a_N\ldots a_1
a^\dagger_j a_l
a_l^\dagger
a^\dagger_1\ldots a^\dagger_{j-1}a^\dagger_{j+1}\dots a_N^\dagger \ket{\emptyset}
\ .
\label{X-H_commutator_5}
\end{align}
Further, since $l\neq j$ ($l\geq N+1$ and $j=1,\ldots N$), $a^\dagger_j$ and $a_l$ can be anticommuted past each other to yield
\begin{align}
\llangle 0|\left[\mathbb X_i,\mathbb H\right]| j,l\rrangle
=& (-1)^{N}
   \bra{j} [\hat x_i,\hat H]\ket{l}
\bra{\emptyset}
a_N\ldots a_1a_l
a^\dagger_j 
a_l^\dagger
a^\dagger_1\ldots a^\dagger_{j-1}a^\dagger_{j+1}\dots a_N^\dagger \ket{\emptyset}
\ ,
\label{X-H_commutator_6}
\end{align}
and subsequently $a^\dagger_j $ and $
a_l^\dagger$ may be anticommuted past each other to bring it to the form
\begin{align}
\llangle 0|\left[\mathbb X_i,\mathbb H\right]| j,l\rrangle
=& (-1)^{N+1}
   \bra{j} [\hat x_i,\hat H]\ket{l}
\bra{\emptyset}
a_N\ldots a_1a_l
a_l^\dagger
a^\dagger_j 
a^\dagger_1\ldots a^\dagger_{j-1}a^\dagger_{j+1}\dots a_N^\dagger \ket{\emptyset}
\label{X-H_commutator_7}\\[1em]
=& (-1)^{N+j}
   \bra{j} [\hat x_i,\hat H]\ket{l}
\bra{\emptyset}
a_N\ldots a_1a_l
a_l^\dagger 
a^\dagger_1\ldots a^\dagger_{j-1} a^\dagger_j a^\dagger_{j+1}\dots a_N^\dagger \ket{\emptyset}
\ ,
\label{X-H_commutator_8}
\end{align}
where in the last step $a^\dagger_j$ was anticommuted past operators to appear between $a^\dagger_{j-1}$ and $a^\dagger _{j+1}$.
\par 
The inner product on the right-hand side of 
(\ref{X-H_commutator_8})
comprises an inner product of $N+1$ different creation operators acting on the ket of the vacuum $\ket{\emptyset}$ ,
with  $N+1$  corresponding annihilation operators acting on the bra of the vacuum $\bra{\emptyset}$.
It can be shown by induction to equal unity, namely
\begin{equation}
\bra{\emptyset}a_N\dots a_1a_1^\dagger \dots a_N^\dagger\ket{\emptyset}=1\ .
\label{aNaNdagger=1}
\end{equation}
To prove (\ref{aNaNdagger=1}) by induction, we start with the $n=1$ case:
\[\bra{\emptyset}a_1
a^\dagger_1 \ket{\emptyset}=
\bra{\emptyset}\{a_1,
a^\dagger_1\} \ket{\emptyset}
-
\bra{\emptyset}a_1^\dagger
a_1 \ket{\emptyset}
=
\bra{\emptyset}1 \ket{\emptyset}
-
\bra{\emptyset}0\ket{\emptyset}
=
\braket{\emptyset|\emptyset}
\]
by (\ref{fxd_numb_partcls_5A}). Hence our claim is true for the $n=1$  case.
For the inductive step, we assume that it is true for $n=N$,
then we write down (\ref{aNaNdagger=1}) for the $n=N+1$ case, viz
\[\bra{\emptyset}a_{N+1}a_N\dots a_1
a^\dagger_1 \dots a_N^\dagger a_{N+1}^\dagger \ket{\emptyset}.\] 
By anticommuting
$a_{N+1}$ to the right $N$ places  and anticommuting $a_{N+1}^\dagger$
to the left $N$ places,   we bring it to the form
\[\bra{\emptyset}a_N\dots a_1 a_{N+1} a_{N+1}^\dagger 
a^\dagger_1 \dots a_N^\dagger  \ket{\emptyset}=
\bra{\emptyset}a_N\dots a_1 \{a_{N+1}, a_{N+1}^\dagger\} 
a^\dagger_1 \dots a_N^\dagger  \ket{\emptyset}
-
\bra{\emptyset}a_N\dots a_1  a_{N+1}^\dagger  a_{N+1} 
a^\dagger_1 \dots a_N^\dagger  \ket{\emptyset}\]
then we invoke (\ref{fxd_numb_partcls_5A})
on the 1st term, and in the 2nd term we anticommute operators
$a_{N+1}^\dagger  $ and $ a_{N+1}$ to obtain
\[\bra{\emptyset}a_N\dots a_1 a_{N+1} a_{N+1}^\dagger 
a^\dagger_1 \dots a_N^\dagger  \ket{\emptyset}=
\bra{\emptyset}a_N\dots a_1   
a^\dagger_1 \dots a_N^\dagger  \ket{\emptyset}
-
\bra{\emptyset}a_{N+1}^\dagger a_N\dots a_1  
a^\dagger_1 \dots a_N^\dagger  a_{N+1}\ket{\emptyset}.\]
The 2nd term vanishes and the 1st term is the $n=N$ case, which is unity by the inductive hypothesis. Therefore,
$\bra{\emptyset}a_N\dots a_1 a_{N+1} a_{N+1}^\dagger 
a^\dagger_1 \dots a_N^\dagger  \ket{\emptyset}=1$ and the claim is proven.
\par
On the  basis of (\ref{aNaNdagger=1}) that was just proven, 
(\ref{X-H_commutator_8}) simplifies to
\begin{align}
\llangle 0|\left[\mathbb X_i,\mathbb H\right]| j,l\rrangle
=& (-1)^{N+j}
   \bra{j} [\hat x_i,\hat H]\ket{l}
\ .
\label{X-H_commutator_9}
\end{align}
Eq.~(\ref{X-H_commutator_9}) was established on the basis that the state $\llangle n|$
comprises a single particle that gets excited from the ground state to a higher state.
It can be shown that the analogous state in which two or more particles are in excited
states do not contribute.
Consider the case where the state $| n\rrangle$
is  the state that differs from the ground state 
(\ref{N_particle_ground_state}) 
by two particles one out of the $N$ particles, say particles $j_1$ and $j_2$
($j_1,j_2=1,\dots N$), 
which get excited and jump to a state of higher energy.
Correspondingly
the creation operators $a^\dagger_{j_1}$, $a^\dagger_{j_2}$
are replaced with $a^\dagger_{l_1}$, $a^\dagger_{l_2}$
respectively ($l_1,l_2=N+1,\dots,\infty$),
such that $| n\rrangle$ has the form
\begin{equation}
|n\rrangle= a^\dagger_1\ldots a^\dagger_{j_1-1}a^\dagger_{j_1+1}\dots
a^\dagger_{j_2-1}a^\dagger_{j_2+1}\dots
a_N^\dagger a_{l_1}^\dagger
a_{l_2}^\dagger
\ket{\emptyset}=:|l_1,l_2;j_1,j_2\rrangle \ ,
\label{N_particle_2_excited_states}
\end{equation}
where in (\ref{N_particle_2_excited_states}), 
$ j_1,j_2=1,\dots,N$ and $l_1,l_2\geq N+1$.
Then by (\ref{X-H_commutator_2}), (\ref{N_particle_2_excited_states})
and (\ref{N_particle_excited_state}),
\begin{align}
\llangle 0|\left[\mathbb X_i,\mathbb H\right]| n\rrangle
=&
\bra{\emptyset}
a_N\ldots a_1
\sum_{k,n} a^\dagger_k    [\hat x_i,\hat H]_{kn} a_n
a^\dagger_1\ldots a^\dagger_{j_1-1}a^\dagger_{j_1+1}\dots
a^\dagger_{j_2-1}a^\dagger_{j_2+1}\dots
a_N^\dagger a_{l_1}^\dagger
a_{l_2}^\dagger\ket{\emptyset}
\nonumber\\[1em]
=&
\sum_{k,n}     [\hat x_i,\hat H]_{kn}
\bra{\emptyset}
a_N\ldots a_1
a^\dagger_k a_n
a^\dagger_1\ldots a^\dagger_{j_1-1}a^\dagger_{j_1+1}\dots
a^\dagger_{j_2-1}a^\dagger_{j_2+1}\dots
a_N^\dagger a_{l_1}^\dagger
a_{l_2}^\dagger\ket{\emptyset}
\nonumber\\[1em]
=&
\sum_{k,n}    \bra{k} [\hat x_i,\hat H]\ket{n}\nonumber\\&
\bra{\emptyset}
a_N\ldots a_1
a^\dagger_k a_n
a^\dagger_1\ldots a^\dagger_{j_1-1}a^\dagger_{j_1+1}\dots
a^\dagger_{j_2-1}a^\dagger_{j_2+1}\dots
a_N^\dagger a_{l_1}^\dagger
a_{l_2}^\dagger\ket{\emptyset}
\ .
\label{X-H_commutator_10}
\end{align}
In this expression $a^\dagger_k a_n$ can be replaced with $\{a^\dagger_k, a_n\}-a_na^\dagger_k=\delta_{nk}-a_na_k^\dagger$
by (\ref{fxd_numb_partcls_5A}) to obtain
\begin{align}
\llangle 0|\left[\mathbb X_i,\mathbb H\right]| n\rrangle
=&
\sum_{k,n}    \bra{k} [\hat x_i,\hat H]\ket{n}\delta_{nk}\nonumber\\&
\bra{\emptyset}
a_N\ldots a_1
a^\dagger_1\ldots a^\dagger_{j_1-1}a^\dagger_{j_1+1}\dots
a^\dagger_{j_2-1}a^\dagger_{j_2+1}\dots
a_N^\dagger a_{l_1}^\dagger
a_{l_2}^\dagger\ket{\emptyset}
\nonumber\\
&-
\sum_{k,n}    \bra{k} [\hat x_i,\hat H]\ket{n}\nonumber\\&
\bra{\emptyset}
a_N\ldots a_1a_n
a^\dagger_k 
a^\dagger_1\ldots a^\dagger_{j_1-1}a^\dagger_{j_1+1}\dots
a^\dagger_{j_2-1}a^\dagger_{j_2+1}\dots
a_N^\dagger a_{l_1}^\dagger
a_{l_2}^\dagger\ket{\emptyset}
\ .
\label{X-H_commutator_11}
\end{align}
The first inner product vanishes.
The ket on the right comprises 
two particles corresponding to the creation operators $a_{l_1}^\dagger$
, 
$a_{l_2}^\dagger$ ($l_1,l_2\geq N+1$), but there are no such particles in the bra on the right,
as there are no corresponding annihilation operators 
$a_{l_1}$
, 
$a_{l_2}$.
The second inner product vanishes for similar reasons.
Say that $a_k^\dagger$ is identified with $a_{j_1}^\dagger$ and $a_n$ with $a_{l_1}$.
There would still be a particle created by $a_{l_2}^\dagger$ in the left-hand ket 
with no corresponding particle in the bra on the right,
and there would still be a particle annihilated by $a_{j_2}$ in the right-hand bra 
with no corresponding particle created in the ket on the left.
The argument holds even if the particles are permuted.
In conclusion, the inner product between the ground state and states with two  excited particles
vanishes.
And an analogous argument shows that the inner product between the ground state and states with more than two  excited particles
vanishes.
\par
Finally, (\ref{X-H_commutator_9}) may be substituted in (\ref{sig1_yav}).
The sum over states labeled by $n$ becomes a sum over $j$, namely a sum over single particles that jump from the ground state to a higher excitation.
The difference in energy between the state $\ket{n}$ and the ground state $\ket{0_k}$ (denoted by $E_n-E_0$ in the denominator),
is equal to the difference in energy between particle $j$ in an excited state and in the ground state, which shall be denoted $E_l-E_j$.
The resulting expression for the topological invariant is
\begin{align}
{\cal N} 
=&   -\frac{ 2\pi i}{A} \sum_{l=N+1}^\infty \sum_{j=1}^N
\frac{
 \bra{j} [\hat x_1,\hat H]\ket{l}
 \bra{l} [\hat x_2,\hat H]\ket{j}
 -
 \bra{j} [\hat x_2,\hat H]\ket{l}
 \bra{l} [\hat x_1,\hat H]\ket{j} }{(E_l-E_j)^2}\ ,
\label{sig1_yav-1}
\end{align} 
and for the conductivity itself
\begin{align}
\sigma 
=&   -\frac{i}{A} \sum^\infty_{l=N+1}\sum_{j=1}^N
\frac{
 \bra{j} [\hat x_1,\hat H]\ket{l}
 \bra{l} [\hat x_2,\hat H]\ket{j}
 -
 \bra{j} [\hat x_2,\hat H]\ket{l}
 \bra{l} [\hat x_1,\hat H]\ket{j} }{(E_l-E_j)^2}\ .
\label{conductivity_yav}
\end{align} 
In ref. \cite{Zubkov:2019amq}
the expression in (\ref{conductivity_yav}) has been shown to be equal to
 \begin{align}
{\cal N} =&    \frac{1}{2 A}  \,\frac{1}{(2\pi)^{2} }\,  \int  d\omega  d^2p  \, d^2x  
\nonumber\\&\epsilon^{jk}\, {\rm tr}
\bigg[   {G}^{(1)}_W(\omega,p,x)\star\! \frac{\partial {Q}^{(1)}_W(\omega,p,x)}{\partial \omega} \star 
G^{(1)}_W(\omega,p,x)
\star 
\frac{\partial  {Q}^{(1)}_W( \omega,p,x)}{\partial p^j} 
\nonumber\\&
\star G^{(1)}_W(\omega,p,x)
\star \frac{\partial  {Q}^{(1)}_W(\omega,p,x)}{\partial p ^k}  \bigg].
\label{proof1_2_A_yav}
\end{align}
\section{Conclusions}\noindent
To conclude, in this paper we propose a topological description of the fractional Hall effect. The corresponding  conductivity (averaged over the system area) in 
units of $e^2/\hbar $ has the form
\begin{equation}
\sigma_H = \frac{1}{2\pi} \frac{\cal N}{K}
\end{equation}
where $K$ is the degeneracy of the ground state while $\cal N$ is the topological invariant composed of the multi  - particle Green functions. 
While the original expression for $\cal N$ contains a summation over all possible numbers of fermionic legs, 
in actual fact  only the term with $2 N_0$ legs contributes, where $N_0$ is the number of electrons in the ground state of the system. 
\par 
The expression for the  topological invariant $\cal N$ contains a generalization of the Wigner transformation of  multi - leg Green functions, 
and a generalization of Moyal product. The form of this expression resembles that of the integer Hall effect. 
The main difference between the two is that now  $\cal N$  is expressed in terms of multi - particle Green functions  instead of  one - particle Green functions.
We have demonstrated that in the absence of interactions, our expression for the conductivity reduces to that of the IQHE (in terms of  one - particle Green functions). 
In the general case of an interacting system, the value of $\cal N$ is discrete and is likely also to be given by an integer. 
This result is an alternative proof of the topological nature of the FQHE. Moreover, 
unlike the topological expression in ref. \cite{11}, our result is valid for the case of varying external  fields. 
As well as the expression in ref. \cite{11}, our expressions in Eqs. (\ref{rez1}) and (\ref{calM2d238_0}) are of no use for practical calculations of the FQHE 
conductivity. Nevertheless  they stand as a rigorous proof of the robustness of ${\cal N}$ with respect to  smooth modification of the system.\par         
M.A.Z. is grateful to Xi Wu for useful discussions during the initial stage of the work on the present paper.  
\appendix
\renewcommand\thesubsection{\thesection.\arabic{subsection}}
\section{Multi particle states}\label{sec_multi_particle_states}\noindent
\subsection{Fermion creation/annihilation operators and one fermion states}\label{sec_ferm_crtn_annhiliatn_ops_one_ferm_states}\noindent
A single-particle state $\ket{p}$ is a momentum eigenstate with momentum $p$,
constructed by acting on $\ket{0}$ with a creation operator as
\begin{equation}
a^\dagger(p)\ket{0}=\ket{p}\ .
\label{one_fermion_2}
\end{equation}
Consistent with this definition the operator  $\int dp\ket{p}\bra{p}$
behaves like an identity operator when it acts on one-particle states:
\begin{align}
\left(\int dp\ \ket{p}\bra{p}\right)\ket{q}=&\int dp\ a^\dagger(p)\ket{0}\bra{0}a(p)a^\dagger(q)\ket{0}
=
\int dp\  a^\dagger (p) \{ a(p),a^\dagger(q)\}\ket{0}
=
\int dp\ a^\dagger (p)\ \delta(p-q)\ket{0}
\nonumber\\
=&
a^\dagger (q)\ket{0}
=
\ket{q}\ .
\label{one_fermion_3}
\end{align}\par 
\subsection{Two fermion states}\label{sec_two_ferm_states}\noindent
A similar reasoning can be used  to 
find an analogous expression for the identity operator 
that acts on a
\emph{two-fermion state}, comprising two identical fermions in two distinguishable states. 
Re-arranging the fermions to be in different states
introduces a minus sign if the permutation is odd.
This is the very  antisymmetric property that characterizes states of more than one fermions.
More generally this property guarantees that the number of fermions in any given state 
is either zero and one. This  phenomenon is the Pauli-exclusion principle.
Accordingly expressions for two-fermion states
must be  anti-symmetric under odd-permutations of fermions between states.
To that extent
the order of terms within expressions must be preserved
as they appear here in the discussion.
Below in \S\ref{sec_n_ferm_states}
a more sophisticated
notation 
fixes the
ordering of fermions between states,
such that
attention to the order of terms when writing expressions
is redundant.
While this notation indeed is needed to
assign fermions to states in the right order,
the proper assignment can be done for two fermions by
merely writing terms in the correct order,
with the advantage that the antisymmetry is more obvious
without extra cumbersome notation.
\par 
A two-fermion state in the momentum representation has the form
\begin{align}
\ket{p_1\ p_2}=&
a^\dagger(p_1)a^\dagger (p_2)\ket{0}
=\frac1{2}\left(\ket{p_1}\ket{p_2}-\ket{p_2}\ket{p_1}\right)\ ,
\label{two_fermion_4}\\
\bra{p_1\ p_2}=&
\bra{0}a(p_2)a  (p_1)\ .
\label{two_fermion_4A}
\end{align}
The first term describes a fermion with momentum $p_1$ in the 1st state
and a fermion 
with momentum $p_2$   in the 2nd state, while
in the second term 
the two fermions have swapped states.
Acting on $\ket{q_1\ q_2 }$, a two-fermion state with momenta $q_1$, $q_2$,
with the operator $\frac1{2}
\int dp_1\ dp_2 \ \ket{p_1\ p_2}\bra{p_1\ p_2}$ 
produces
\begin{align}
&\frac1{2}
\int dp_1\ dp_2 \ \ket{p_1\ p_2}\braket{p_1\ p_2|q_1\ q_2}
\nonumber\\
=& \frac1{2}
\int dp_1\ dp_2 \ \ket{p_1\ p_2}\bra{0}
a (p_2)a (p_1) 
a^\dagger(q_1)a^\dagger(q_2) 
\ket{0}
\label{two_fermion_5}
\end{align}
Generally speaking 
\[\begin{aligned}
a(p_2)a(p_1)a^\dagger(q_1)a^\dagger(q_2)\ket{0}
=&
a(p_2)\{ a(p_1),a^\dagger(q_1)\}a^\dagger(q_2)\ket{0}
-
a(p_2)  a^\dagger(q_1) a(p_1) a^\dagger(q_2)\ket{0}
\\
=&
\{a(p_2),a^\dagger(q_2)\}\delta(p_1-q_1)\ket{0}
-
\{a(p_2),  a^\dagger(q_1) \}\{a(p_1) ,a^\dagger(q_2)\}\ket{0}
\\
=&
\delta(p_1-q_1)\delta(p_2-q_2)\ket{0}
-
\delta(p_2-q_1)\delta(p_1-q_2)\ket{0}
\end{aligned},
\]
such that (\ref{two_fermion_5}) becomes
\begin{align}
&\frac1{2}
\int dp_1\ dp_2 \ \ket{p_1\ p_2}\braket{p_1\ p_2|q_1\ q_2}
\nonumber\\
=& \frac1{2}
\int dp_1\ dp_2 \ \ket{p_1\ p_2}
\bra{0}
\bigg(\delta(p_1-q_1)\delta(p_2-q_2)
-
\delta(p_2-q_1)\delta(p_1-q_2)
\bigg)
\ket{0}
\nonumber\\[1em]
=&
\frac1{2}
\left(
\ket{q_1\ q_2}-\ket{q_2\ q_1}\right)\braket{0|0}
\nonumber\\[1em]
=&
\ket{q_1\ q_2}\ ,
\label{two_fermion_6}
\end{align}
where the final step follows from the very 
antisymmetric property of
multi-fermion states,  namely $\ket{q_1\ q_2}=-\ket{q_2\ q_1}$.
\par 
The upshot is that the identity operator for the group of two-fermion states
is
\begin{equation}
\mathds 1=\frac1{2}\int dp_1\,dp_2\,\ket{p_1\ p_2}\bra{p_1\ p_2 }\ .\end{equation}
\par 
\subsection{$N$ fermion states}\label{sec_n_ferm_states}\noindent
Before moving on to states of more than two fermions,
new notation is needed to
represent the permutation of particles between different states,
which closely follows the convention in
\cite{Dirac_permutation_notation}.
A given state of $N$ fermions is represented by the  ket
\begin{equation}
\ket{\psi^{i_1}\ \psi^{i_2}\dots\psi^{i_N}}=\frac1{N!}\sum_{i_1\dots i_N}\epsilon_{i_1\dots i_N}\ket{\psi^{i_1}}\otimes \ket{\psi^{i_2}}\otimes\dots \otimes\ket{\psi^{i_N}}
=
A\ket{\psi^{1}}\otimes \ket{\psi^{2}}\otimes\dots \otimes\ket{\psi^{N}}\ .
\label{n_ferm_state_1}
\end{equation}
Indices  ${}^{s_1}$,${}^{s_{2}}$,$\cdots$,${}^{s_N}={}^{(1)}$,${}^{(2)}$,$\cdots$ ${}^{(N)}$
label one-particle base ket-vectors.
In the broad scheme it is an abstract label not necessarily 
assigned to be any specific physical observable (e.g. an eigenvalue of momentum).
The  base ket-vector labels  ${}^{s_1}$,${}^{s_{2}}$,$\cdots$,${}^{s_N}$
assign an order to the one-particle kets.
This ordering does not refer to any physical configuration of the particles themselves, rather it
enables an antisymmetric sum over configurations for fermions (and a symmetric sum for bosons)
without the need to write the order of terms explicitly like in \S\ref{sec_two_ferm_states} with just two fermions.
On the right the operator $A$ produces an antisymmetric sum over configurations of 
$N$ different base ket-vectors.\par 
Such an antisymmetric sum 
possesses the required properties of fermion states:
states vanish if more than one identical fermion occupies a given state,
and a single fermion state 
has  odd-integer spin.
The former is the Pauli exclusion principle and the latter is a result of the spin-statistics theorem.
Conversely, mutli-boson states are represented by a symmetric sum that carries the required property that
any number of identical bosons can co-exist in a given state, and by the spin-statistics theorem,
bosons carry integer spin.
Bosons and fermions do not literally assume a particular order in nature any more than they are confined to a particular location,
like classical particles.
The ordering described here is purely a mathematical ordering of terms to give multi-particle states the correct 
properties: Fermi-Dirac statistics for fermions and Bose-Einstein statistics for bosons.\par 
To ensure that at most one fermion lies in any given state,
the following anti-commutation relations are assumed:
\begin{equation}
\{a(p_1),a(p_2)\}=0,\qquad 
\{a^\dagger(p_1),a^\dagger(p_2)\}=0,\qquad 
\{a(p_1),a^\dagger(p_2)\}=\delta(p_1-p_2)\ .
\label{n_ferm_state_3}
\end{equation}
An $N$-fermion state  can be expressed as an antisymmetric tensor product  of single-particle momentum  eigenstates     as
\begin{equation}
\ket{p_1 \ldots p_N}=
\frac1{N!}\sum_{i_1\cdots i_N}
\epsilon^{i_1\cdots i_N}\ket{p_{i_1}}\otimes \dots \otimes \ket{p_{i_N}}
=  
a^\dagger(p_{ 1})\cdots a^\dagger(p_{N})\ket{0}\ .
\label{n_ferm_state_4}
\end{equation}
It is clear from (\ref{n_ferm_state_4}) that
having a label for the particle itself and a separate label for the basis ket
means that assigning fermions 
to basis kets in an antisymmetric way 
is effortless with no need to pay attention to the order of terms.
The anti-symmetric tensor is contracted on indices of  basis kets.
The order of assignment of  $N$ particles 
into $N$ different kets is different between different terms,
where two terms differing by an odd-permutation have opposite signs.
\par
An operator that behaves like an identity on $N$ fermion states
is achieved by the same approach as
\S\ref{sec_two_ferm_states}. Suppose that
the identity is
\begin{equation}
\frac1{N!}
\int \prod^N_{n=1}dp_{k}\ \ket{p_1\dots p_N}\bra{p_1\dots p_N}\label{id_on_N_fermion_states}\end{equation}
with $\ket{p_1\dots p_N}$ given by (\ref{n_ferm_state_4}).
Acting on the analogous state $\ket{q_1\dots q_N}$ produces
\begin{align}
&\frac1{N!}\left(\int \prod^N_{k=1}dp_{k}\ \ket{p_1\dots p_N}\bra{p_1\dots p_N}\right)\ket{q_1\dots q_N}
\nonumber\\
=& \frac1{N!}
\int \prod^N_{k=1}dp_{k}\ \ket{p_1\dots p_N}
\bra{0}
a (p_{N})\dots a (p_{1})
a^\dagger(q_{1})\cdots a^\dagger(q_{N})\ket{0}\ .
\label{n_ferm_state_5}
\end{align}
A way to write 
$\bra{0}
a (p_{N})\dots a (p_{1})
a^\dagger(q_{1})\cdots a^\dagger(q_{N})\ket{0}$
that makes the integrals 
easily solvable is needed.
The following identity can be proved that serves this purpose.
\begin{mylist}[leftmargin=*,series=tobecont]
\item\label{thm_inner_prod_multi_ferm_states_1}\thry{Inner product of multi-fermion states
\uppercase{I}}{
\begin{equation}
\bra{0}a (p_{N})\dots a (p_{1})
a^\dagger(q_{1})\cdots a^\dagger(q_{N})\ket{0}
=\sum_{i_1\dots i_N}\epsilon^{i_1\dots i_N}\delta(p_1-q_{i_1})\delta(p_2-q_{i_2})\dots 
\delta(p_N-q_{i_N})\ .
\label{n_ferm_state_6}
\end{equation}}
\prf{\,\\\noindent
The most straightforward way to prove this statement is by induction.
We start by verifying it for $N=2$. The left-hand side is
\begin{align*}
\bra{0}a (p_{2})a (p_{1})
a^\dagger(q_{1})  a^\dagger(q_{2})\ket{0}
=&
\bra{0}a (p_{2})\{a(p_1),q^\dagger(q_1)\}
a^\dagger(q_{2})\ket{0}
-\bra{0}a (p_{2})
a^\dagger(q_{1})a (p_{1})  a^\dagger(q_{2})\ket{0}
\nonumber\\[1em]
=&
\delta(p_1-q_1)
\bra{0}\{a (p_{2}),
a^\dagger(q_{2})\}\ket{0}
- 
\bra{0}\{a (p_{2}),
a^\dagger(q_{1}) \}\{a(p_1),a^\dagger(q_2)\} \ket{0}
\nonumber\\[1em]
=&
\delta(p_1-q_1)
\delta(p_2-q_2)
-\delta(p_1-q_2)\delta(p_2-q_1)
\nonumber\\[1em]
=&
\sum_{i_1 i_2}\epsilon^{i_1i_2}\delta(p_1-q_{i_1})\delta(p_2-q_{i_2})\ .
\end{align*}
But this is precisely the right of (\ref{n_ferm_state_6}) for $N=2$.
The implication is that (\ref{n_ferm_state_6}) is true for $N=2$.
Now for the inductive step of the prove. Assume that it is true for $N-1$.
We anticommute $a(p_1)$ one place to the right to obtain
\begin{align*}
\bra{0}a (p_{N})\dots a (p_{1})
a^\dagger(q_{1})\cdots a^\dagger(q_{N})\ket{0}
=&\phantom{+}\delta(p_1-q_1)
\bra{0}a (p_{N})\dots a(p_2)a^\dagger(q_2)\cdots a^\dagger(q_{N})\ket{0}
\nonumber\\
&-
\bra{0}a (p_{N})\dots a (p_{2})a^\dagger(q_{1})  a (p_{1})
a^\dagger (q_{2})\cdots a^\dagger(q_{N})\ket{0}\ .
\end{align*}
In the second line we 
anticommute $a(p_1)$ one place to the right again to obtain 
\begin{align*}
\bra{0}a (p_{N})\dots a (p_{1})
a^\dagger(q_{1})\cdots a^\dagger(q_{N})\ket{0}
=&\phantom{+}\delta(p_1-q_1)
\bra{0}a (p_{N})\dots a(p_2)a^\dagger(q_2)\cdots a^\dagger(q_{N})\ket{0}
\nonumber\\
&-\delta(p_1-q_2)
\bra{0}a (p_{N})\dots a (p_{2})a^\dagger(q_{1})  
a^\dagger (q_{3})\cdots a^\dagger(q_{N})\ket{0}
\nonumber\\
&+
\bra{0}a (p_{N})\dots a (p_{2})a^\dagger(q_{1})  a^\dagger(q_{2})  
a (p_{1})
a^\dagger (q_{3})\cdots a^\dagger(q_{N})\ket{0}\ ,
\end{align*}
and we continue this process to eventually end up with
\begin{align*}
&\bra{0}a (p_{N})\dots a (p_{1})
a^\dagger(q_{1})\cdots a^\dagger(q_{N})\ket{0}\nonumber\\
=&\phantom{+}\delta(p_1-q_1)
\bra{0}a (p_{N})\dots a(p_2)a^\dagger(q_2)\cdots a^\dagger(q_{N})\ket{0}
\nonumber\\
&-\delta(p_1-q_2)
\bra{0}a (p_{N})\dots a (p_{2})a^\dagger(q_{1})  
a^\dagger (q_{3})\cdots a^\dagger(q_{N})\ket{0}
\nonumber\\
&+\delta(p_1-q_3)
\bra{0}a (p_{N})\dots a (p_{2})a^\dagger(q_{1})  
a^\dagger (q_{2}) a^\dagger (q_{4})\cdots a^\dagger(q_{N})\ket{0}
\nonumber\\
&\vdots
\nonumber\\
&+(-1)^{N-1}\delta(p_1-q_N)
\bra{0}a (p_{N})\dots a (p_{2})a^\dagger(q_{1})  a^\dagger(q_{2})  
\cdots a^\dagger(q_{N-1})\ket{0}\ ,
\end{align*}
The term next to the delta functions in each line is precisely the left of
(\ref{n_ferm_state_6}) for $N-1$, and since
(\ref{n_ferm_state_6})  is assumed true for $N-1$, the right-hand side can be substituted to yield
\begin{align}
&\bra{0}a (p_{N})\dots a (p_{1})
a^\dagger(q_{1})\cdots a^\dagger(q_{N})\ket{0}\nonumber\\[1em]
=&\phantom{-}\sum_{i_2\dots i_N}\delta(p_1-q_1)
\epsilon^{i_2\dots i_N}\delta(p_2-q_{i_2})\delta(p_3-q_{i_3})\dots\delta(p_N-q_{i_N})\bigg|_{i_k\neq 1} 
\nonumber\\
&-\sum_{i_2\dots i_N}\delta(p_1-q_2)
\epsilon^{i_2\dots i_N}\delta(p_2-q_{i_2})\delta(p_3-q_{i_3})\dots\delta(p_N-q_{i_N})\bigg|_{i_k\neq 2} 
\nonumber\\
&+\sum_{i_2\dots i_N}\delta(p_1-q_3)
\epsilon^{i_2\dots i_N}\delta(p_2-q_{i_2})\delta(p_3-q_{i_3})\dots\delta(p_N-q_{i_N})\bigg|_{i_k\neq 3}
\nonumber\\
&\vdots
\nonumber\\
&+(-1)^{N-1}\sum_{i_2\dots i_N}\delta(p_1-q_N)
\epsilon^{i_2\dots i_N}\delta(p_2-q_{i_2})\delta(p_3-q_{i_3})\dots\delta(p_N-q_{i_N})\bigg|_{i_k\neq N} \ .
\label{n_ferm_state_11}\tag{*}
\end{align}
The exclusion of a specific $i_k$ is excluded from the contraction of the  Levi-Civita symbol
can be expressed as, for example,
$$\sum_{i_2\dots i_N}\epsilon^{i_2\dots i_N}\delta(p_2-q_{i_2})\dots \delta(p_N-q_{i_N})\big|_{i_k\neq 1} 
=\sum_{i_2\dots i_N}
\epsilon^{1i_2\dots i_N}\delta(p_2-q_{i_2})\dots \delta(p_N-q_{i_N}),$$ where the Levi-Civita symbol on the left has $N-1$ indices 
and that on the right has $N$ indices. Hence (\ref{n_ferm_state_11}) becomes 
\begin{align*}
&\bra{0}a (p_{N})\dots a (p_{1})
a^\dagger(q_{1})\cdots a^\dagger(q_{N})\ket{0}\nonumber\\[1em]
=&\phantom{-}\sum_{i_2\dots i_N}\delta(p_1-q_1)
\epsilon^{1i_2\dots i_N}\delta(p_2-q_{i_2})\delta(p_3-q_{i_3})\dots\delta(p_N-q_{i_N})
\nonumber\\
&-\sum_{i_2\dots i_N}\delta(p_1-q_2)
\epsilon^{2i_2\dots i_N}\delta(p_2-q_{i_2})\delta(p_3-q_{i_3})\dots\delta(p_N-q_{i_N})
\nonumber\\
&+\sum_{i_2\dots i_N}\delta(p_1-q_3)
\epsilon^{3i_2\dots i_N}\delta(p_2-q_{i_2})\delta(p_3-q_{i_3})\dots\delta(p_N-q_{i_N})
\nonumber\\
&\vdots
\nonumber\\
&+(-1)^{N-1}\sum_{i_2\dots i_N}\delta(p_1-q_N)
\epsilon^{Ni_2\dots i_N}\delta(p_2-q_{i_2})\delta(p_3-q_{i_3})\dots\delta(p_N-q_{i_N})
\nonumber\\[1em]
=&\sum^N_{k=1}(-1)^{k-1}\sum_{i_2\dots i_N}\epsilon^{ki_2\dots i_N}\delta(p_1-q_k)
\delta(p_2-q_{i_2})\delta(p_3-q_{i_3})\dots\delta(p_N-q_{i_N})\ .
\end{align*}
In each term in the sum, we 
move the index ${}^k$ in the Levi-Civita symbol $k-1$ places to the right and insert 
the accompanying  factor of $(-1)^{k-1}$,
we
re-label dummy indices appropriately and we
replace the sum over $k$ with the Einstein summation convention to end up with
\begin{equation*}
\bra{0}a (p_{N})\dots a (p_{1})
a^\dagger(q_{1})\cdots a^\dagger(q_{N})\ket{0}\nonumber\\[1em]
=\sum_{i_1\dots i_N}\epsilon^{i_1i_2\dots i_N}\delta(p_1-q_{i_1})
\delta(p_2-q_{i_2})\delta(p_3-q_{i_3})\dots\delta(p_N-q_{i_N})\ .
\end{equation*}
This is precisely the right-hand side of (\ref{n_ferm_state_6}). This completes the proof of the assertion in 
(\ref{n_ferm_state_6}) by induction.}
\item\label{thm_inner_prod_multi_ferm_states_2}\thry{Inner product of multi-fermion states
\uppercase{II}}{
\begin{equation}
\bra{0}a (p_{N})\dots a (p_{1})
a^\dagger(q_{1})\cdots a^\dagger(q_{N'})\ket{0}
=\sum_{i_1\dots i_N}\epsilon^{i_1\dots i_N}\delta(p_1-q_{i_1})\delta(p_2-q_{i_2})\dots 
\delta(p_N-q_{i_N})\ \delta_{NN'}.
\label{n_ferm_state_6_again}
\end{equation}}
\prf{\,\\\noindent
There are three possibilities:
\begin{enumerate}[label=(\roman*)]
\item $N'=N$
\item $N'>N$
\item $N'<N$.
\end{enumerate}
Case (i): If $N=N'$ the theorem is proved by theorem \ref{thm_inner_prod_multi_ferm_states_1}.\par 
Case (ii): $N'>N$.
The proof of this statement is by induction and follows  similar lines as the proof of theorem
\ref{thm_inner_prod_multi_ferm_states_1}.
We start by proving it for 
$N=2$:
\begin{align*}
&\bra{0}a (p_{2})a (p_{1})
a^\dagger(q_{1})  a^\dagger(q_{2})\dots a^\dagger_{N'}\ket{0}
\nonumber\\
=&
\bra{0}a (p_{2})\{a(p_1),q^\dagger(q_1)\}
a^\dagger(q_{2})\dots a^\dagger_{N'}\ket{0}
-\bra{0}a (p_{2})
a^\dagger(q_{1})a (p_{1})  a^\dagger(q_{2})\dots a^\dagger_{N'}\ket{0}
\nonumber\\[1em]
=&
\delta(p_1-q_1)
\bra{0}\{a (p_{2}),
a^\dagger(q_{2})\}a^\dagger (q_3)\dots a^\dagger_{N'}\ket{0}
- 
\bra{0}\{a (p_{2}),
a^\dagger(q_{1}) \}\{a(p_1),a^\dagger(q_2)\}a^\dagger (q_3)\dots a^\dagger_{N'} \ket{0}\nonumber\\
&+
\bra{0}\{a (p_{2}),
a^\dagger(q_{1}) \}a^\dagger(q_2) a(p_1) a^\dagger (q_3)\dots a^\dagger_{N'} \ket{0}\nonumber\\
\nonumber\\[1em]
=&
\delta(p_1-q_1)
\delta(p_2-q_2)\bra{0}a^\dagger (q_3)\dots a^\dagger_{N'}\ket{0}
- 
\delta(p_2-q_1)
\delta(p_1-q_2)
\bra{0} a^\dagger (q_3)\dots a^\dagger_{N'} \ket{0}\nonumber\\
&+
\delta(p_2-q_1)
\bra{0} a^\dagger(q_2) a(p_1) a^\dagger (q_3)\dots a^\dagger_{N'} \ket{0}\nonumber\\
\nonumber\\[1em]
=&\begin{cases}
\delta(p_1-q_1)
\delta(p_2-q_2)
-\delta(p_2-q_1)
\delta(p_1-q_2)&N'=2 \\
0&N'>2\end{cases}\nonumber\\
=&
\begin{cases}
\epsilon^{i_1i_2}\delta(p_1-q_{i_1})\delta(p_2-q_{i_2})&N'=2 \\
0&N'>2\end{cases}
\ .
\end{align*}
But this is precisely the right of (\ref{n_ferm_state_6_again}) for $N=2$.
The implication is that (\ref{n_ferm_state_6_again}) is true for $N=2$.
\par
Now for the inductive step of the prove. We assume that it is true for $N-1$.
We anticommute $a(p_1)$ one place to the right to obtain
\begin{align*}
&\bra{0}a (p_{N})\dots a (p_{1})
a^\dagger(q_{1})\cdots a^\dagger(q_{N})a^\dagger(q_{N+1})\dots a^\dagger(q_{N'})\ket{0}\nonumber\\
=&\phantom{+}\delta(p_1-q_1)
\bra{0}a (p_{N})\dots a(p_2)a^\dagger(q_2)\cdots a^\dagger(q_{N})a^\dagger(q_{N+1})\dots a^\dagger(q_{N'})\ket{0}
\nonumber\\
&-
\bra{0}a (p_{N})\dots a (p_{2})a^\dagger(q_{1})  a (p_{1})
a^\dagger (q_{2})\cdots a^\dagger(q_{N})a^\dagger(q_{N+1})\dots a^\dagger(q_{N'})\ket{0}\ .
\end{align*}
In the second line we
anticommute $a(p_1)$ one place to the right again to obtain 
\begin{align*}
&\bra{0}a (p_{N})\dots a (p_{1})
a^\dagger(q_{1})\cdots a^\dagger(q_{N})\ket{0}\nonumber\\
=&\phantom{+}\delta(p_1-q_1)
\bra{0}a (p_{N})\dots a(p_2)a^\dagger(q_2)\cdots a^\dagger(q_{N})a^\dagger(q_{N+1})\dots a^\dagger(q_{N'})\ket{0}
\nonumber\\
&-\delta(p_1-q_2)
\bra{0}a (p_{N})\dots a (p_{2})a^\dagger(q_{1})  
a^\dagger (q_{3})\cdots a^\dagger(q_{N})a^\dagger(q_{N+1})\dots a^\dagger(q_{N'})\ket{0}
\nonumber\\
&+
\bra{0}a (p_{N})\dots a (p_{2})a^\dagger(q_{1})  a^\dagger(q_{2})  
a (p_{1})
a^\dagger (q_{3})\cdots a^\dagger(q_{N})a^\dagger(q_{N+1})\dots a^\dagger(q_{N'})\ket{0}\ ,
\end{align*}
and we continue this process to eventually end up with
\begin{align*}
&\bra{0}a (p_{N})\dots a (p_{1})
a^\dagger(q_{1})\cdots a^\dagger(q_{N})a^\dagger(q_{N+1})\dots a^\dagger(q_{N'})\ket{0}\nonumber\\
=&\phantom{+}\delta(p_1-q_1)
\bra{0}a (p_{N})\dots a(p_2)a^\dagger(q_2)\cdots a^\dagger(q_{N})a^\dagger(q_{N+1})\dots a^\dagger(q_{N'})\ket{0}
\nonumber\\
&-\delta(p_1-q_2)
\bra{0}a (p_{N})\dots a (p_{2})a^\dagger(q_{1})  
a^\dagger (q_{3})\cdots a^\dagger(q_{N})a^\dagger(q_{N+1})\dots a^\dagger(q_{N'})\ket{0}
\nonumber\\
&+\delta(p_1-q_3)
\bra{0}a (p_{N})\dots a (p_{2})a^\dagger(q_{1})  
a^\dagger (q_{2}) a^\dagger (q_{4})\cdots a^\dagger(q_{N})a^\dagger(q_{N+1})\dots a^\dagger(q_{N'})\ket{0}
\nonumber\\
&\vdots
\nonumber\\
&+(-1)^{N-1}\delta(p_1-q_N)
\bra{0}a (p_{N})\dots a (p_{2})a^\dagger(q_{1})  a^\dagger(q_{2})  
\cdots a^\dagger(q_{N-1})\ a^\dagger(q_{N+1})\dots a^\dagger(q_{N'})\ket{0}\ ,
\end{align*}
The term next to the delta functions in each line is precisely the left of
(\ref{n_ferm_state_6}) for $N-1$, $N'-1$ and since
(\ref{n_ferm_state_6})  is assumed true for $N-1$,$N'-1$ the right-hand side can be substituted to yield
\begin{align}
&\bra{0}a (p_{N})\dots a (p_{1})
a^\dagger(q_{1})\cdots a^\dagger(q_{N})a^\dagger(q_{N+1})\dots a^\dagger(q_{N'})\ket{0}\nonumber\\[1em]
=&\phantom{-}\sum_{i_2\dots i_N}\delta(p_1-q_1)
\epsilon^{i_2\dots i_N}\delta(p_2-q_{i_2})\delta(p_3-q_{i_3})\dots\delta(p_N-q_{i_N})\bigg|_{i_k\neq 1} \delta_{N-1,N'-1}
\nonumber\\
&-\sum_{i_2\dots i_N}\delta(p_1-q_2)
\epsilon^{i_2\dots i_N}\delta(p_2-q_{i_2})\delta(p_3-q_{i_3})\dots\delta(p_N-q_{i_N})\bigg|_{i_k\neq 2} \delta_{N-1,N'-1}
\nonumber\\
&+\sum_{i_2\dots i_N}\delta(p_1-q_3)
\epsilon^{i_2\dots i_N}\delta(p_2-q_{i_2})\delta(p_3-q_{i_3})\dots\delta(p_N-q_{i_N})\bigg|_{i_k\neq 3}\delta_{N-1,N'-1}
\nonumber\\
&\vdots
\nonumber\\
&+(-1)^{N-1}\sum_{i_2\dots i_N}\delta(p_1-q_N)
\epsilon^{i_2\dots i_N}\delta(p_2-q_{i_2})\delta(p_3-q_{i_3})\dots\delta(p_N-q_{i_N})\bigg|_{i_k\neq N} \delta_{N-1,N'-1}\ .
\label{n_ferm_state_11}\tag{*}
\end{align}
The exclusion of a specific $i_k$  from the contraction of the  Levi-Civita symbol
can be expressed as, for example,
$\sum_{i_2\dots i_N}\epsilon^{i_2\dots i_N}\delta(p_2-q_{i_2})\dots \delta(p_N-q_{i_N})\big|_{i_k\neq 1} 
=
\sum_{i_2\dots i_N}\epsilon^{1i_2\dots i_N}\delta(p_2-q_{i_2})\dots \delta(p_N-q_{i_N})$, where the Levi-Civita symbol on the left has $N-1$ indices 
and that on the right has $N$ indices. Hence (\ref{n_ferm_state_11}) becomes 
\begin{align*}
&\bra{0}a (p_{N})\dots a (p_{1})
a^\dagger(q_{1})\cdots a^\dagger(q_{N})\ket{0}\nonumber\\[1em]
=&\phantom{-}\sum_{i_2\dots i_N}\delta(p_1-q_1)
\epsilon^{1i_2\dots i_N}\delta(p_2-q_{i_2})\delta(p_3-q_{i_3})\dots\delta(p_N-q_{i_N})\delta_{N ,N' }
\nonumber\\
&-\sum_{i_2\dots i_N}\delta(p_1-q_2)
\epsilon^{2i_2\dots i_N}\delta(p_2-q_{i_2})\delta(p_3-q_{i_3})\dots\delta(p_N-q_{i_N})\delta_{N ,N' }
\nonumber\\
&+\sum_{i_2\dots i_N}\delta(p_1-q_3)
\epsilon^{3i_2\dots i_N}\delta(p_2-q_{i_2})\delta(p_3-q_{i_3})\dots\delta(p_N-q_{i_N})\delta_{N ,N' }
\nonumber\\
&\vdots
\nonumber\\
&+(-1)^{N-1}\sum_{i_2\dots i_N}\delta(p_1-q_N)
\epsilon^{Ni_2\dots i_N}\delta(p_2-q_{i_2})\delta(p_3-q_{i_3})\dots\delta(p_N-q_{i_N})\delta_{N ,N' }
\nonumber\\[1em]
=&\sum^N_{k=1}(-1)^{k-1}\sum_{i_2\dots i_N}\epsilon^{ki_2\dots i_N}\delta(p_1-q_k)
\delta(p_2-q_{i_2})\delta(p_3-q_{i_3})\dots\delta(p_N-q_{i_N})\delta_{N ,N' }\ .
\end{align*}
In each term in the sum,  we
move the index ${}^k$ in the Levi-Civita symbol $k-1$ places to the right,  we insert 
the accompanying  factor of $(-1)^{k-1}$, we
re-label dummy indices appropriately and we
replace the sum over $k$ with the Einstein summation convention to end up with
\begin{equation*}
\bra{0}a (p_{N})\dots a (p_{1})
a^\dagger(q_{1})\cdots a^\dagger(q_{N})\ket{0}\nonumber\\[1em]
=\sum_{i_1\dots i_N}\epsilon^{i_1i_2\dots i_N}\delta(p_1-q_{i_1})
\delta(p_2-q_{i_2})\delta(p_3-q_{i_3})\dots\delta(p_N-q_{i_N})\delta_{N ,N' }\ .
\end{equation*}
This is precisely the right-hand side of (\ref{n_ferm_state_6_again}). This completes the proof of the assertion in 
(\ref{n_ferm_state_6_again}) by induction.}
\end{mylist}
Having verified (\ref{n_ferm_state_6}), we substitute it in 
(\ref{n_ferm_state_5}) to obtain
\begin{align}
&\frac1{N!}\left(\int \prod^N_{k=1}dp_{k}\ \ket{p_1\dots p_N}\bra{p_1\dots p_N}\right)\ket{q_1\dots q_N}
\nonumber\\
=& \frac1{N!}
\int \prod^N_{k=1}dp_{k}\ \ket{p_1\dots p_N}
\sum_{i_1\dots i_N}\epsilon^{i_1i_2\dots i_N}\delta(p_1-q_{i_1})
\delta(p_2-q_{i_2})\delta(p_3-q_{i_3})\dots\delta(p_N-q_{i_N})\ .
\label{n_ferm_state_14}
\end{align}
As promised, in this form it is now a straightforward matter to evaluate the $p_k$ integrals.
The result is
\begin{equation}
\frac1{N!}\left(\int \prod^N_{k=1}dp_{k}\ \ket{p_1\dots p_N}\bra{p_1\dots p_N}\right)\ket{q_1\dots q_N}
=\frac1{N!}\sum_{i_1\dots i_N}
\epsilon^{i_1i_2\dots i_N}\ket{q_{i_1}\dots q_{i_N}}\ .
\label{n_ferm_state_15}
\end{equation}
From  the definition in (\ref{n_ferm_state_4}), namely
$\ket{q_1\dots q_N}=a^\dagger(q_1)\dots a^\dagger (q_N)\ket{0}$,
it is obvious,
that 
\begin{equation}
\sum_{i_1\dots i_N}\epsilon^{i_1i_2\dots i_N}\ket{q_{i_1}\dots q_{i_N}}
=\sum_{i_1\dots i_N}\epsilon^{i_1i_2\dots i_N}a^\dagger(q_{i_1})\dots a^\dagger( q_{i_N})\ket{0}
=
N!\ket{q_1\dots q_N},
\label{n_ferm_state_16}
\end{equation}
i.e. the contraction of the Levi-Civita on  $\ket{q_{i_1}\dots q_{i_N}}$, which is already completely antisymmetric by definition,
simply results in a sum over $N!$ permutations. 
The conclusion is that
\begin{equation}
\frac1{N!}\left(\int \prod^N_{k=1}dp_{k}\ \ket{p_1\dots p_N}\bra{p_1\dots p_N}\right)\ket{q_1\dots q_N}
=
\ket{q_1\dots q_N}\ ,
\label{n_ferm_state_17}
\end{equation}
i.e. the operator $\frac1{N!}\int \prod^N_{k=1}dp_{k}\ \ket{p_1\dots p_N}\bra{p_1\dots p_N} $ acts as an identity on $N$ fold multi-fermion states
defined in (\ref{n_ferm_state_4}).
\par 
A corollary of (\ref{n_ferm_state_6}) is that 
\begin{align}
\braket{q_1\dots q_N|p_1\dots p_N}=&\bra{0}a(q_N)\dots a(q_1)a^\dagger (p_1)\dots a^\dagger (p_N)\ket{0}
\nonumber\\[1em]
=&\sum_{i_1\dots i_N}\epsilon^{i_1\dots i_N}\delta(p_1-q_{i_1})\dots \delta(p_N-q_{i_N})\ .
\label{n_ferm_state_18}
\end{align}
\begin{mylist}[leftmargin=*,resume=tobecont]
\item\label{thry_anticom_arbt_num_fermion_ops_1}\thry{Anticommutator of an  arbitrary number of fermion operators
}{
\begin{equation}
\{b,a_1\dots a_N\}=\sum^N_{k=1}(-1)^{k-1}a_1\dots a_{k-1}\{b,a_k\}a_{k+1}\dots a_N
+(1+(-1)^{N})a_1\dots a_N b\ .
\label{anti_comm_1}
\end{equation}}
\prf{\,\\\noindent
The most straightforward way to prove this statement is by induction.
We start by verifying it for $N=2$:
\begin{align*}
\{b,a_1a_2\}=&
ba_1a_2+a_1a_2b
\nonumber\\
=&
\{b,a_1\}a_2-a_1 ba_2+a_1a_2b
\nonumber\\
=&
\{b,a_1\}a_2-a_1 \{b,a_2\}+2a_1a_2b
\ .
\end{align*}
But this is precisely the right of (\ref{anti_comm_1}) for $N=2$.
The implication is that (\ref{anti_comm_1}) is true for $N=2$.
Now for the inductive step of the prove. We assume that it is true for $N=n-1$.
\begin{equation*}
\{b,a_1\dots a_n\}=
ba_1\dots a_n+a_1\dots a_n b=
\{b,a_1\dots a_{n-1}\}a_n-a_1\dots a_{n-1}ba_n+a_1\dots a_n b
\end{equation*}
But if (\ref{anti_comm_1}) is true for $N=n-1$, then 
\begin{align*}
\{b,a_1\dots a_n\}
=&
\sum^{n-1}_{k=1}(-1)^{k-1}a_1\dots a_{k-1}\{b,a_k\}a_{k+1}\dots a_{n-1}   a_n
\nonumber\\
&+(1+(-1)^{n-1})a_1\dots a_{n-1}ba_n
-a_1\dots a_{n-1}ba_n+a_1\dots a_n b
\nonumber\\[1em]
=&
\sum^{n-1}_{k=1}(-1)^{k-1}a_1\dots a_{k-1}\{b,a_k\}a_{k+1}\dots a_{n-1}   a_n
+ (-1)^{n-1} a_1\dots a_{n-1}b a_n
+a_1\dots a_{n-1} a_nb
\nonumber\\[1em]
=&
\sum^{n-1}_{k=1}(-1)^{k-1}a_1\dots a_{k-1}\{b,a_k\}a_{k+1}\dots a_{n-1}   a_n
\nonumber\\&
+ (-1)^{n-1} a_1\dots a_{n-1}\{b,a_n\}
- (-1)^{n-1} a_1\dots a_{n-1}a_n b
+a_1\dots a_n b
\nonumber\\[1em]
=&
\sum^{n}_{k=1}(-1)^{k-1}a_1\dots a_{k-1}\{b,a_k\}a_{k+1}\dots a_{n-1}   a_n
+ (-1)^{n} a_1\dots a_{n-1}a_n b
+a_1\dots a_n b
\nonumber\\[1em]
=&
\sum^{n}_{k=1}(-1)^{k-1}a_1\dots a_{k-1}\{b,a_k\}a_{k+1}\dots a_{n-1}   a_n
+(1+(-1)^n)a_1\dots a_n b\ .
\end{align*}
But this is none other than (\ref{anti_comm_1})  for $N=n$. 
Hence, if (\ref{anti_comm_1}) holds for $N=n-1$ it must also be true for $N=n$. 
This proves (\ref{anti_comm_1}), by induction.}\end{mylist}
\subsection{Derived identities involving the projection operator onto $N>1$ particle states}\label{sec_ids_involving_PiN}\noindent
The projection operator onto $N$ particle states is defined as
\begin{equation}\hat \Pi_N=\frac1{N!}\int dp_1 ... dp_N \ket{p_1 ...p_N}\bra{p_1...p_N}\ .\label{projectio_operator_appendix}\end{equation}
The 
form of the projection operator that shall be assumed is 
\begin{equation}
\hat \Pi_N=\frac1{N!}\int dx_1\dots dx_N \, a^\dagger(x_1)\dots a^\dagger(x_N)\ket{0}\bra{0}a (x_N)\dots a (x_1)\ .
\label{the_form_of_PiN_appendix}
\end{equation}
This is consistent with the requirement that, given a state $\ket{\psi}=\ket{x_1\dots x_{N'}}=a^\dagger(x_1)\dots
a^\dagger(x_N')\ket{0}$,
then $\hat \Pi_N\ket{\psi}=\delta_{NN'}\ket{\psi}$,
which is easily shown to be true by invoking theorem \ref{thm_inner_prod_multi_ferm_states_1}.
Based on (\ref{the_form_of_PiN}) and the form of the Hamiltonian in (\ref{top_inv_var_partcl_numb_3}),
then it can be shown that the two commute:
\begin{equation}
[\hat H,\hat \Pi_N]=0.\label{H_commutes_with_PiN_appendix}
\end{equation}
To show that they commute we substitute their explicit forms:
\begin{align}
[\hat H,\hat \Pi_N]=&
\int dX\, {\mathscr H}_0(X) \frac1{N!} 
\int dx_1\dots dx_N \, 
[a^\dagger (X) a(X),a^\dagger(x_1)\dots a^\dagger(x_N)\ket{0}\bra{0}a (x_N)\dots a (x_1)]
\nonumber\\[1em]
&+
\int dX\,dY\, {\mathscr V}(X-Y) \frac1{N!} 
\int dx_1\dots dx_N \,\nonumber\\&
[a^\dagger (X) a(X)a^\dagger (Y) a(Y),a^\dagger(x_1)\dots a^\dagger(x_N)\ket{0}\bra{0}a (x_N)\dots a (x_1)]
\nonumber\\[2em]
=&
\int dX\, {\mathscr H}_0(X) \frac1{N!} 
\int dx_1\dots dx_N \, 
\bigg\{
a^\dagger (X) a(X)a^\dagger(x_1)\dots a^\dagger(x_N)\ket{0}\bra{0}a (x_N)\dots a (x_1)
\nonumber\\
&-
a^\dagger(x_1)\dots a^\dagger(x_N)\ket{0}\bra{0}a (x_N)\dots a (x_1)
a^\dagger (X) a(X)
\bigg\}
\nonumber\\[1em]
&+
\int dX\,dY\, {\mathscr V}(X-Y) \frac1{N!} 
\int dx_1\dots dx_N \,\nonumber\\&
\bigg\{
a^\dagger (X) a(X)
a^\dagger (Y) a(Y)
a^\dagger(x_1)\dots a^\dagger(x_N)\ket{0}\bra{0}a (x_N)\dots a (x_1)
\nonumber\\
&-
a^\dagger(x_1)\dots a^\dagger(x_N)\ket{0}\bra{0}a (x_N)\dots a (x_1)
a^\dagger (X) a(X)
a^\dagger (Y) a(Y)
\bigg\}
\label{Pi_commutes_with_H_1_appendix}
\end{align}
Note that 
\begin{equation}
a^\dagger (X) a(X)a^\dagger(x_1)\dots a^\dagger(x_N)\ket{0}
=
\sum^N_{i=1}\delta(X-x_i)a^\dagger(x_1)\dots a^\dagger (x_{i-1})a^\dagger (X) a^\dagger (x_{i+1})\dots a^\dagger(x_N)\ket{0}
\label{Pi_commutes_with_H_2_appendix}\end{equation}
\begin{equation}
\bra{0}a (x_N)\dots a (x_1)a^\dagger (X) a(X) 
=
\sum^N_{i=1}\delta(X-x_i)\bra{0}a (x_N)\dots a(x_{i+1})a(X)a(x_{i-1})\dots a(x_1)
\label{Pi_commutes_with_H_3_appendix}\end{equation}
{\small \begin{align}&a^\dagger (Y) a(Y)
a^\dagger(x_1)\dots a^\dagger(x_N)\ket{0}\nonumber\\
=&
\sum^N_{\substack{i,j=1\\i\neq j}}\delta(X-x_i)\delta(Y-x_j)
a^\dagger(x_1)\dots a^\dagger (x_{i-1})a^\dagger (X) a^\dagger (x_{i+1})
\dots 
a^\dagger (x_{j-1})a^\dagger (Y) a^\dagger (x_{j+1})
\dots a^\dagger(x_N)\ket{0}
\label{Pi_commutes_with_H_4_appendix}\end{align}}\noindent
{\small \begin{align}&\bra{0}a (x_N)\dots a (x_1)a^\dagger (X) a(X)a^\dagger (Y) a(Y)\nonumber\\
=&
\sum^N_{\substack{i,j=1\\i\neq j}}\delta(X-x_i)\delta(Y-x_j)\bra{0}a (x_N)
\dots 
a (x_{j-1})a (Y) a (x_{j+1})
\dots a(x_{i+1})a(X)a(x_{i-1})\dots a(x_1)
\label{Pi_commutes_with_H_5_appendix}
\end{align}}\noindent
such that by substituting (\ref{Pi_commutes_with_H_2_appendix})--(\ref{Pi_commutes_with_H_5_appendix})
into (\ref{Pi_commutes_with_H_1_appendix}) and integrating over $X$ and $Y$ the result is
\begin{align}
[\hat H,\hat \Pi_N]
=&
\sum^N_{i=1}{\mathscr H}_0(x_i) \frac1{N!} 
\int dx_1\dots dx_N \,\nonumber\\&
\bigg\{
a^\dagger(x_1)\dots a^\dagger(x_N)\ket{0}\bra{0}a (x_N)\dots a (x_1)
-
a^\dagger(x_1)\dots a^\dagger(x_N)\ket{0}\bra{0}a (x_N)\dots a (x_1)
\bigg\}
\nonumber\\
&+
\sum^N_{\substack{i,j=1\\i\neq j}} {\mathscr V}(x_i-x_j) \frac1{N!} 
\int dx_1\dots dx_N \,\nonumber\\&
\bigg\{
a^\dagger(x_1)\dots a^\dagger(x_N)\ket{0}\bra{0}a (x_N)\dots a (x_1)
-
a^\dagger(x_1)\dots a^\dagger(x_N)\ket{0}\bra{0}a (x_N)\dots a (x_1)
\bigg\}
\nonumber\\[2em]
=0
\label{Pi_commutes_with_H_6_appendix}
\end{align}
\section{Miscellaneous identities for Weyl symbols with $N>1$}\label{sec_ids_weyl_symbols}\noindent
The \emph{Moyal product} of the Weyl symbols of two
operators $\hat A$ and $\hat B$ is defined as
\begin{align}
&A_W(\{x_a\},\{p_a\}) \star B_W(\{x_a\},\{p_a\}) \nonumber\\=&
A_W(\{x_a\},\{p_a\})\exp\left[\frac{i}{2} \sum^N_{a=1}\sum^2_{i=1}\left( 
\overleftarrow{\f\PD{\PD x_a^i}}\ \overrightarrow{\f\PD{\PD p^i_{a}}}
-
\overleftarrow{\f\PD{\PD p_a^i}}\ \overrightarrow{\f\PD{\PD x^i_{a}}}
\right )\right] B_W(\{x_a\},\{p_a\})\ .
\label{moyal_prod_6}
\end{align} 
The functional trace of a Weyl symbol of an operator on single-particle states is defined as
\cite{Fialkovsky_1}
\begin{equation}
\Tr A_W(  x,  p )\equiv\frac1{(2\pi)^D}
\int dx \, dp \ \tr A_W(  x,  p )\ .
\label{misc_m3}
\end{equation}
The analogous expression for $N$ identical particles is
\begin{equation}
\Tr A_W(x_1,\dots x_N,p_1,\dots ,p_N)\equiv
\frac1{(2\pi)^{ND}}
\int dx_1\dots dx_N \, dp_1\dots dp_N \ \tr A_W(x_1,\dots x_N,p_1,\dots ,p_N )\ .
\label{misc_m3A}
\end{equation}\par 
The Wigner transformation of the operator $\hat A$ that acts on 
one-particle states is
\begin{equation}
A_W(p,x) = \int  dq  \ e^{iq  x }  \langle   p +\tfrac{q}{2}|\hat{A} | p -\tfrac{q }{2} \rangle\ ,
\label{AW_1}
\end{equation}
The Wigner transformation of the product of two operators, by analogy with 
(\ref{AW_1}), is
\begin{equation}
\left(AB\right)_W(p,x) = \int  dq  \ e^{iq  x }  \langle   p +\tfrac{q}{2}|\hat{A} \hat B| p -\tfrac{q }{2} \rangle
=
\int  dq \, dQ \ e^{iq  x }  \langle   p +\tfrac{q}{2}|\hat{A}\ket{Q}\bra{Q} \hat B| p -\tfrac{q }{2} \rangle\ ,
\label{ABW_1}
\end{equation}
where the analogue of (\ref{id_on_N_fermion_states}) for $N=1$ was substituted  ($\int dQ\ket{Q}\bra{Q}=1$ when acting on one-particle states).
Through the change of variables $q=u+v$, $Q=p-\frac u{2}+\frac v{2}$
with the associated Jacobian
$\PD(q,Q)/\PD(u,v)=1$,
(\ref{ABW_1}) takes the form
\begin{align}
\left(AB\right)_W(p,x) =&
\int  du \, dv \ e^{i(u+v)  x }  \langle   p +\tfrac{u}{2}+\tfrac{v}{2}|\hat{A}\ket{p-\tfrac{u}2+\tfrac{v}2}\bra{p-\tfrac{u}2+\tfrac{v}2}
\hat B| p-\tfrac{u}2-\tfrac{v}2 \rangle
\nonumber\\
=&
\int  du  \, e^{iux }  \langle   p +\tfrac{u}{2}|\hat{A}\ket{p-\tfrac{u}2}
\exp\left(\frac{v}2\overleftarrow{\PD_p}-\frac u{2}\overrightarrow{\PD_p}\right)
\int  dv  \, e^{ivx }\bra{p+\tfrac{v}2}
\hat B| p-\tfrac{v}2 \rangle
\nonumber\\
=&
\int  du  \, e^{iux }  \langle   p +\tfrac{u}{2}|\hat{A}\ket{p-\tfrac{u}2}
\exp\left(\frac{i}2\overleftarrow{\PD_p}\overrightarrow{\PD_x}-\frac{i}2\overleftarrow{\PD_x}\overrightarrow{\PD_p}\right)
\int  dv  \, e^{ivx }\bra{p+\tfrac{v}2}
\hat B| p-\tfrac{v}2 \rangle
\nonumber\\
=&
A_W(x,p)\star B_W(x,p)
\ .
\label{ABW_2}
\end{align}
Now to generalize this result to   
operators on $N>1$ particle states,
the Wigner transformation of the operator $\hat{A}$ is defined as a function of $2N+1$ variables 
$\omega, p_a, x_a$
($a=1,...,N$),  
in terms of its matrix elements in momentum space:
\begin{equation}
\int dp_1\dots dp_N\frac1{(2\pi)^{ND}}\int dx_1\dots dx_N A_W(\{x_a\},\{p_a\})=\frac1{N!}\int
dp_1\dots dp_N\braket{\{p_a\}|\hat A|\{p_a\}}\ .\label{ABW_2A}
\end{equation}
And
\begin{equation}
A_W(\omega,\{p_a\},\{x_a\}) = \frac1{N!}\int \left(\prod\limits_{a=1}^N dq_a \ e^{iq_a x_a}\right)  
\langle  \{ p_a +\tfrac{q_a}{2}\}|\hat{A} |\{ p_a -\tfrac{q_a }{2}\} \rangle\ .
\label{misc_m2}
\end{equation}
 \sloppy
The extra factor of $1/N!$ ensures that 
the $N$-particle extension of the result in (\ref{ABW_2})
is the same.
That is, given two operators $\hat A$ and $\hat B$ that act on $N$-particle states,
the Wigner transformation of the product of the two operators is the $N$-particle generalization 
of (\ref{ABW_1}):
\begin{align}
\left(AB\right)_W(\{p_a\},\{x_a\}) 
=& \frac1{N!}\int  \left(\prod^N_{a=1} dq_a  \ e^{iq_a  x_a }\right)  
\langle  \{ p_a +\tfrac{q_a}{2}\}|\hat{A} \hat B| \{p_a -\tfrac{q_a }{2} \}\rangle
\nonumber\\
=&\frac1{N!^2}
\int \left(\prod^N_{a=1} dq_a\,dQ_a  \ e^{iq_a  x_a }\right)  
\langle   \{p_a +\tfrac{q_a}{2}\}|\hat{A}\ket{\{Q_a\}}\bra{\{Q_a\}} \hat B| \{p_a -\tfrac{q_a }{2}\} \rangle\ .
\label{ABW_N}
\end{align}
Note the presence of the extra factor of $1/N!$ in the second equality, coming from 
(\ref{id_on_N_fermion_states}).
Following the same steps as in (\ref{ABW_2}), \eq{ABW_N}
yields
\begin{align}
\left(AB\right)_W(\{p_a\},\{x_a\}) 
=&\frac1{N!}
\int  \left(\prod^N_{a=1} du_a  \ e^{iu_a  x_a }\right)    \langle  \{ p_a +\tfrac{u_a}{2}\}|\hat{A}\ket{\{p_a-\tfrac{u_a}2\}}
\nonumber\\
&
\exp\left[\frac{i}{2} \sum^N_{a=1}\sum^2_{i=1}\left( 
\overleftarrow{\f\PD{\PD x_a^i}}\ \overrightarrow{\f\PD{\PD p^i_{a}}}
-
\overleftarrow{\f\PD{\PD p_a^i}}\ \overrightarrow{\f\PD{\PD x^i_{a}}}
\right )\right] 
\nonumber\\
&
\frac1{N!}
\int  \left(\prod^N_{a=1} dv_a  \ e^{iv_a  x_a }\right)
\bra{\{p_a+\tfrac{v_a}2\}}
\hat B|\{ p_a-\tfrac{v_a}2\} \rangle
\nonumber\\
=&
A_W(\{p_a\},\{x_a\}) \star B_W(\{p_a\},\{x_a\}) \ ,
\label{ABW_N2}
\end{align}
as required.
\par 
From the definition
of
(\ref{misc_m2})
an important property of $A_W(\omega,\{p_a\},\{x_a\})$
emerges.
Take for example, the appropriate expression for $N=2$, viz
\begin{equation}
A_W(\omega,p_1,p_2,x_1,x_2)= \frac1{2}\int dq_1\, dq_2\, e^{iq_1x_1+i q_2 x_2}\,\langle p_1+\tfrac{q_1}2,p_2+\tfrac{q_2}2\big|\hat A\big|
p_1-\tfrac{q_1}2,p_2-\tfrac{q_2}2\rangle
\label{change_of_order}
\end{equation}
paying attention to the order of variables in the ket 
$\big|p_1-\tfrac{q_1}2,p_2-\tfrac{q_2}2\rangle$ and similarly for the bra vector
$\langle p_1+\tfrac{q_1}2,p_2+\tfrac{q_2}2\big|$.
It then follows from the definition in (\ref{n_ferm_state_4})
that exchanging the order of variables results in a change in sign:
\begin{equation}
\big|p_2-\tfrac{q_2}2,p_1-\tfrac{q_1}2\rangle=-\big|
p_1-\tfrac{q_1}2,p_2-\tfrac{q_2}2\rangle\ ,\qquad 
\langle p_2+\tfrac{q_2}2,p_1+\tfrac{q_1}2\big|
=-
\langle p_1+\tfrac{q_1}2,p_2+\tfrac{q_2}2\big|\ .
\end{equation}
Under such a change of order of variables in the two-state vectors the expression in
(\ref{change_of_order}) 
becomes
\begin{align}
A_W(\omega,p_1,p_2,x_1,x_2)
=&\frac{
(-1)^2}2\int dq_1\, dq_2\, e^{iq_1x_1+i q_2 x_2}\,\langle p_2+\tfrac{q_2}2,p_1+\tfrac{q_1}2\big|\hat A\big|
p_2-\tfrac{q_2}2,p_1-\tfrac{q_1}2\rangle\ ,
\label{change_of_order_1}
\end{align}
with two minus signs that enter in the form on the right that cancel.
Now by relabeling the dummy integration variables we obtain
\begin{align}
A_W(\omega,p_1,p_2,x_1,x_2)
=&
\int dq_2\, dq_1\, e^{iq_2x_1+i q_1 x_2}\,\langle p_2+\tfrac{q_1}2,p_1+\tfrac{q_2}2\big|\hat A\big|
p_2-\tfrac{q_1}2,p_1-\tfrac{q_2}2\rangle\ \nonumber\\
=&
\int dq_1\, dq_2\, e^{i q_1 x_2+iq_2x_1}\,\langle p_2+\tfrac{q_1}2,p_1+\tfrac{q_2}2\big|\hat A\big|
p_2-\tfrac{q_1}2,p_1-\tfrac{q_2}2\rangle\ ,
\label{change_of_order_2}
\end{align}
which, upon comparison with (\ref{change_of_order}),
is found to be  precisely the same expression but with the variables 
$p_1$, $p_2$ interchanged and 
the variables
$x_1$, $x_2$ interchanged. The upshot is that
\begin{equation}
A_W(\omega,p_1,p_2,x_1,x_2)
=
A_W(\omega,p_2,p_1,x_2,x_1)\ .
\label{change_of_order_3}
\end{equation}
It can be shown using analogous examples generalized to the case of states greater than two particles that 
\begin{equation}
A_W\left(\omega,p_1,p_2,\dots, p_N,x_1,x_2,\dots x_N\right)
=
A_W\left(\omega,p_{\sigma(1)},p_{\sigma(2)},\dots ,p_{\sigma(N)},x_{\sigma(1)},x_{\sigma(2)},\dots ,x_{\sigma(N)}\right)\ .
\label{change_of_order_4}
\end{equation}
where $\sigma(i)$ is the number in position $i$
under a permutation $\sigma$.
\par 
The Weyl symbol of a product of operators, 
$(AB)_W(x,p)$
is defined as
\cite{Fialkovsky_2}
\begin{equation}
(AB)_W(\{x_a\},\{p_a\}):=A_W(\{x_a\},\{p_a\}) \star B_W(\{x_a\},\{p_a\}) 
\label{misc_m1}
\end{equation} 
where the Moyal product or $\star$ product  of the Weyl symbols of two operators $\hat A$ and $\hat B$ is defined
as \cite{Fialkovsky_3}
\begin{align}
&A_W(\{x_a\},\{p_a\}) \star B_W(\{x_a\},\{p_a\}) \nonumber\\=&
A_W(\{x_a\},\{p_a\})\exp\left[\frac{i}{2} \sum^N_{a=1}\sum^2_{i=1}\left( 
\overleftarrow{\f\PD{\PD x_a^i}}\ \overrightarrow{\f\PD{\PD p^i_{a}}}
-
\overleftarrow{\f\PD{\PD p_a^i}}\ \overrightarrow{\f\PD{\PD x^i_{a}}}
\right )\right] B_W(\{x_a\},\{p_a\})\ .
\label{fxd_numb_partcls_8}
\end{align} 
where the subscript $a=1,\dots ,N$ distinguishes between variables that belong to the $N$ different particles and 
the label $i=1,2=x,y$ refers to the component each variable in the lattice plane.
To clarify 
the expression in 
(\ref{fxd_numb_partcls_8})
take the case of $N=2$:
{\small \begin{align}
&\left(AB\right)_W(\{x_a\},\{p_a\}) \nonumber\\
=&
\left(AB\right)_W(x_1,x_2,p_1,p_2) \nonumber\\
=&
A_W(x_1,x_2,p_1,p_2) \star B_W(x_1,x_2,p_1,p_2)
\nonumber\\
=&
A_W(x_1,x_2,p_1,p_2) 
\exp\left[\frac{i}{2}  \sum^2_{i=1}\left( 
\overleftarrow{\f\PD{\PD x_1^i}}\ \overrightarrow{\f\PD{\PD p^i_{1}}}
+
\overleftarrow{\f\PD{\PD x_2^i}}\ \overrightarrow{\f\PD{\PD p^i_{2}}}
-
\overleftarrow{\f\PD{\PD p_1^i}}\ \overrightarrow{\f\PD{\PD x^i_{1}}}
-
\overleftarrow{\f\PD{\PD p_2^i}}\ \overrightarrow{\f\PD{\PD x^i_{2}}}
\right)\right] 
B_W(x_1,x_2,p_1,p_2)\ .
\end{align} }\noindent
But thanks to the property (\ref{change_of_order_3}),
$A_W(x_1,x_2,p_1,p_2) =A_W(x_2,x_1,p_2,p_1) $
and similarly
$B_W(x_1,x_2,p_1,p_2) =B_W(x_2,x_1,p_2,p_1) $,
hence, since the interchange of the variables 
$p_1,p_2$  and interchange of the variables
$x_1,x_2$
inside the $\star$ operator does not change the operator, such that 
\begin{equation}
\left(AB\right)_W(x_1,x_2,p_1,p_2) =\left(AB\right)_W(x_2,x_1,p_2,p_1)\ .\label{change_of_order_5} 
\end{equation}
(\ref{change_of_order_5}) is the invariance property analogous to the one found in 
(\ref{change_of_order_3}) for just one Weyl symbol.
Similar arguments using 
(\ref{change_of_order_4})
leads to a generalization of (\ref{change_of_order_5}) for $N>2$, namely
\begin{equation}
\left(AB\right)_W\left(
p_1,p_2,\dots ,p_N,x_1,x_2,\dots ,x_N
\right) =\left(AB\right)_W\left(p_{\sigma(1)},p_{\sigma(2)},\dots ,p_{\sigma(N)},x_{\sigma(1)},x_{\sigma(2)},\dots ,x_{\sigma(N)}\right)\ .\label{change_of_order_6} 
\end{equation}
\par 


\begin{thebibliography}{10}

\bibitem{vonKlitzing:1980pdk}
K.~von Klitzing, G.~Dorda, and M.~Pepper.
\newblock {New method for high accuracy determination of the fine structure
  constant based on quantized Hall resistance}.
\newblock {\em Phys. Rev. Lett.}, 45:494--497, 1980.

\bibitem{Thouless:1982zz}
D.~J. Thouless, M.~Kohmoto, M.~P. Nightingale, and M.~den Nijs.
\newblock {Quantized Hall Conductance in a Two-Dimensional Periodic Potential}.
\newblock {\em Phys. Rev. Lett.}, 49:405--408, 1982.

\bibitem{Tong:QHE}
David Tong.
\newblock {\it `` The Quantum Hall Effect.''}.
\newblock \url{http://www.damtp.cam.ac.uk/user/tong/qhe/two.pdf}.

\bibitem{Girvin:99}
Steven~M. Girvin.
\newblock The quantum hall effect: Novel excitations and broken symmetries.
\newblock 1999.

\bibitem{goerbig2009quantum}
M.~O. Goerbig.
\newblock Quantum hall effects.
\newblock 2009.

\bibitem{Witten:2015aoa}
Edward Witten.
\newblock {Three lectures on topological phases of matter}.
\newblock {\em Riv. Nuovo Cim.}, 39(7):313--370, 2016.

\bibitem{Nayak:04}
Chetan Nayak.
\newblock {Quantum Condensed Matter Physics - Lecture}.
\newblock {\em Cornell}, 2013.

\bibitem{Volovik:2003fe}
G.~E. Volovik.
\newblock {\em {The Universe in a helium droplet}}, volume 117.
\newblock 2006.

\bibitem{Ishikawa:1986wx}
Kenzo Ishikawa and Toyoki Matsuyama.
\newblock {Magnetic Field Induced Multi Component {QED} in Three-dimensions and
  Quantum Hall Effect}.
\newblock {\em Z. Phys. C}, 33:41, 1986.

\bibitem{volovik:90}
Grigory Volovik.
\newblock The gravitational-topological chern-simons term in a film of
  superfluid 3ha.
\newblock {\em Soviet Journal of Experimental and Theoretical Physics Letters},
  01 1990.

\bibitem{Coleman:1985zi}
Sidney~R. Coleman and Brian~Russell Hill.
\newblock {No More Corrections to the Topological Mass Term in QED in
  Three-Dimensions}.
\newblock {\em Phys. Lett. B}, 159:184--188, 1985.

\bibitem{Lee:1985pg}
Taejin Lee.
\newblock {The Absence of Radiative Corrections From Higher Order Loops to
  Topological Mass in (2+1)-dimensional Electrodynamics}.
\newblock {\em Phys. Lett. B}, 171:247--250, 1986.

\bibitem{Zhang:2019mpf}
C.~X. Zhang and M.~A. Zubkov.
\newblock {Influence of interactions on the anomalous quantum Hall effect}.
\newblock {\em J. Phys. A}, 53(19):195002, 2020.
\newblock [Erratum: J.Phys.A 54, 329501 (2021)].

\bibitem{Zubkov:2019amq}
M.~A. Zubkov and Xi~Wu.
\newblock {Topological invariant in terms of the Green functions for the
  Quantum Hall Effect in the presence of varying magnetic field}.
\newblock {\em Annals Phys.}, 418:168179, 2020.
\newblock [Erratum: Annals Phys. 430, 168510 (2021)].

\bibitem{Fialkovsky:2019nso}
Ignat~V. Fialkovsky and Mikhail~A. Zubkov.
\newblock {Elastic Deformations and Wigner\textendash{}Weyl Formalism in
  Graphene}.
\newblock {\em Symmetry}, 12(2):317, 2020.

\bibitem{Zhang:2019zqa}
C.~X. Zhang and M.~A. Zubkov.
\newblock {Hall Conductivity as the Topological Invariant in the Phase Space in
  the Presence of Interactions and a Nonuniform Magnetic Field}.
\newblock {\em Pisma Zh. Eksp. Teor. Fiz.}, 110(7):480--481, 2019.

\bibitem{Suleymanov:2020wcb}
M.~Suleymanov and M.~A. Zubkov.
\newblock {Chiral separation effect in nonhomogeneous systems}.
\newblock {\em Phys. Rev. D}, 102(7):076019, 2020.



\bibitem{Volovik:88}
Volovik.
\newblock {\it ``An analog of the quantum Hall effect in a superfluid 3He
  film''}.
\newblock JETP {\bf 67}, 9 (1988), zhETF, Vol. 94, No. 3(9), 123.

\bibitem{Volovik1}
see \S 21.2.1 in \cite{Volovik:2003fe}.

\bibitem{Fialkovsky:2019dmc}
I.~V. Fialkovsky, M.~Suleymanov, Xi~Wu, \textdaggerdbl{}. C.~X. Zhang, and
  M.~A. Zubkov.
\newblock {Hall conductivity as topological invariant in phase space}.
\newblock {\em Phys. Scripta}, 95(6):064003, 2020.

\bibitem{Suleymanov:2018hkm}
M.~Suleymanov and M.~A. Zubkov.
\newblock {Wigner\textendash{}Weyl formalism and the propagator of Wilson
  fermions in the presence of varying external electromagnetic field}.
\newblock {\em Nucl. Phys. B}, 938:171--199, 2019.
\newblock [Erratum: Nucl.Phys.B 946, 114674 (2019)].

\bibitem{2} R. Kubo, H. Hasegawa, and N. Hashitsume, Journal of the Physical Society of Japan 14, 56 (1959).

\bibitem{3} Q. Niu, D. J. Thouless, and Y.-S. Wu, Quantized Hall conductance as a topological invariant, Phys. Rev. B 31, 3372
(1985).

\bibitem{4} B. L. Altshuler, D. Khmel'nitzkii, A. I. Larkin, and P. A. Lee, Phys.Rev.B , 5142 (1980).

\bibitem{5} B. L. Altshuler and A. G. Aronov, Electron-electron inter-action in disordered systems (Editors: A (L. Efros, M. Pollak,
Elsevier, North Holland, Amsterdam, 1985).

\bibitem{6} J. E. Avron, R. Seiler, and B. Simon, Homotopy and quantization in condensed matter physics, Phys. Rev. Lett. 51, 51
(1983).

\bibitem{7} E. Fradkin, Field Theories of Condensed Matter Physics (Addison Wesley Publishing Company, 1991).

\bibitem{8} Y. Hatsugai, Topological aspects of the quantum Hall effect, J. Phys. Condens. Matter 9, 2507 (1997).

\bibitem{9}
X.-L. Qi, T. L. Hughes, and S.-C. Zhang, Topological field theory of time-reversal invariant insulators, Phys. Rev. B 78,
195424 (2008).

\bibitem{10}
R. M. Kaufmann, D. Li, and B. Wehefritz-Kaufmann, Notes on topological insulators, Rev. Math. Phys. 28, 1630003
(2016), arXiv:1501.02874 [math-ph].

\bibitem{11}
 Q. Niu, D. J. Thouless, and Y. Wu, Phys, Rev. B 31, 3372 (1985).

\bibitem{Groenewold:1946kp}
H.~J. Groenewold.
\newblock {On the Principles of elementary quantum mechanics}.
\newblock {\em Physica}, 12:405--460, 1946.

\bibitem{Moyal:1949sk}
J.~E. Moyal.
\newblock {Quantum mechanics as a statistical theory}.
\newblock {\em Proc. Cambridge Phil. Soc.}, 45:99--124, 1949.

\bibitem{Dirac_permutation_notation}
See \cite{Dirac:67} \S59 p.225. Here the discussion is about boson states but
  the notation, described in detail here, applies also to fermion states, as
  described later on in \cite{Dirac:67} \S65 p.248-259. Dirac writes
  $\ket{\alpha^a_1\ \alpha^b_2\ \dots\ \alpha^g_{u'}}$, each $\alpha$
  corresponding to a particle, where the suffixes $1,2,3,\dots, u'$ label the
  particles themselves, while $a,b,c\dots,g$ denote the indices $\,{}^{(1)}$,
  $\,{}^{(2)}$,$\,{}^{(3)}$, $\dots$ in the basic kets for one particle, or in
  equivalent terms $a,b,c\dots,g$ label the actual states in which the
  particles lie.

\bibitem{Dirac:67}
P.A.M. Dirac.
\newblock {\em {\it\, ``The principles of quantum mechanics''\,}}.
\newblock Oxford Science Publications, 1967.

\bibitem{Dirac_1}
See for example \cite{Dirac:67} p.79 Eq.(61).

\bibitem{Dirac_5}
See for example \cite{Dirac:67} p.231 Eq.(29).

\bibitem{Dirac_braket_notation}
See \cite{Dirac:67} p.79 for a detailed description of bra and ket notation, in
  particular how to interpret $\psi(p)$ and $\ket{\psi(p)}$.

\bibitem{Fialkovsky_1}
I.~V. Fialkovsky, M.~Suleymanov, Xi~Wu, \textdaggerdbl{}. C.~X. Zhang, and
  M.~A. Zubkov.
\newblock {Hall conductivity as topological invariant in phase space}.
\newblock {\em Phys. Scripta}, 95(6):064003, 2020.
\newblock See Eq.(10).

\bibitem{Fialkovsky_2}
I.~V. Fialkovsky, M.~Suleymanov, Xi~Wu, \textdaggerdbl{}. C.~X. Zhang, and
  M.~A. Zubkov.
\newblock {Hall conductivity as topological invariant in phase space}.
\newblock {\em Phys. Scripta}, 95(6):064003, 2020.
\newblock See Eq.(5).

\bibitem{Fialkovsky_3}
I.~V. Fialkovsky, M.~Suleymanov, Xi~Wu, \textdaggerdbl{}. C.~X. Zhang, and
  M.~A. Zubkov.
\newblock {Hall conductivity as topological invariant in phase space}.
\newblock {\em Phys. Scripta}, 95(6):064003, 2020.
\newblock See Eq.(9).

\bibitem{Dirac_4}
See \cite{Dirac:67} p.97.

\end{thebibliography}
\end{document}